\documentclass[]{aa}
\usepackage[utf8]{inputenc}

\newcommand{\sepa}{s}
\newcommand{\kthree}{\vec{\underline{k}}}
\newcommand{\lthree}{\vec{\underline{l}}}
\newcommand{\jthree}{\vec{\underline{j}}}
\newcommand{\mthree}{\vec{\underline{m}}}
\newcommand{\kpthree}{\vec{\underline{k}^{\prime}}}
\newcommand{\lpthree}{\vec{\underline{l}^{\prime}}}

\newcommand{\xthree}{\vec{\underline{x}}}
\newcommand{\xpthree}{\vec{\underline{x}^{\prime}}}
\newcommand{\dd}{{\rm d}}
\newcommand{\emithree}{\epsilon}
\newcommand{\fovarea}{\mathcal{S}_{\mathcal{A}}}

\title{Towards mapping turbulence in the intra-cluster medium}
\subtitle{I. Sample variance in spatially-resolved X-ray line diagnostics}
	\author{Nicolas Clerc\inst{1} \and Edoardo Cucchetti\inst{1} \and Etienne Pointecouteau\inst{1} \and Philippe Peille\inst{2}}
   \institute{IRAP, Universit{\'e} de Toulouse, CNRS, UPS, CNES, Toulouse, France\\
              \email{nicolas.clerc@irap.omp.eu}
              \and
              CNES, 18 Avenue Edouard Belin, 31400 Toulouse, France\\
             }

   \date{Received April 2019; accepted July 2019.}

\abstract{
   X-ray observations of galaxy clusters provide insights on the nature of gaseous turbulent motions, their physical scales and on the fundamental processes they are related to. Spatially-resolved, high-resolution spectral measurements of X-ray emission lines provide diagnostics on the nature of turbulent motions in emitting atmospheres. Since they are acting on scales comparable to the size of the objects, the uncertainty on these physical parameters is limited by the number of observational measurements, through sample variance.
}{
We propose a different and complementary approach for the computation of sample variance to repeating numerical simulations (i.e.~Monte-Carlo sampling) by introducing new analytical developments for lines diagnosis.
}{
      We consider the model of a "turbulent gas cloud", consisting in isotropic and uniform turbulence described by a universal Kolmogorov power-spectrum with random amplitudes and phases in an optically thin medium. Following a simple prescription for the 4-term correlation of Fourier coefficients, we derive generic expressions for the sample mean and variance of line centroid shift, line broadening and projected velocity structure function. We perform a numerical validation based on Monte-Carlo simulations for two popular models of gas emissivity based on the $\beta$-model.
}{
    Generic expressions for the sample variance of line centroid shifts and broadening in arbitrary apertures are derived and match the simulations within their range of applicability. Generic expressions for the mean and variance of the structure function are provided and verified against simulations. An application to the \emph{Athena}/X-IFU (\emph{Advanced Telescope for High-ENergy Astrophysics}/X-ray Integral Field Unit) and \emph{XRISM}/Resolve (\emph{X-ray Imaging and Spectroscopy Mission}) instruments forecasts the potential of sensitive, spatially-resolved spectroscopy to probe the inertial range of turbulent velocity cascades in a Coma-like galaxy cluster.
}{
    The formulas provided are of generic relevance and can be implemented in forecasts for upcoming or current X-ray instrumentation and observing programs.
}

\begin{document}

\maketitle

\section{Introduction}

Galaxy clusters form by accretion of matter along filaments of the cosmic web, either continuously or episodically through major and minor merger events. The baryonic gas flowing along filamentary structures and falling into their deep gravitational wells acquires kinetic energy that is transformed into thermal energy, magnetic field amplification and cosmic ray acceleration in the intra-cluster medium, by a succession of shocks, large-scale motions and dissipation by turbulent processes \citep{ryu2008, zhuravleva2014, gaspari2014, miniati2015, gaspari2018, vazza2018}. Observational signatures of these phenomena are rare and difficult to obtain. The most promising and efficient diagnostics are issued from spectroscopic observations of the hot intra-cluster gas which permeates the entire volume of massive halos and emits copious amounts of X-ray light.

Focusing mainly on X-ray emission lines extracted along a single line-of-sight, \citet{inogamov2003} demonstrated that departures from Gaussian line shapes carry important indications on the nature of large-scale turbulence in the intra-cluster medium. The authors extended formalism to two-dimensional diagnostics by introducing the correlation function of the projected velocity field and calculating its scaling relative to fundamental parameters such as the turbulent injection and dissipation scales. Applying these findings to simple, but realistic configurations of the intra-cluster medium, \citet{zhuravleva2012} calculated exact expressions for emission line diagnostics such as centroid shift, broadening and two-dimensional correlation function. They evaluated the associated sampling uncertainty (also called 'sample variance' or 'sampling variance') by multiple Monte-Carlo realisations of the velocity field and showed that it can dominate the overall error budget in presence of large-scale turbulence. \citet{zuhone2016} could evaluate the contribution of sample variance and statistical errors for the well-defined observational case of the Coma cluster, thereby demonstrating the impact of the observational strategy on this source of uncertainty. Using numerically simulated clusters instead, \citet{roncarelli2018} performed end-to-end simulations to derive expected values of the indicators of turbulence issued from emission line measurements, postponing calculation of sample variance to a later stage by means of multiple realisations.

In this work we propose a formal approach to the problem of sample variance by considering the ideal case of an arbitrary, optically-thin gas distribution in which uniform and isotropic turbulent motions take place. This study is motivated by the intent to obtain reliable and fast estimates of this specific class of uncertainties and to identify key parameters impacting them. We will consider three popular diagnostics extracted from a continuum-free, isolated spectral line in the X-ray wavebands: line centroid shift (hereafter $C$), line broadening ($S$) and projected velocity structure function ($SF$). The latter is defined as the squared difference of projected velocities averaged among all points separated by a distance $\sepa$ on sky. Instrumental characteristics and signal-to-noise considerations related to, e.g.~the exposure time or the energy resolution, are deliberately excluded and addressed in a separate work (Cucchetti et al., in press, hereafter paper~II) The results of the present work are therefore instrument-independent to some extent.

Among these indicators, the structure function appears as a very promising diagnostic of turbulence since it takes advantage of spatially-resolved spectroscopic observations, as enabled by Integral Field Units. It is also the least intuitive of all three. Effects such as heterogenous sampling, non-stationarity, anisotropies, etc. reflect diversely in the modelling of $SF$. Interestingly, the structure function as a mathematical tool has received extensive interest in multiple fields of research involving spatial statistics, notably geostatistics and Earth science, under the name 'variogram' \citep[e.g.][]{matheron1965, matheron1973, cressie1985, haslett1997, armstrong1998, corstanje2008}. In the field of astronomy and astrophysics where its use is comparatively less widespread, it is involved in various works under both terms 'structure function' and 'variogram', to analyse data either in one dimension \citep[e.g.][for stellar variability]{roelens2017}, two \citep[e.g.][for Cosmic Microwave Background]{cayon2010} or three and more dimensions \citep[e.g.][for galaxy clustering]{martinez2010}.

We first introduce the derivation of the average and variance of the centroid shift and line broadening for measurements of an X-ray spectral line along a single line-of-sight, together with a numerical validation (Section~\ref{sect:1d_statistics}). Section~\ref{sect:2d_statistics} generalises these results to the case of three-dimensional turbulent fields and the extension of the line diagnostics to two dimensions, thereby treating the case of the structure function. We perform a numerical validation of these results in Section~\ref{sect:num_valid}. We finally discuss our results in Section~\ref{sect:discussion} and highlight two specific cases matching the future X-ray instruments \emph{XRISM}/Resolve \citep{ishisaki2018} and \emph{Athena}/X-IFU \citep{barret2016, barret2018}. We report most of the details on calculations and their discussions in the appendices, to which the reader can refer for more details.

The convention in our notations is as follows: the line-of-sight direction is denoted by $x$ and the plane-of-sky coordinate is $\vec \theta=(y,z)$. Units of these coordinates are physical (kpc) since in practice the angular distance at the redshift of the object is known. Three-dimensional vectors are underlined to differentiate them from two-dimensional vectors. The velocity $v$ (units km s$^{-1}$) is the component of the gas velocity projected along the line-of-sight. All following definitions and derivations (e.g.~turbulent velocity dispersion, power-spectrum, etc.) are relative to this line-of-sight component. We denote with brackets $\langle . \rangle$ the sample average of the estimators and random variables. We will decompose the velocity field in Fourier coefficients with discrete indices (involving the discrete summation sign $\sum_k$). The emissivity and geometrical shapes will be treated with their continuous Fourier transforms (involving the continuous summation sign $\int \dd k$). This distinction will often be purely formal: this is the choice made for clarifying the calculations. One- and three-dimensional Fourier transforms are indicated with a tilde (e.g.~$\widetilde{\rho}$), two-dimensional transforms with a hat (e.g.~$\widehat{\mathcal{W}}$).

%%%%%%%%%%%%%%%%%%%%%%%%%%%%%%%%%%%%%%%%%%%%%%%%%%%%%%%%%%%%%%%%%%%%
\section{Measured velocity dispersion along single line-of-sight}
	\label{sect:1d_statistics}

In this section we assume the velocity structure diagnostics are issued from the measurement of an emission line profile (e.g.~iron XXV at $\sim 6.5$~keV) along a given line of sight. Measuring a line profile is a complex task involving tools and methods developed under a certain set of observational conditions (binning of the spectra, level of noise, background subtraction, continuum subtraction, etc.) In order to illustrate our findings, we adopt a simplified approach where the analysis applies to a continuum- and background-subtracted spectrum with no source of noise nor uncertainty and no systematic (corresponding to a virtually infinite exposure time with a perfectly calibrated instrument) . Only one emission line is investigated, thus we neglect blending with neighbouring lines. Importantly, we do not provide a prescription for the measurement process itself (Gaussian or more complex fit, non-parametric fit, full spectrum fitting, etc.) Instead we model such measurement as a calculation of the zero-th, first and second moments of the line energy distribution $I_l(E)$ integrated along a line-of-sight $\vec\theta_0$, such that:
\begin{align*}
        F(\vec\theta_0) & =  \int I_{l}(\vec\theta_0;E) \dd E \\
        \delta E(\vec\theta_0)  & =  \left( F^{-1} \int E I_{l}(\vec\theta_0,E) \dd E \right) - E_0 \\
        \Sigma^2(\vec\theta_0) &  =   F^{-1} \int \left(E-\delta E -E_0\right)^2 I_{l}(\vec\theta_0;E)  \dd E
\end{align*}
where $\delta E$ and $\Sigma$ are the observed centroid shift (relative to reference energy $E_0$) and width (or broadening) of the line. $F$ is a normalization factor, namely the flux in the line. We introduce the gas velocity field along the line-of-sight fixed by $\vec \theta_0$ with $v(x) \equiv v(x,\vec\theta_0)$. We define $C = c \delta E/E_0$ and $S^2 = c^2 \Sigma^2/E_0^2$ the observed centroid shift and width in velocity space.

    \subsection{Emission along the line of sight}

At microscopic level (i.e. below the turbulent dissipation scale in the medium), emission is assumed to follow a thermally broadened line profile. We neglect any additional broadening such as natural (Lorentzian) and assign each ion a rest-frame Gaussian emission profile. Assuming purely collisional origin of the emission line, the amplitude of the line is assumed to scale with emissivity $\epsilon(\vec{r}) \propto n_{Fe} n_e = n_e^2$. The coefficient of proportionality may depend on the local property of the medium (metallicity, temperature, etc.)

In the following we assume that the turbulent dissipation scale $L_{diss}$ is large enough for each volume of size $(L_{diss})^3$ to contain a significantly large number of line emitters: it is practically always fulfilled, even for the tenuous intra-cluster medium with typical density $10^{-3}$\,cm$^{-3}$ and kpc-scale injection scales. Accounting for the Doppler shift in energy $v(x)/c$ and emissivity $\epsilon(x)$, we therefore model the total emission at each point by:
\begin{equation}
\label{eq:linedef}
    \frac{\dd I_{l}(E)}{\dd x} = \frac{\epsilon(x) c}{E_0\sigma_{th}(x) \sqrt{2 \pi}} \exp \left( -\frac{1}{2} \left[\frac{E-E_0 (1+v(x)/c)}{E_0 \sigma_{th}(x)/c}\right]^2 \right)
\end{equation}

Assuming an optically thin medium and reordering the summation over velocities we can write:
\begin{align}
    F(\vec\theta_0) & =  \int \epsilon(x) \dd x \nonumber \\
    C(\vec\theta_0) & =  F^{-1} \int \epsilon(x) v(x) \dd x \nonumber \\
    S^2(\vec\theta_0) & =  F^{-1} \int \epsilon(x) \sigma_{th}^2(x) \dd x  + \frac{F^{-2}}{2}  \iint G(x,x^{\prime}) \dd x \dd x^{\prime}  \label{eq:dispersion}\\
    G(x,x^{\prime}) & =  \epsilon(x) \epsilon(x^{\prime}) \left[v(x)-v(x^{\prime})\right]^2\nonumber
\end{align}

These integrals extend along the line-of-sight indexed by $x$, that we assume to range in a large interval $[-L,L]$. In principle $v(x)$ encapsulates the effect of hydrodynamical motions (turbulence) and bulk motions of the gas. 
Without loss of generality, we assume bulk motions over a small area are known and subtracted from the measurements; we therefore set its contribution to zero.

    \subsection{Turbulent velocity field}

We describe the turbulent velocity field $v$ by its Fourier series expansion, with positive and negative values of $k$:
\begin{equation*}
    v(x) = \sum_{k} V_k \exp{\left( \frac{i 2 \pi k x}{L} \right)}
\end{equation*}

The coefficients $V_k$ are complex random variables, defined as $V_k = V_{-k}^* = |V_k| e^{i \psi_k}$. Here $\psi_k$ is a random phase and $|V_k|$ a random modulus, supposed independent from each other. In the following, we define for convenience $\omega = 2\pi/L$. We note that $\langle v \rangle = 0$, leading to $V_0=0$. Averaging over multiple random realisations provides:
\begin{equation}
\label{eq:coeffaverage}
    \langle V_j V_k \rangle = \delta_{j,-k} P(k) 
\end{equation}
where $P(k) = P(-k) = \langle |V_k|^2 \rangle$ is the power-spectrum of the turbulent velocity field and $\delta_{ij}=1$ if $i=j$, 0 if $i \neq j$.

Our hypothesis of uniform turbulence implies that the normalization of the power spectrum matches the square of the turbulent velocity dispersion $\sigma_{turb}$ at any given point $x$. It is defined through the calculation of the second moment $\sigma_{turb} = \sqrt{\langle v^2 \rangle}$, where averaging occurs over random phases and moduli. Therefore (taking e.g. $x=0$) $\sigma_{turb}^2 = \sum_k P(k)$. This definition does not involve the profile of the emissivity, in contrast to e.g.~\citet{zuhone2016}.

One simple assumption for the distribution of moduli \citep[e.g.][]{inogamov2003} consists in non-random coefficients, leaving only phases as random.

Another popular and physically motivated assumption (although not systematically required in the rest of this paper) is the Rayleigh distribution \citep[e.g.][]{zuhone2016}:
\begin{equation*}
|V_k| \sim \mathcal{P}(\nu) \dd \nu = \frac{2 \nu}{P(k)} e^{-\nu^2/P(k)} \dd \nu
\end{equation*}
which reflects that the turbulent velocity is a Gaussian random field. Introducing $R_k = P(k)^2 - \mathrm{Var}(|V_k|^2)$, we obtain under the assumption of Rayleigh-distributed moduli: $R_k=0$ for all $k$.

    \subsection{Statistics of the centroid shift and line broadening}

Calculations in App.~\ref{app:1d_calc} provide the following formulas for the centroid shift:
\begin{equation}\label{eq:c1d_mean}
	\langle C \rangle = 0
\end{equation}
and its variance:
\begin{equation}\label{eq:c1d_var}
    \mathrm{Var}(C) = \langle C^2 \rangle = F^{-2} \sum_{k} P_{\epsilon}(k) P(k)
\end{equation}
This expression is identical to Eq.~(A9) in \citet{zhuravleva2012}, since $P_{\epsilon}(k) = |\widetilde{\epsilon}(k)|^2$ is the Fourier power spectrum of the (unnormalised, one-dimensional) emissivity $\epsilon$.

As for the line broadening, we obtain in App.~\ref{app:1d_calc}:
\begin{equation}\label{eq:s1d_mean}
    \langle S^2 \rangle = \overline{\sigma_{th}^2} + \sigma_{turb}^2 - \langle C^2 \rangle
\end{equation}
The horizontal bar denotes averaging of the thermal component along the line-of-sight.
Again this expression is similar to Eq.~B4 in \citet{zhuravleva2012}. Also interesting is the contribution of the last term, indicating that averaging many (independent) measurements of broadening measurements generally provides a biased estimate of the (thermal+turbulent) broadening. The bias is zero only in cases where the turbulent power-spectrum and the emissivity power spectrum act on distinct spatial scales.
Finally, the variance can be written:
\begin{multline}\label{eq:s1d_var}
\mathrm{Var}(S^2) = 2 \sum_{j,k} P(k) P(j) \left| \frac{\widetilde{\epsilon}(j+k)}{F}-\frac{\widetilde{\epsilon}(j) \widetilde{\epsilon}(k)}{F^2} \right|^2 \\
- \sum_k R_k \times \left\{ \left| \frac{\widetilde{\epsilon}(2k)}{F} - \frac{\widetilde{\epsilon}(k)^2}{F^2} \right|^2 + 2 \left[ 1-\frac{P_{\epsilon}(k)}{F^2} \right]^2 \right\}
\end{multline}

Conveniently, if the moduli $|V_k|$ are Rayleigh-distributed, the term under the second $k$-sum vanishes.

	\subsection{Numerical validation}

\begin{figure}
    \centering
	    \includegraphics[width=\linewidth]{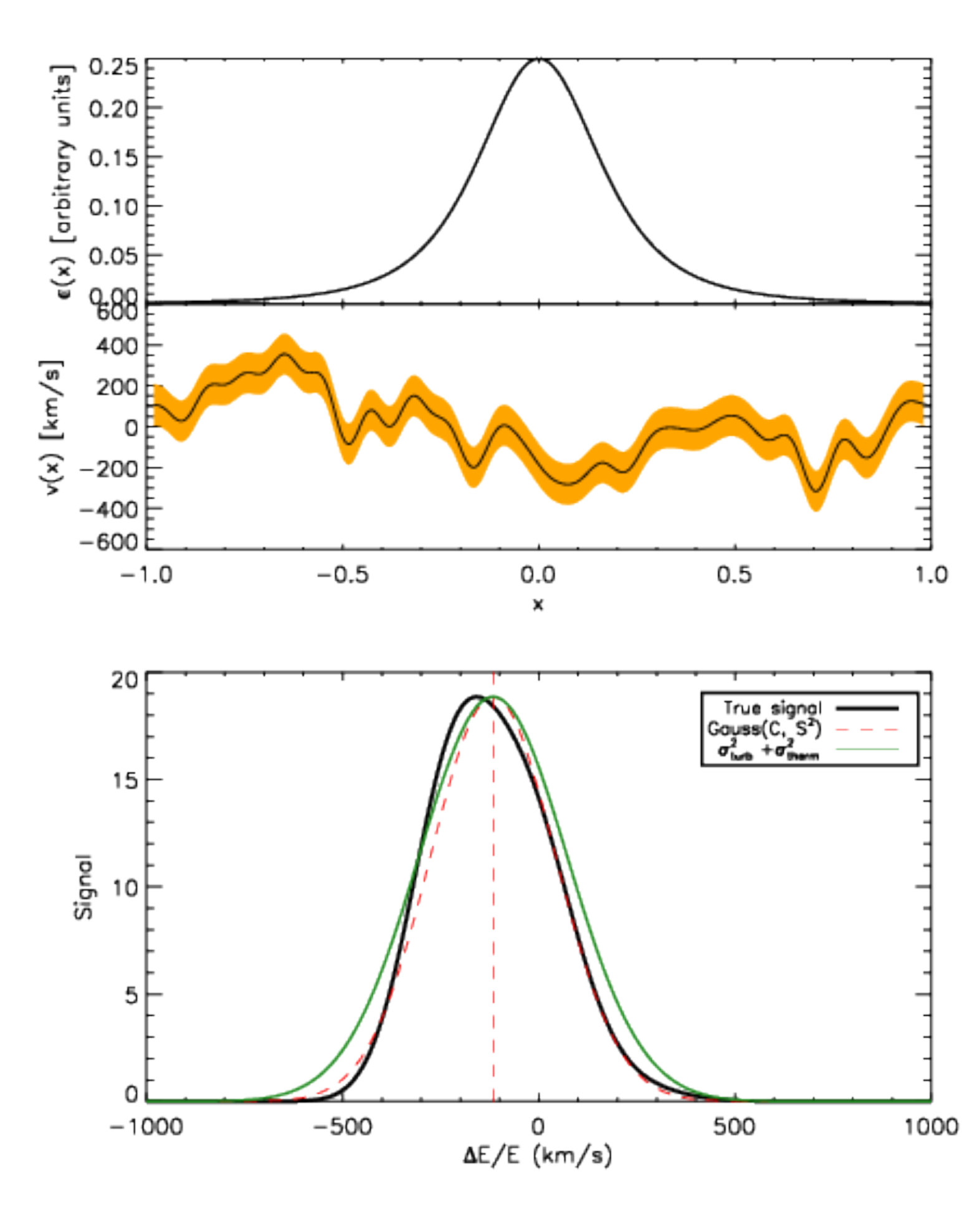}
    \caption{One realisation of a unidimensional turbulent velocity field (middle panel) along the spatial axis $x$ (in arbitrary units) with parameters $\alpha=1/3$, $k_{min}=1$ ($L_{inj} = 2$), $k_{max}=20$ ($L_{diss}=0.1$), $\sigma_{turb} = 160$~km\,s$^{-1}$ and $\sigma_{th}=100$~km\,s$^{-1}$ (materialized by the yellow shading). The emissivity profile (top panel) of the gas corresponds to a $\beta$ density model with core-radius $r_c=0.2$ and at a distance $\theta=0.2$ from the centre. The lower panel shows the resulting line profile as a thick black line. The "best-fit" Gaussian centred on $C$ (vertical line) of width $S$ is shown as a dashed red line. The green thin curve shows the Gaussian centred on the line centroid of width the geometrical mean of the thermal and turbulent broadening.}
    \label{fig:line_single_los}
\end{figure}

A verification of the equations previously derived is a relatively quick task with modern computing resources. We considered a power spectrum in the form $P(k) \propto 1/k^{2 \alpha+1}$ in the inertial range $[k_{min},k_{max}]$ and zero outside of it. The slope is $\alpha=1/3$. We simulated the line profile resulting from the projection through a (isothermal, isometallic) $\beta$-model density profile \citep{cavaliere1978} with $\beta=2/3$, core-radius $r_c$ and distance $\theta$ to the cluster centre:
\[
\epsilon(x) \propto \left(1+\frac{x^2+\theta^2}{r_c^2}\right)^{-3\beta}
\]
where $L=2$, $r_c=0.2$ and $\theta=0.2$. The moduli $V_k$ may be constant (non-random) or follow a Rayleigh distribution. The Fourier coefficients of the emissivity are given by \citet{zuhone2016} (see also App.~\ref{app:fourier_eta}): $\widetilde{\epsilon}(k)/F = \exp (- \omega |k| c ) (1 + \omega |k| c)$ with $c^2 =r_c^2+\theta^2$. An example of such realisation is shown in Fig.~\ref{fig:line_single_los}. Only phases are random in this illustration. This configuration and random realisation are specifically chosen to highlight the possible discrepancy between the value of the broadening $S^2$ and the simple estimate $\sigma_{turb}^2+\overline{\sigma_{th}^2}$. Indeed, purely by chance, most of the points where velocity is high are located in low-emissivity regions, therefore their contribution to line broadening is weak.

\begin{figure*}
    \centering
    \begin{tabular}{cc}
	    \includegraphics[width=0.45\linewidth]{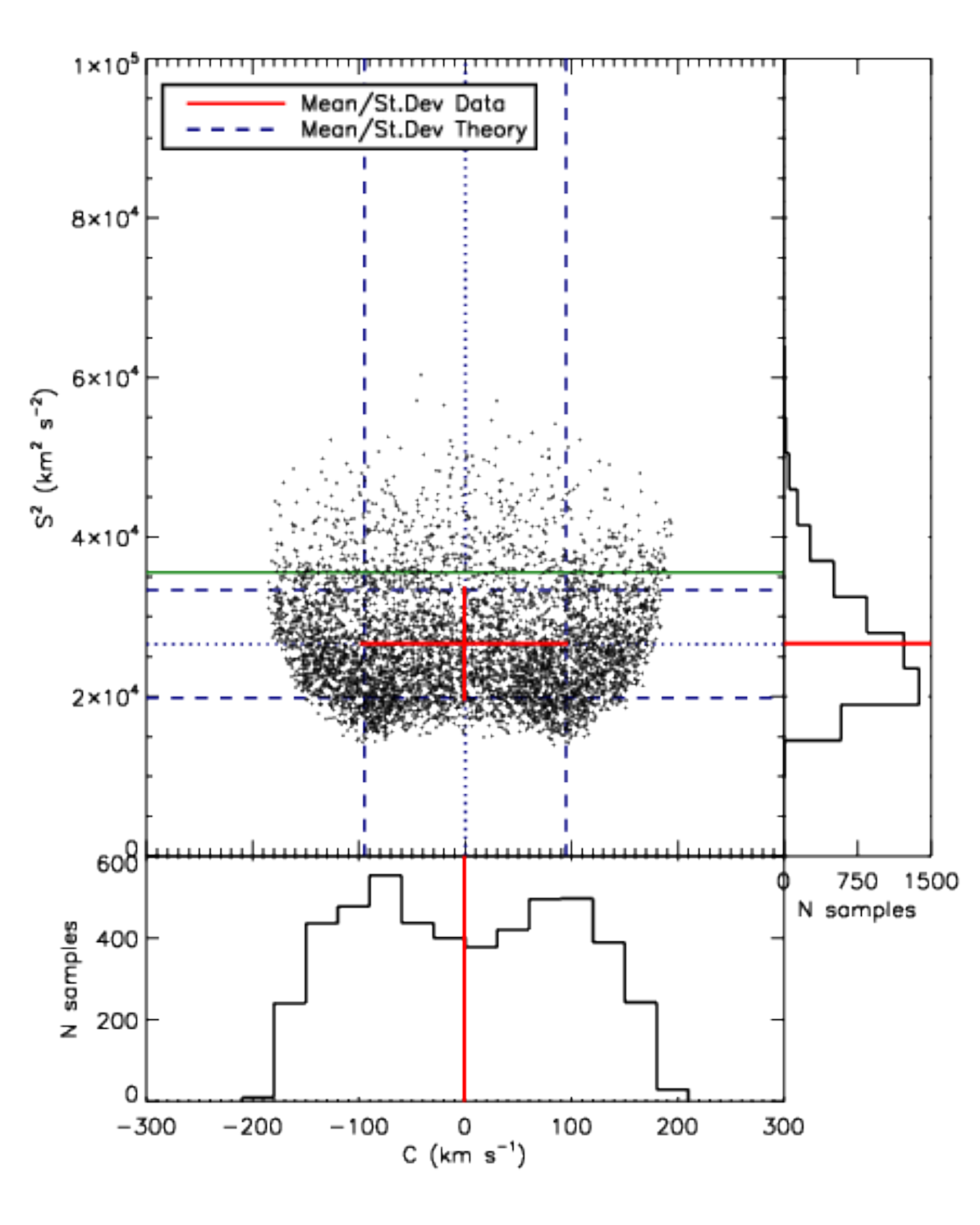}	& \includegraphics[width=0.45\linewidth]{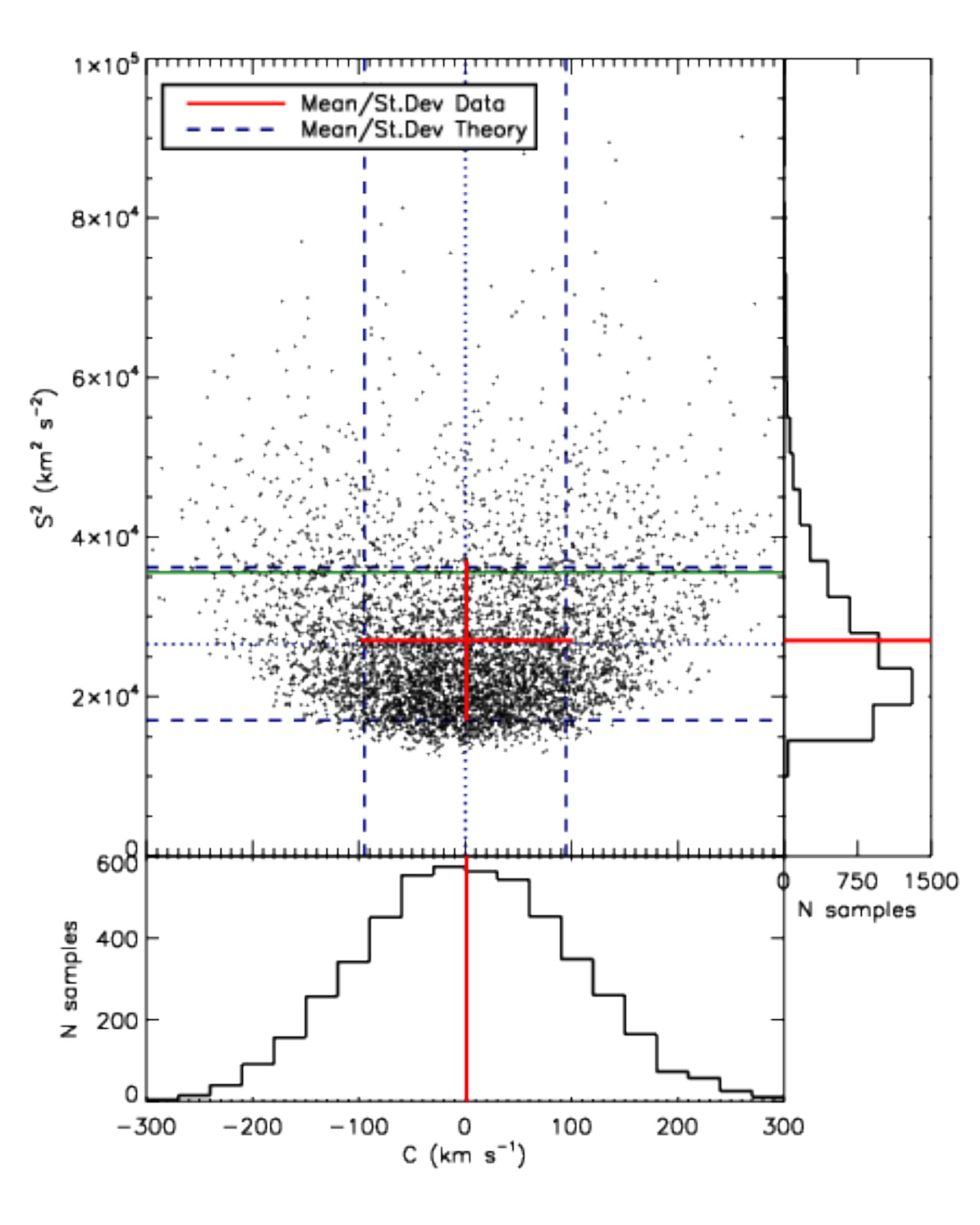}\\
    \end{tabular}
    \caption{The line centroid $C$ versus line width (squared) $S^2$ of an emission line along a single line-of-sight and 5000 realisations of a 1-dimensional turbulent velocity field ($\sigma_{turb}=160$\,km/s, $k_{min}=1$, $k_{max}=20$, $\sigma_{th}=100$\,km/s). Left panel assumes only random phases while right panel also includes randomly distributed moduli (Rayleigh distribution). Points show measurements, red cross is the measured mean and standard deviations along both axes. The blue lines represent the results obtained from analytical calculations (Eq.~\ref{eq:c1d_mean}, \ref{eq:c1d_var}, \ref{eq:s1d_mean} and~\ref{eq:s1d_var}). The plain green line shows the location of the geometric mean of the turbulent and thermal dispersions: the presence of turbulent motions on scales comparable to that of the cluster makes such estimate a biased one.}
    \label{fig:single_los}
\end{figure*}

Figure~\ref{fig:single_los} shows the excellent agreement between the theoretical and simulated quantities after 5000 random realisations of the velocity field. It is important to notice the strong non-gaussianity of $S^2$ and $C$ in general. The configuration chosen for this simulation clearly illustrates that $\langle S^2 \rangle < \overline{\sigma_{th}^2}+\sigma_{turb}^2$. This is a consequence of the injection scale ($L_{inj} \sim L$) being much larger than the typical cluster characteristic scale (here $r_c=L/10$), leading to non-gaussian line shapes \citep{inogamov2003}.

%%%%%%%%%%%%%%%%%%%%%%%%%%%%%%%%%%%%%%%%%%%%%%%%%%%%%%%%%%%%%%%%%
%%%%%%%%%%%%%%%%%%%%%%%%%%%%%%%%%%%%%%%%%%%%%%%%%%%%%%%%%%%%%%%%%
%%%%%%%%%%%%%%%%%%%%%%%%%%%%%%%%%%%%%%%%%%%%%%%%%%%%%%%%%%%%%%%%%
%%%%%%%%%%%%%%%%%%%%%%%%%%%%%%%%%%%%%%%%%%%%%%%%%%%%%%%%%%%%%%%%%
\section{Two-dimensional characterisation of the velocity field}
	\label{sect:2d_statistics}

Observations and diagnostics of the intra-cluster medium rarely rely on single line-of-sight measurements. Instead, due to instrumental resolution limits and signal-to-noise considerations, line centroid shifts and broadening are measured from spectra collected over well-defined 2-dimensional regions, sometimes denoted as "bins" or "pixels". A popular diagnostic tool in the field of astrophysical turbulence \citep[e.g.][]{lis1998, esquivel2007, anorve2019} is the two-dimensional structure function, loosely speaking a 2-d correlation function analysis of the line centroid shift map. Obtaining analytical expressions of the sample variance of these estimators requires the formalism above to be extended and to account for the 3-dimensional structure of the velocity field and the emissivity field.
A supplementary difficulty one has to face is the non-stationarity of the projected velocity field: even though the 3-d velocity field is homogeneous (stationary) in the medium, spatial variations of the emissivity in general break this property.

    \subsection{The tridimensional velocity field\label{sect:3dfield}}

Similarly as in previous section, we define the centroid shift and line width measured over a spectral line being a sum over all individual line-of-sights selected within a region. We introduce the window function $\mathcal{W}(\vec{\theta}) = \mathcal{W}(y,z)$ equals to 1 (one) for selected line-of-sights and zero elsewhere. The measured spectral parameters of the line write:
\begin{align*}
        F(\mathcal{W})  & =  \int I_{l}^{\mathcal{W}}(E) \dd E \\
        \delta E(\mathcal{W})  & =  \left( F^{-1} \int E \times I_{l}^{\mathcal{W}}(E) \dd \vec{\theta} \dd E \right) - E_0 \\
        \Sigma^2(\mathcal{W})  &  =   F^{-1} \int \left(E-\delta E -E_0\right)^2 I_{l}^{\mathcal{W}}(E)  \dd E
\end{align*}
by defining:
\[
I_{l}^{\mathcal{W}}(E) = \int I_{l}(\vec{\theta}, E) \mathcal{W}(\vec{\theta}) \dd \vec{\theta}
\]
All results from previous sections are obviously recovered with $\mathcal{W}(\vec{\theta}) = \delta(\vec{\theta}-\vec{\theta}_0)$.

Introducing $v(\xthree) \equiv v(x,y,z)$ the line-of-sight component of the velocity, and operating the following substitutions in Eq.~\ref{eq:linedef}: $\epsilon(x) \rightarrow \emithree (\xthree)$, $v(x) \rightarrow v(\xthree)$, we can rewrite the observed "aperture" flux, centroid and velocity dispersion as:
\begin{align*}
    F_{\mathcal{W}} & =  \int \dd \vec \theta \, \mathcal{W}(\vec \theta) \int \dd x \, \emithree(x, \vec{\theta}) =  \int \mathcal{W} F \dd \vec\theta \\ 
    C_{\mathcal{W}} & =  F_{\mathcal{W}}^{-1} \int \dd \vec \theta \, \mathcal{W}(\vec \theta) \int  \emithree(x, \vec{\theta}) v(x, \vec{\theta}) \dd x \\
    	& = F_{\mathcal{W}}^{-1} \int \mathcal{W} F C \dd \vec\theta  \label{eq:centroid3d}\\
    S^2_{\mathcal{W}} & = F_{\mathcal{W}}^{-1} \int \dd \xthree \, \sigma_{th}^2(\xthree) \mathcal{W}(\vec \theta) \emithree(\xthree)  + \frac{ F_{\mathcal{W}}^{-2}}{2} \int \dd \xthree \dd \xpthree  \, G(\xthree, \xpthree) \\
    	& = F_{\mathcal{W}}^{-1} \int  \mathcal{W} F \left(S^2 + C^2 \right) \dd \vec \theta - C_{\mathcal{W}}^2 \\
    G(\xthree, \xpthree) & =  \mathcal{W}(\vec \theta) \emithree(\xthree) \mathcal{W}(\vec \theta^{\prime}) \emithree(\xpthree) \left[v(\xthree)-v(\xpthree)\right]^2
\end{align*}

Integrations over $x$ range over an arbitrary large interval $[-L,L]$ and over all possible values of the plane-of-sky position $\vec \theta$.
The velocity can be written in terms of its Fourier decomposition with $\kthree = (k_x,k_y,k_z) = (k_x, \vec \xi)$:
\begin{equation*}
v(\xthree) = \sum_{\kthree} V_{\kthree} \exp \left( i \omega \kthree \cdot \xthree \right)
\end{equation*}
and we note $P_{3D}(\kthree) = \langle | V_{\kthree} |^2 \rangle = P_{3D}(k)$. We have the following relations:
\begin{align*}
    V_{-\kthree} & =  V_{\kthree}^* \\
    \langle V_{\kthree} V_{\kpthree} \rangle & =  \delta_{\kthree; -\kpthree} P_{3D}(k)
\end{align*}
Similarly as in the 1-dimensional case (Sect.~\ref{sect:1d_statistics}), we introduce $R_{\kthree} = P_{3D}(\kthree)^2 - Var\left(|V_{\kthree}|^2\right)$, such that $R_{\kthree}=0$ for Rayleigh-distributed moduli. A minimal assumption on the 4-term bracket $\langle V_{\jthree} V_{\kthree} V_{\lthree} V_{\mthree} \rangle$ is necessary and our ansatz is explicitly provided in App.~\ref{app:sf_covariance}.

%%%%%%%%%%%%%%%%%%%%%%%%%%%%%%%%%%%%%%%%%%%%%%%%%%%%%%%%%%%%%%%%%

    \subsection{Statistics of the aperture line centroid}
We find that the average of the velocity shift measurements over several realisations is 0:
\begin{equation}\label{eq:avg_centroidshift3d}
	\langle C_{\mathcal{W}} \rangle=0
\end{equation}

The calculations are actually very similar to the one-dimensional case and we refer to App.~\ref{app:1d_calc} for details. By using the Fourier decomposition of the velocity field, the variance in centroid shifts measurements reads:
\begin{equation}\label{eq:var_centroidshift3d}
    \langle C_{\mathcal{W}}^2 \rangle = \frac{1}{F_{\mathcal{W}}^2} \sum_{\kthree} P_{3D}(k) |c_{\emithree.\mathcal{W}}(\kthree)|^2 
\end{equation}

Here $c_{\emithree.\mathcal{W}}(\kthree)$ is the Fourier coefficient of the product $\emithree(\xthree) \mathcal{W}(y,z)$.
This expression differs from Eq.~(E7) in \citet{zhuravleva2012} because we do not assume $\emithree$ being independent of the line-of-sight. If instead $\emithree(x,y,z) = \epsilon(x)$ in the domain of $\mathcal{W} \neq 0$, then $c_{\emithree . \mathcal{W}} (\kthree) = \widetilde{\epsilon}(k_x)\widehat{\mathcal{W}}(\vec \xi)$; therefore we can rewrite our finding under the factorised form:
\[
    \langle C_{\mathcal{W}}^2 \rangle = \frac{1}{F_{\mathcal{W}}^2} \sum_{\kthree} P_{3D}(k) P_{\mathcal{W}}(\vec \xi) P_{\epsilon}(k_x)
\]
with $P_{\mathcal{W}}$ being the power-spectrum of the window function $\mathcal{W}$. This expression is applicable considering for instance small, pencil-beam, window functions or, equally interesting, narrow annular window functions, if the emissivity shows a circular symmetry.
We provide in App.~\ref{app:fourier_eta} a detailed calculation of the function $c_{\emithree.\mathcal{W}}$ for the case of the isothermal, isometallicity $\beta$-model gas density.

%%%%%%%%%%%%%%%%%%%%%%%%%%%%%%%%%%%%%%%%%%%%%%%%%%%%%%%%%%%%%%%%%
    \subsection{Statistics of the aperture line broadening}

       % Expectation value 
The calculation of the average of $S_{\mathcal{W}}^2$ over multiple realisations of the turbulent field follows similar steps as in the 1-dimensional case presented before and we find:

\begin{equation*}
	\langle S_{\mathcal{W}}^2 \rangle = \overline{\sigma_{th}^2} + \sigma_{turb}^2 - F_{\mathcal{W}}^{-2} \sum_{\kthree} P_{3D}(k) |c_{\emithree.\mathcal{W}}(\kthree)|^2
\end{equation*}

The bar indicates the average of the thermal broadening over the cluster volume defined by $\mathcal{W}$. With these notations the relation found in the 1-d case still holds:
\begin{equation}\label{eq:avg_broadening3d}
\langle S_{\mathcal{W}}^2 \rangle + \langle C_{\mathcal{W}}^2 \rangle = \overline{\sigma_{th}^2} + \sigma_{turb}^2
\end{equation}
        
	% Variance
Finally the variance of the line broadening writes:
\begin{multline}\label{eq:var_broadening3d}
\mathrm{Var}(S_{\mathcal{W}}^2) = \\
2 \sum_{\kthree, \kpthree} P_{3D}(\kthree) P_{3D}(\kpthree) \left| \frac{c_{\emithree .\mathcal{W}} (\kthree + \kpthree) }{F_{\mathcal{W}}} - \frac{ c_{\emithree . \mathcal{W}}(\kthree) c_{\emithree . \mathcal{W}}(\kpthree) }{F_{\mathcal{W}}^2} \right|^2 \\
 - \sum_{\kthree} R_{\kthree} \times \left\{ \left| \frac{c_{\emithree . \mathcal{W}}(2 \kthree)}{F_{\mathcal{W}}} - \frac{c_{\emithree . \mathcal{W}}(\kthree)^2}{F_{\mathcal{W}}^2} \right|^2 \right. \\
 \left. + 2 \left[ 1- \frac{|c_{\emithree . \mathcal{W}}(\kthree)|^2}{F_{\mathcal{W}}^2} \right]^2\right\}
\end{multline}
which reduces to the first term only in case of Rayleigh-distributed moduli.

%%%%%%%%%%%%%%%%%%%%%%%%%%%%%%%%%%%%%%%%%%%%%%%%%%%%%%%%%%%%%%%%%
    \subsection{Statistics of the structure function}

We define the structure function as the integral:
\begin{equation*}
    SF(\sepa) = \frac{1}{N_p(\sepa)} \int_{d(\mathcal{W},\mathcal{W}^{\prime})=\sepa} \left| C_{\mathcal{W}^{\prime}} - C_{\mathcal{W}} \right|^2 \dd N_p
\end{equation*}

This expression simply describes an average over all pairs of regions (called 'bins' or 'pixels') $(\mathcal{W},\mathcal{W}^{\prime})$ separated by a distance\footnote{There is quite a latitude in choosing the definition of distance, either considering geometrical centres of each region or flux-weighted barycentres, etc.} $d(\mathcal{W},\mathcal{W}^{\prime})=\sepa$. Here $N_p(\sepa)$ is the number of such pairs of regions. For instance, considering single line-of-sight measurements and the Euclidian distance between two points on sky, i.e.~$\mathcal{W}(\vec\theta) \equiv \delta(\vec\theta-\vec\theta_0)$, we recover the standard formulation \citep[e.g.][]{zuhone2016}:
\[
SF (\sepa) = \frac{1}{N_p(\sepa)} \int_{\vec\theta_{0}, |\vec r|=\sepa} \left| C(\vec\theta_0 + \vec r) - C(\vec\theta_0) \right|^2 \dd N_p
\]
The integration runs over an arbitrary large, but bounded region of sky $\mathcal{A}$ of total area $\fovarea$. In the following we consider $\mathcal{A}(\vec \theta)$ as a function taking value 1 in the analysis region and 0 outside. There, $N_p(\sepa)$ needs to be interpreted as the integral $\int \dd N_p$ for all $(\vec \theta_0, |\vec r|=\sepa$).

Such defined, $SF(\sepa)$ is a random variable that depends on the particular realisation of the velocity field and we can therefore compute its mean and variance across several realisations, hereafter called $\mathrm{sf}(\sepa)$ and $\sigma^2_\mathrm{sf}(\sepa)$.

% % % % % % % % % % % %
    \subsubsection{Expected value $\mathrm{sf}(\sepa)$}

Under the assumption that finite-size (i.e~border) effects are negligible, we find for the most general expression of the emissivity field (see App.~\ref{app:sf_general}) :
\begin{align}
\label{eq:sf_average}
\mathrm{sf}(\sepa)  & = \nonumber \\
& 2 K \sum_{\kthree} P_{3D}(k)  \int \dd \vec \xi^{\prime} P_{\rho}(k_x,\vec \xi^{\prime}) \left[ 1 - J_0\left(\left| \vec \xi + \vec \xi^{\prime} \right| \omega \sepa \right) \right] \nonumber \\
& = 2 \left(\frac{\omega}{2\pi}\right)^2 \int \left[ 1 - J_0\left( \omega \left| \vec \xi \right| \sepa \right)\right] P_{2D}(\vec \xi)  \dd \vec \xi 
\end{align}
with $\rho(x,y,z) = \emithree(x,\vec \theta)/F(\vec \theta)$, $K=\omega^2/(4 \pi^2 \fovarea)$, $J_0$ being the Bessel function of the first kind and order 0. The function $P_{2D}$ is the 2D power spectrum of the centroid shift map, which expressions are properly defined and derived in App.~\ref{app:p2d_general}. One must be careful that the power-spectrum $P_{\rho}$ involved in these expressions is that of the normalised emissivity field $\rho$, i.e.~the 3-d emissivity $\emithree$ divided by the "flux map" $F(\vec \theta)$. It is strongly dependent on the choice of analysis domain $\mathcal{A}$.

In the special case where the two-dimensional spectrum is isotropic this expression takes the following form:
\begin{equation*}
\mathrm{sf}(\sepa) \simeq 4 \pi \left(\frac{\omega}{2\pi}\right)^2 \int_0^{+\infty} P_{2D}(\xi) \left(1-J_0(\omega \xi \sepa ) \right) \xi \dd \xi
\end{equation*}

The latter equation resembles \citet[][]{zuhone2016}, their eq.~29. However this result implicitly includes the shape of the domain of analysis through $P_{2D}$, which effectively acts as a high-pass spatial filter. We recall here the assumptions leading to this result: i)~centroid shift measurements are performed along individual line-of-sights, ii)~isotropy of the two-dimensional power-spectrum $P_{2D}$, iii)~the averaging domain allows all possible orientations of the pair vector $\vec r$ and iv)~the sum over modes $\vec \xi$ can be written as an integral.

We provide in App.~\ref{app:sf_general} a generic formula to correct the above expressions for border effects. This involves calculation of the number of pairs enclosed within the analysis domain and those crossing its frontier, both dependent on the separation length $\sepa$ and the exact shape of the domain. These are easily calculated for a circular domain of analysis of radius $R$ and we provide the equations in the appendix.

We also provide in App.~\ref{app:filter_field} a prescription to account for pixelization of the centroid map with pixels of arbitrary size and shapes. Provided such pixels are small with respect to the typical scales of the surface brightness fluctuations, a correction is obtained by multiplying $P_{2D}$ by the two-dimensional power-spectrum of the pixel shape $P_{\ell}$. As shown in appendix, this prescription should not be used in combination with the correction formula for border effects, especially if pixels are of sizeable length compared to the analysis domain. We do not provide here a complete analytical formulation accounting simultaneously for border effects and pixelization; it may be more advantageous in such case to numerically estimate the average structure function from its primary definition involving the $C_{\mathcal{W}}$'s.

Nevertheless, an exact solution for $\mathrm{sf}(\sepa)$ is obtained in case of a stationary 2-dimensional velocity field -- e.g.~if $\emithree(x,y,z)=\epsilon(x)$ -- by replacing $P_{2D}$ by $P_{2D}^{\infty}$ in Eq.~\ref{eq:sf_average}, that is the power spectrum computed in the limit of an infinitely extended analysis domain (see~App.~\ref{app:p2d_general} for details.) Such formulation then matches exactly that proposed by \citet{zuhone2016}. The above prescription for pixel binning then also becomes exact and raises no issue due to a finite region of analysis. These properties are used in App.~\ref{app:sf_general} and~\ref{app:filter_field} to validate our correction formulas and to stress their limitations.

% % % % % % % % % % % %
    \subsubsection{Variance $\sigma^2_\mathrm{sf}(\sepa)$}
    
A full calculation of the covariance
\[
\Sigma_{ij} = \mathrm{Cov}\left(SF(\sepa_i), SF(\sepa_j)\right)
\]
between structure functions measured at different scales is provided in App.~\ref{app:sf_covariance} under the assumption of negligible finite-size effects. The complete formula is given in Eq.~\ref{eq:covar_sf} and involves integrals of the Fourier transform $\widetilde{\rho}$ of the normalised emissivity field. 
Because of the relative position of the analysis region and the emissivity distribution, it is in general not possible to factor their respective contributions in the expression of the variance. However, we can study a simpler, practical case where the emissivity is independent on the line-of-sight direction within the given analysis field-of-view. This is for instance the case for an observation pointing at the outskirts of a nearby galaxy cluster or towards the core of a "flat" galaxy cluster (e.g.~Coma). This particular case writes~$\rho(x,\vec \theta) = \epsilon(x)/F$. This leads to decoupling the calculation of "geometrical" terms (i.e. the shape and location of the instrumental field-of-view) and "fluctuation" terms (the coupling between the cluster emissivity and the turbulent velocity spectrum).
In App.~\ref{eq:covar_sf} we obtain the following simple formula, under the supplementary hypothesis of a very large analysis region, i.e.~for $\fovarea^{1/2} \gg (\sepa, L_{inj},...$):
\begin{multline}\label{eq:varsf_xbeta_fovlarge}
\Sigma_{ij} \simeq 16 \pi \left(\frac{\omega}{2\pi}\right)^2  \int \left[ \frac{1}{\fovarea} P_{2D}^{\infty}(\xi) ^2 - \left(\frac{\omega}{2\pi}\right)^2 Q_{2D}^{\infty}(\xi)^2\right] \\
\times \left(1-J_0(\omega \xi \sepa_i)\right) \left(1-J_0(\omega \xi \sepa_j)\right) \xi \dd \xi
\end{multline}
where $Q_{2D}^{\infty} = 0$ for Rayleigh-distributed moduli. The diagonal term of this quantity is then: $\Sigma_{ii} = \sigma^2_\mathrm{sf}(\sepa)$. Similarly as for the calculation of the average $\mathrm{sf}(\sepa)$ in case of pixelized data, one has to multiply $P^{\infty}_{2D}$ by the power-spectrum of the elementary pixel shape.
Eq.~\ref{eq:varsf_xbeta_fovlarge} in the Rayleigh regime is the expression we will validate in the next section, keeping in mind the series of assumption made to obtain this simple formulation.

%%%%%%%%%%%%%%%%%%%%%%%%%%%%%%%%%%%%%%%%%%%%%%%%%%%%%%%%%%%%%%%%%
	\section{Numerical validation}
	\label{sect:num_valid}

We performed a set of numerical experiments to validate the equations derived previously. This requires generation of multiple velocity boxes in three dimensions. Given the high computational demand, only a selected set of cases are treated.

% % % % %

	\subsection{Dataset of velocity cubes}
We created a series of velocity boxes with characteristics indicated in Table~\ref{table:simu_char}. The smooth and continuous velocity power-spectrum takes a form similar to that of \citet{zuhone2016}, namely:
\begin{equation}
\label{eq:def_pkzuhone}
P_{3D}(\kthree) = C_n e^{-(k/k_{diss})^2} k^{\alpha} e^{-(k_{inj}/k)^2}
\end{equation}
with $C_n$ a normalization constant having units such that $\int P_{3D}(\kthree) \dd \kthree = \sigma_{turb}^2$ and $\alpha=-11/3$ typical of a Kolmogorov turbulence spectrum \citep{kolmogorov1941}. Scales $k_{diss}=1/L_{diss}$ and $k_{inj}=1/L_{inj}$ represent the dissipation and injection frequencies, respectively. Moduli of the Fourier coefficients are drawn from a Rayleigh distribution.

\begin{table*}
\centering
		\begin{tabular}{cccccccc}
\multicolumn{2}{c}{3-d box size}	&	Inject. scale	&	Dissip. scale	&	Slope ($\alpha$)	&	$C_n$	&	$\sigma_{turb}$	&	N. realisations	\\
(pixel)$^3$	&	Mpc$\times$(kpc)$^2$	&	(kpc)	&	(kpc)	&	&	(*)	&	(km/s)	&	\\
\hline
\hline
$1936 \times 242^2$	&	$4.2	\times	520^2$	&	100	&	10	&	$-11/3$	&	807.9	&	448.3	&	100	\\
$1936 \times 242^2$	&	$4.2	\times	520^2$	&	200	&	10	&	$-11/3$	&	428.8	&	443.7	&	100	\\
$1936 \times 242^2$	&	$4.2	\times	520^2$	&	300	&	10	&	$-11/3$	&	307.1	&	442.3	&	100	\\
\hline
		\end{tabular}
	\caption{\label{table:simu_char}Numerical realisations of a 3-dimensional velocity cube used for validating the analytic calculations of line centroid shift, line broadening and structure function sample variances. The size of the box in the line-of sight direction is 8 times larger than the transverse (plane-of-sky) box size. (*: units km$^2$ s$^{-2}$ kpc$^{\alpha+3}$.)}
\end{table*}

\begin{figure*}
    \centering
    \includegraphics[width=\linewidth]{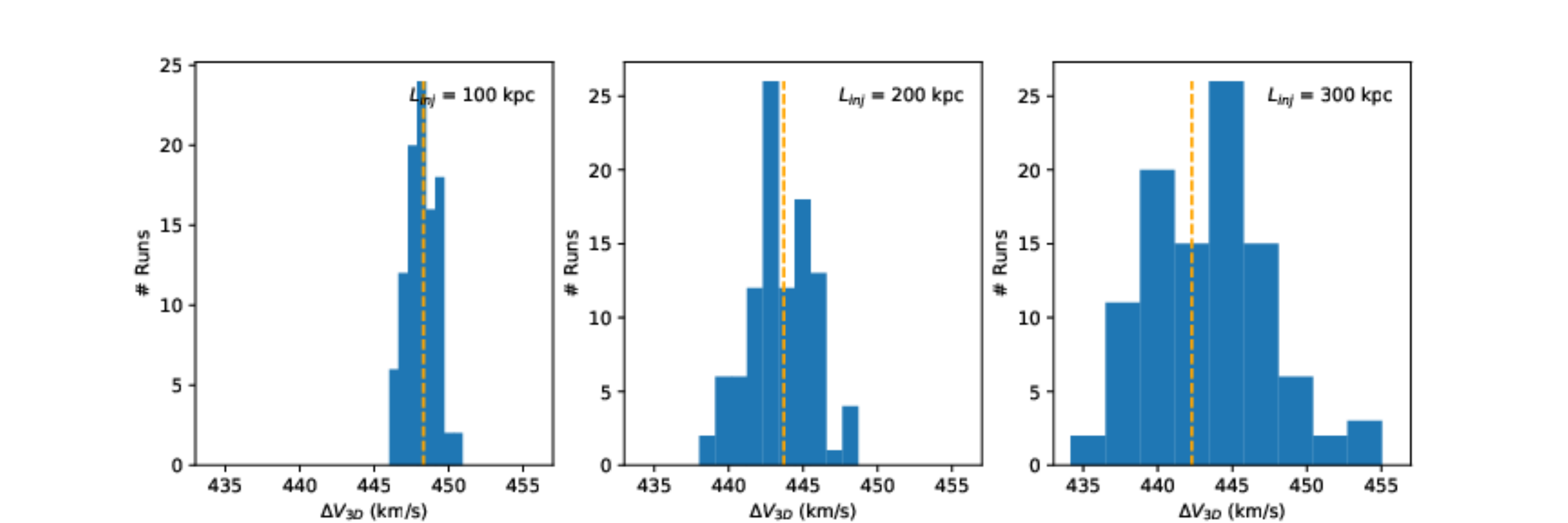}
    \caption{The 3-d velocity dispersion $\Delta V_{3D}$ in each of the 100 numerical realisations for the three configurations in Table~\ref{table:simu_char} is shown as blue histograms. Vertical dashed line indicates the exact value of $\sigma_{turb}$ from integration of the input turbulent power-spectrum (Eq.~\ref{eq:def_pkzuhone}). The increased numerical dispersions in those values as $L_{inj}$ increases, as a consequence of the finite simulation box size.}
    \label{fig:simu_cubestat}
\end{figure*}

The computationally demanding Fast Fourier transforms (FFT) were distributed across 10 processors using 2DECOMP\&FFT\footnote{\url{http://www.2decomp.org}} \citep{2dcomp}. The histograms of the (3-d) velocity standard deviation in each of the three configurations is shown in Fig.~\ref{fig:simu_cubestat}, this is an indicator of the goodness of the simulated field. Clearly, as the injection scale increases the box becomes too small for the periodic boundary condition to apply during the FFT.
The third panel indicates an additional "noise" of order 5-10~km/s in the $L_{inj}=300$~kpc run, which we attribute to aliasing effects. This extra numerical scatter needs to be reminded while comparing analytic results to simulations.

\begin{figure*}
    \centering
    \includegraphics[width=0.95\linewidth]{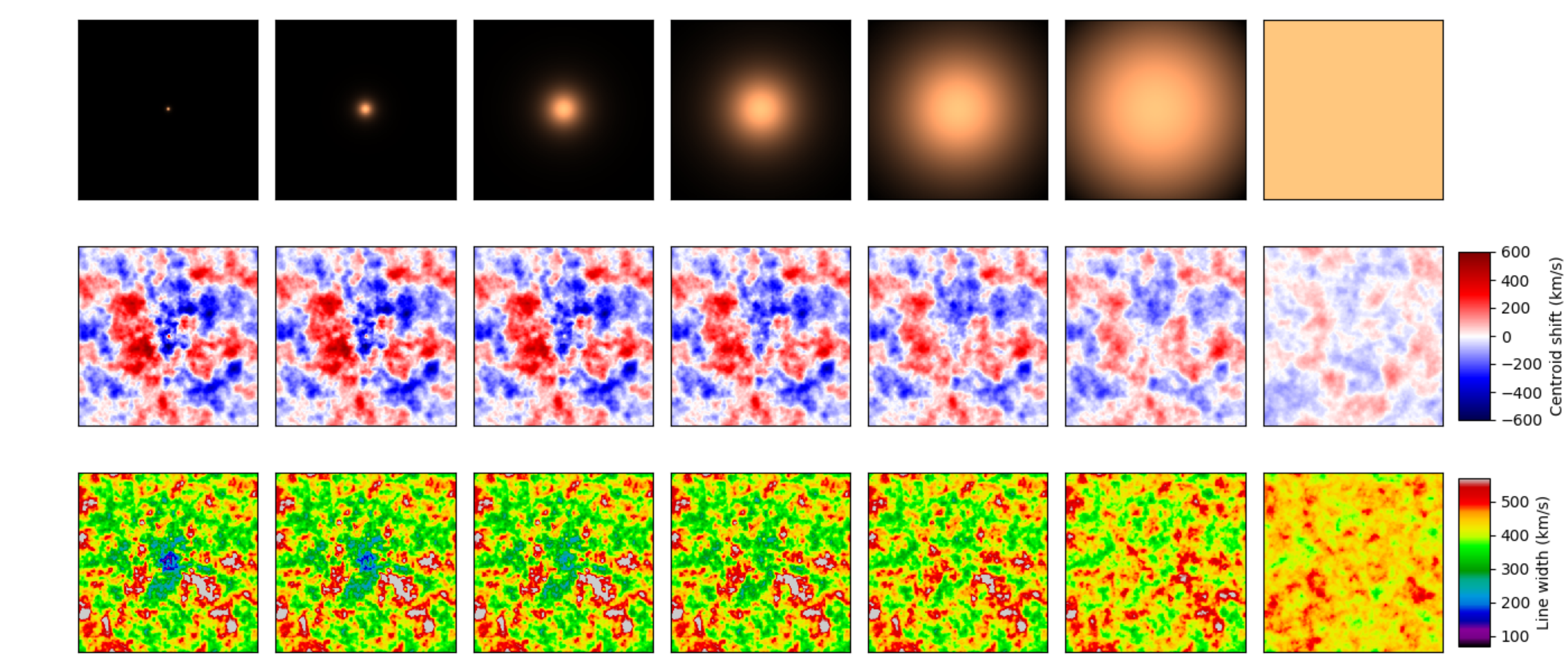}
    \caption{Projection of a single realisation of a 3-d velocity field (injection scale at 200 kpc) with several emissivity models (top row). All but the last column correspond to spherical $\beta$-models with core-radii 4, 21, 54, 107, 215 and 429 kpc (from left to right). The rightmost column corresponds to a constant emissivity in the entire simulation box. The size of each panel is 520~kpc on a side. Middle row shows the centroid shift ($C$) and bottom row shows the line width ($\sqrt{S^2}$). Particularly noticeable is the decrease in contrast (or power) as the core-radius increase and the small line broadening seen through a small cluster core (e.g.~bottom-left figure).}
    \label{fig:simu_linj200}
\end{figure*}

The emissivity of the galaxy cluster gas is taken as the square of a (isothermal, isometallic) gas density considered either as a spherical $\beta$-model (hereafter \emph{beta}) or as a $\beta$-model along the line-of-sight and constant over the plane of the sky (hereafter \emph{Xbeta}). The $\beta$ parameter is held at a value $2/3$ while the core-radius takes value in $\{4, 21, 54, 107, 215, 429\}$ kpc. The normalization of the emissivity plays no objective role in this study, since no signal-to-noise consideration is made. Figure~\ref{fig:simu_linj200} shows one example of the line centroid and line width maps for a 200~kpc injection scale and the various \emph{beta} emissivity models, free of any uncertainty other than numerical noise. In all numerical simulations there is no thermal broadening ($\sigma_{th}=0$~km\,s$^{-1}$ hereafter). In the following we present the comparisons with the $L_{inj}=100$~kpc simulation only. Validation of the other two runs is extensively presented in App.~\ref{app:valid_200-300} for completeness.

% % % % %

	\subsection{Centroid and line broadening\label{sect:valid_c_s2}}

We first carry out the validation of Equations~\ref{eq:avg_centroidshift3d}, \ref{eq:var_centroidshift3d}, \ref{eq:avg_broadening3d} and~\ref{eq:var_broadening3d}, which provide analytical representations of the sample average and variance of the line centroid shift and line broadening (more specifically, the square of the line width) measured in arbitrary apertures.
We limit this validation exercise to circular apertures centred on a galaxy cluster and allow their sizes to vary.

These analytical expressions involve calculation of the 3-d function $c_{\emithree . \mathcal{W}}$: we provide in App.~\ref{app:fourier_eta} the analytical formulas for both considered emissivity models and for circular apertures. Calculation of this function for spherical $\beta$-models demands slightly more computing time than for the \emph{Xbeta} model.

Equation~\ref{eq:var_broadening3d} requires integration over 6 scalar variables. Taking advantage of the isotropy of the velocity power spectrum and the 2-d rotational invariance of this specific configuration, this can be reduced to five integration variables only (e.g.~$k_x, k_x^{\prime}, \xi, \xi^{\prime}$ and one angle $\phi$). This integral is evaluated by Monte-Carlo sampling distributed over 40 computing cores by means of the MCQUAD library\footnote{Available in package SciKit-Monaco, \url{https://pypi.org/project/scikit-monaco/}}. The number of samplings is $2.10^6$ and $2.10^5$ for the \emph{Xbeta} and \emph{beta} emissivity models respectively and we monitor and store the statistical uncertainties out of the numerical sampler.

\begin{figure*}
    \centering
    \includegraphics[width=\linewidth]{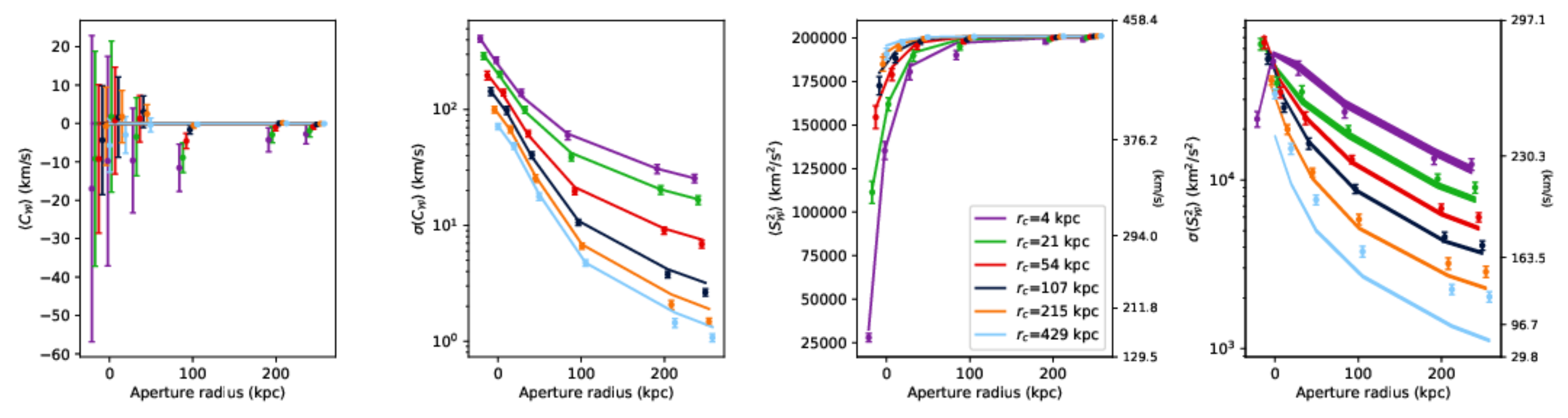}
    \caption{Numerical validation of equations~\ref{eq:avg_centroidshift3d}, \ref{eq:var_centroidshift3d}, \ref{eq:avg_broadening3d} and~\ref{eq:var_broadening3d}, i.e.~the expected value and sample variance of the line centroid shift $C_{\mathcal{W}}$ (first and second panel) and the expected value and sample variance of the line width $\sqrt{S^2_\mathcal{W}}$ (third and fourth panel). Plain lines show the analytical calculations, data points are measured on 11 numerical realisations of a turbulent field (errors estimated via bootstrap) with $L_{inj}=100$\,kpc. The calculations are performed assuming measurements in circular apertures $\mathcal{W}$ of various radii (x-axis). The emissivity model is \emph{Xbeta} with core-radii indicated in legend. The uncertainty on the analytical results for $\sigma(S^2)$ (materialized by the line widths in the last panel) is due to limitations of the numerical integrator used to evaluate Eq.~\ref{eq:var_broadening3d}.}
    \label{fig:linestat_linj100_Xbeta}
\end{figure*}

\begin{figure*}
    \centering
    \includegraphics[width=\linewidth]{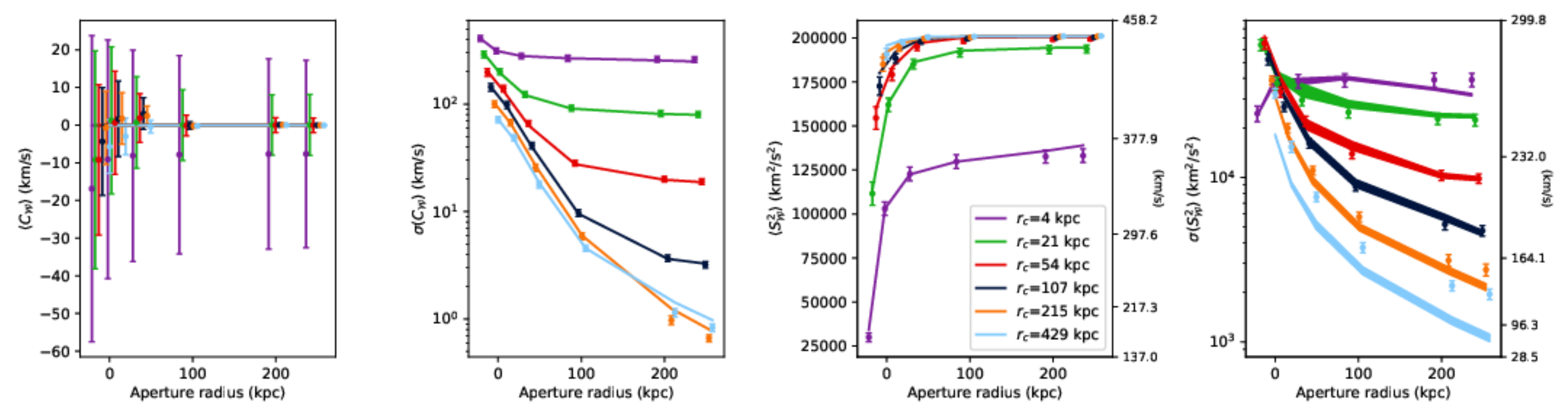}
    \caption{Similar figure as Fig.~\ref{fig:linestat_linj100_Xbeta} for a spherical $\beta$-model emissivity (\emph{beta}). The numerical uncertainties are slightly larger (in the 4th panel, compared to Fig.~\ref{fig:linestat_linj100_Xbeta}) due to a lower accuracy in the numerical integration of Eq.~\ref{eq:var_broadening3d}.}
    \label{fig:linestat_linj100_beta}
\end{figure*}

Figures~\ref{fig:linestat_linj100_Xbeta} and~\ref{fig:linestat_linj100_beta} show the comparison between analytical calculations (plain lines) and numerical simulations (dots with error bars). They correspond to the \emph{Xbeta} and \emph{beta} emissivity models respectively, using the same 100 velocity boxes with $L_{inj}=100$~kpc. Each dot corresponds to a calculation using the 100 velocity realisations and a given core-radius size and a given aperture size. Error bars are derived from bootstrap resampling. Because we always used the same 100 simulations, the deviations to the expected trend appear correlated: this is for instance striking in the left-most panel showing $\langle C \rangle$. This behaviour is likely to disappear with a higher number of realisations.

In any case these figures demonstrate a very good agreement between analytical calculations and simulations. The only exception is the case of very large core-radii ($r_c=429$~kpc), for both emissivity models. This is a consequence of the simulation box being too small in the x-direction (4240~kpc along the line-of-sight). This causes a non-negligible sharp cut-off in the simulated $\beta$-emissivity profile, not accounted for by the analytical equations.

\begin{figure}
    \centering
    \includegraphics[width=\linewidth]{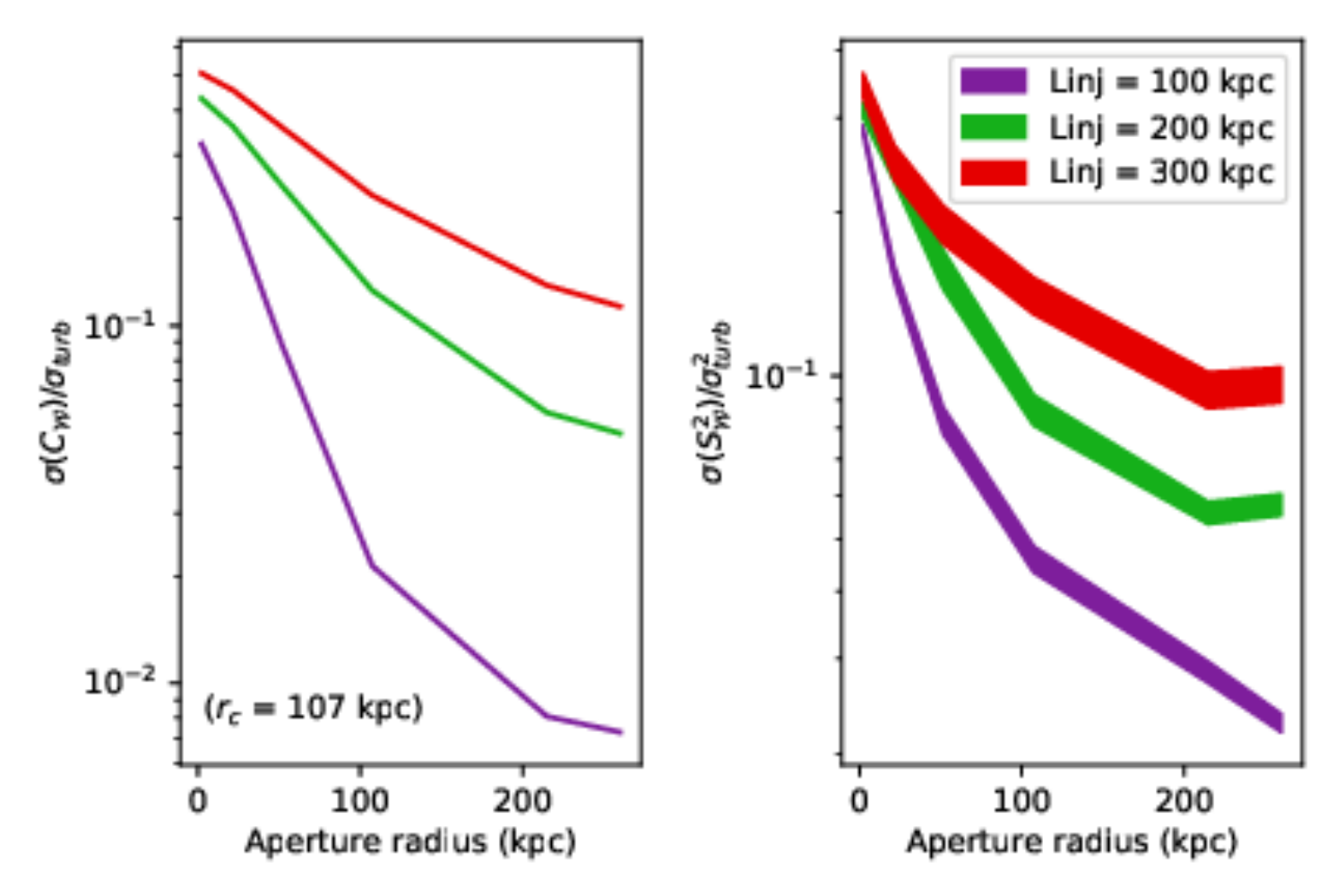}
    \caption{The "sample variance" associated to line measurements -- centroid shift (left) and broadening (right) -- in apertures of growing sizes for clusters presenting identical emissivity models (a spherical $\beta$-model with core-radius 107~kpc.) These curves are predicted analytically by Eq.~\ref{eq:var_centroidshift3d} and~\ref{eq:var_broadening3d}. They have been normalized to the value of the 3-d turbulent velocity dispersion $\sigma_{turb}$. The different shapes of the curves as the aperture radius is growing can be used as a diagnostic to discriminate between various injection scales.}
    \label{fig:linestat_linjALL_beta}
\end{figure}

The sample variance of the centroid shift and the line width exhibit large variations with respect to the aperture radius. Emission line diagnostics in growing apertures for a selection of 'look-alike' galaxy clusters has interesting potential to reveal the properties of the underlying  turbulent power-spectrum. This is illustrated in Fig.~\ref{fig:linestat_linjALL_beta} where we vary the injection scale from 100 to 300 kpc for a given $\beta$-model ($r_c=107$~kpc). It is out of the scope of this paper to provide forecasts on the constraining power of this method, which must also include measurement uncertainties and limitations related to the availability of samples.

% % % % %

	\subsection{Structure function}

We restrict the numerical validation to that of Eq.~\ref{eq:sf_average} and~\ref{eq:varsf_xbeta_fovlarge} for computational reasons.
First, we consider an emissivity model of type \emph{Xbeta} and a large enough analysis region compared to the typical separation and pixelization of the line centroid map. The same $3 \times 100$ simulated boxes are projected and pixelized in square regions of size $\ell \times \ell$ where $\ell = 4, 9, 17, 34, 69$~kpc. Since the two-dimensional velocity field is stationary, it is fine to use $P_{\ell} P_{2D}^{\infty}$. This replacement is justified because in this configuration the surface brightness is constant over the analysis domain. In what follows the analysis domain is a circular aperture of diameter 520~kpc. Figure~\ref{fig:cmap_binning} illustrates how pixelization acts on a simulated centroid shift map: it indeed is very close to a convolution or 'smoothing'.
Similar maps are created for all pixel sizes and core-radii for all simulated boxes. The geometrical centres of the pixels are used to compute the structure function, as the arithmetic mean of the squared centroid gradients at pre-defined separations $\sepa$ (within a range $\delta \sepa$). The average of the structure functions and the standard deviations at each $\sepa$ provide the numerical indicators to be compared to Eq.~\ref{eq:sf_average} and~\ref{eq:varsf_xbeta_fovlarge}.

Analytical calculation of $P_{2D}$ is performed according to Eq.~\ref{eq:p2d_calcul}, by 2-dimensional convolution\footnote{Making use of the FFT convolution implemented in the {\tt signal.convolve} function of Numpy/Scipy.} of the 3-d velocity power-spectrum $P_{3D}$ and the function $P_{\rho}$ at each frequency $k_x$ and eventually summing over those frequencies. The whole procedure is distributed over 40 processors working in parallel. Analytical expressions for $P_{\rho}$ are given in App.~\ref{app:fourier_rho} (particularly Eq.~\ref{eq:prho_smallA}) for the emissivity models relevant to our validation procedure. The calculation of $P_{2D}^{\infty}$ is much more straightforward, see Eq.~\ref{eq:p2dinfini_calcul}.

\begin{figure*}
    \centering
    	\includegraphics[width=\linewidth]{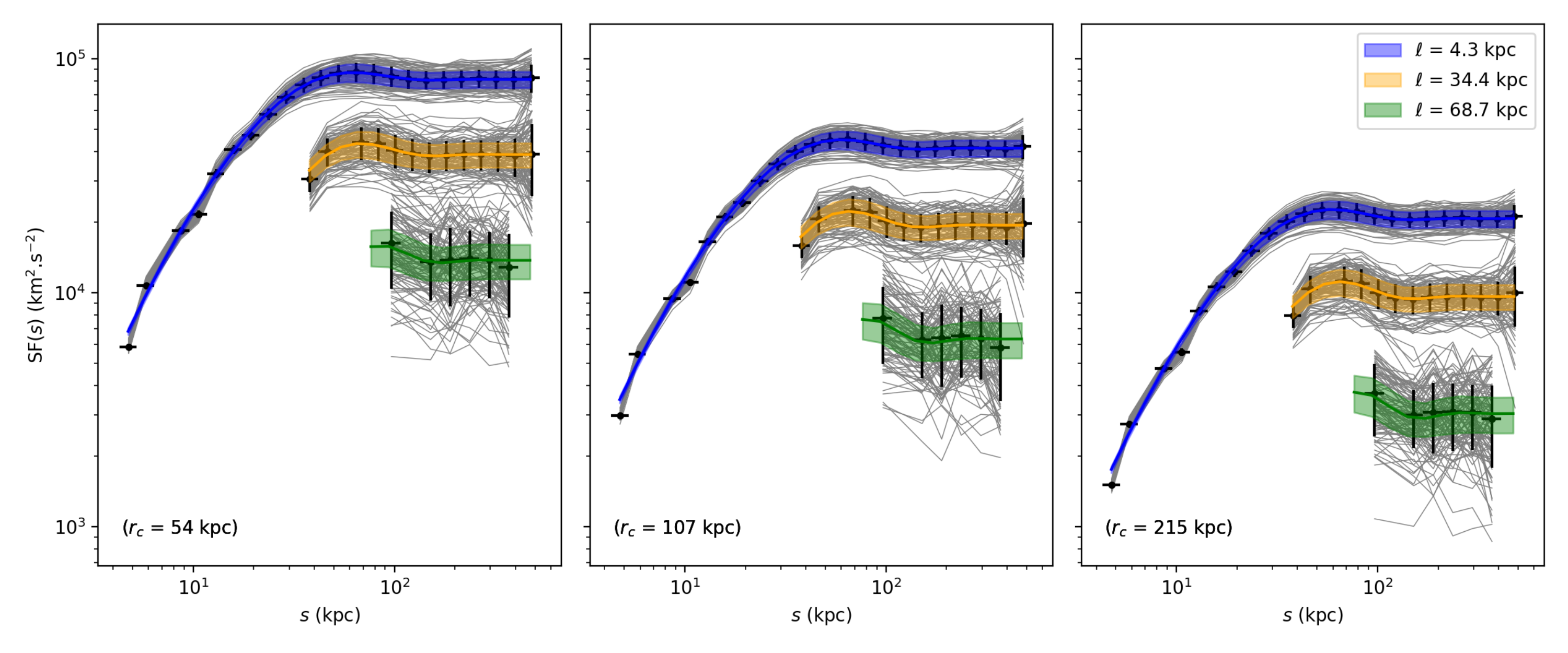}
   	\caption{Comparison of numerical and analytical structure functions and their sample variance for various cluster sizes (i.e.~various core-radii ($r_c$) of the $\beta$-model) and various pixel sizes ($\ell$). The data points and thick error bars show the sample mean and standard deviation of the 100 realisations (individually represented as thin grey lines) for each of the considered configurations. The coloured curves and shaded areas represent the analytical calculations following Eqs.~\ref{eq:sf_average} and~\ref{eq:varsf_xbeta_fovlarge}. The emissivity model is \emph{Xbeta} and the region of analysis is a circle of diameter 520~kpc (as diplayed in Fig.~\ref{fig:cmap_binning}). The turbulent power-spectrum is that of Table~\ref{table:simu_char} with injection scale 100~kpc.}
    \label{fig:sf_compa_Xbeta}
\end{figure*}

A comparison between the analytical and numerical results is illustrated in Fig.~\ref{fig:sf_compa_Xbeta} for a given turbulent power spectrum (injection scale at 100~kpc) and various values for the core-radius and pixel size. The results from the 100 realisations are displayed as thin grey lines and their distribution at each $\sepa$ is likely not Gaussian.

The analytical and numerical values for the sample variance are in very good agreement for all 9 configurations, which is a very encouraging result given the various assumptions involved in both cases.

\begin{figure*}
    \centering
    	\includegraphics[width=\linewidth]{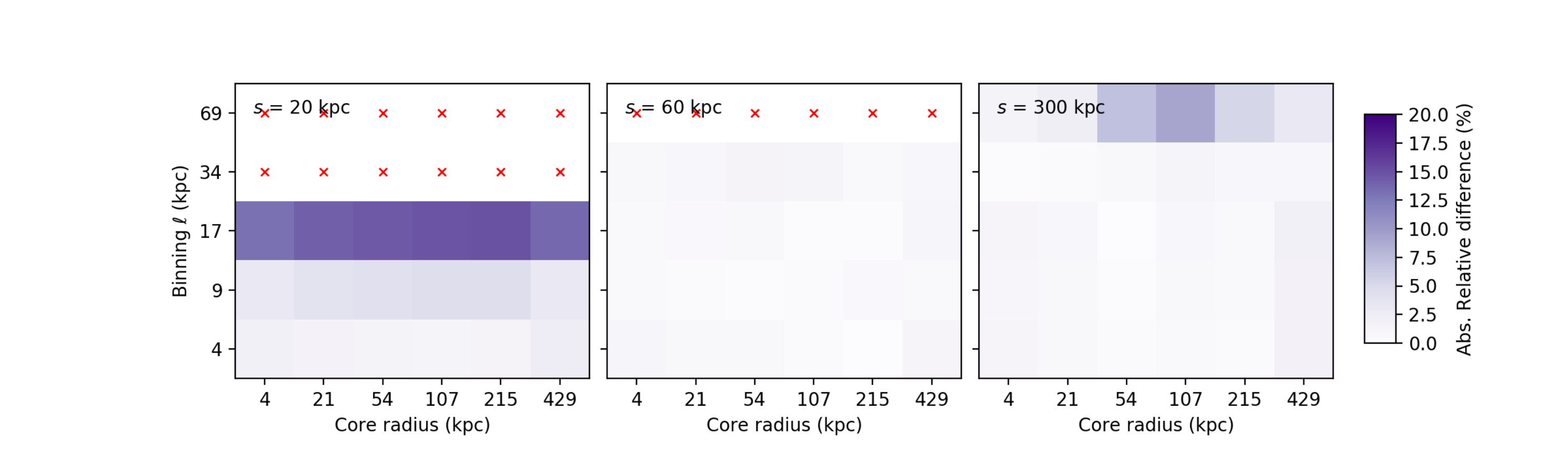}
   	\caption{Representation of the absolute relative difference between the numerical and analytical estimates of the sample mean of the structure function, $\langle SF \rangle$, at three distinct separations $\sepa$. Each coloured square corresponds to one experiment based on the same 100 realisations of the velocity field with $L_{inj}=100$~kpc and various binning sizes (y-axis) and $\beta$-models core-radii (x-axis). Small red crosses indicate locations where the binning size is larger than $\sepa$.}
    \label{fig:reldiff_sf_mean}
\end{figure*}

\begin{figure*}
    \centering
    	\includegraphics[width=\linewidth]{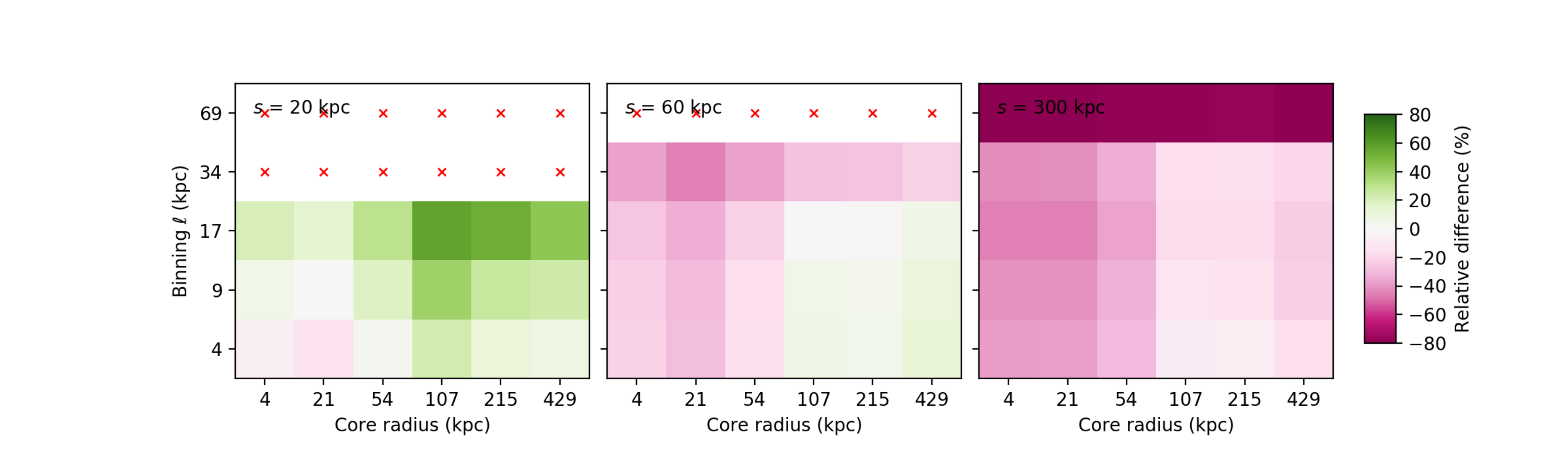}
   	\caption{Similar as Fig.~\ref{fig:reldiff_sf_mean}, but for the sample variance of the structure function, $\mathrm{Var}(SF)$. Positive values indicate higher predicted variance compared to that measured in the numerical validation procedure. Although some of these numbers are high at face value, it is important to recall the assumptions leading to the chosen analytical formula and the noise inherent in our set of numerical simulations (see text).}
    \label{fig:reldiff_sf_var}
\end{figure*}

A thorough assessment of the agreement between the analytical calculations and the numerical validation is summarised in Figures~\ref{fig:reldiff_sf_mean} and~\ref{fig:reldiff_sf_var}. They show the relative difference (expressed in percent) between the analytical and the numerical computations for the expected value of the structure function and its variance respectively. The injection scale is $L_{inj}=100$~kpc, as in Fig.~\ref{fig:sf_compa_Xbeta}. Three separations are illustrated: $\sepa=20$~kpc (close to the dissipation scale), $\sepa=60$~kpc (within the inertial range) and $\sepa=300$~kpc (past the inertial range).

Regarding the sample mean, the analytical model (Eq.~\ref{eq:sf_average}) performs well within 20~\% of the numerical experiment. Keeping in mind the limited number of realisations (100 samples), this result appears satisfactory. A degradation of the prediction accuracy arises as the binning size increases. This is attributed to numerical approximation in computing $P_{\ell}$ and higher levels of sampling noise in the simulation (larger pixels imply fewer $\sepa$-pairs). The slight decrease in accuracy at larger core-radii was already pointed out in Sect.~\ref{sect:valid_c_s2}, as a result of the simulation box size.

As for the sample variance (Fig.~\ref{fig:reldiff_sf_var}), the relative differences between analytical and numerical results show somewhat higher values, as is expected for second order statistics. In general our formula tends to overpredict by a few tens of percent the observed variance of the structure function at small separations ($\sepa=20$~kpc) as a result of numerical uncertainties both in the simulations and the evaluation of integrals. At large separations ($\sepa=300$~kpc) and for the largest pixel size, the analytic formula underpredicts the variance by up to 80\%. The analysis region $\mathcal{A}$ indeed cannot be considered as infinitely large any longer, making simplification of Eq.~\ref{eq:covar_sf} into Eq.~\ref{eq:varsf_xbeta_fovlarge} less accurate.

\begin{figure*}
    \centering
    	\includegraphics[width=\linewidth]{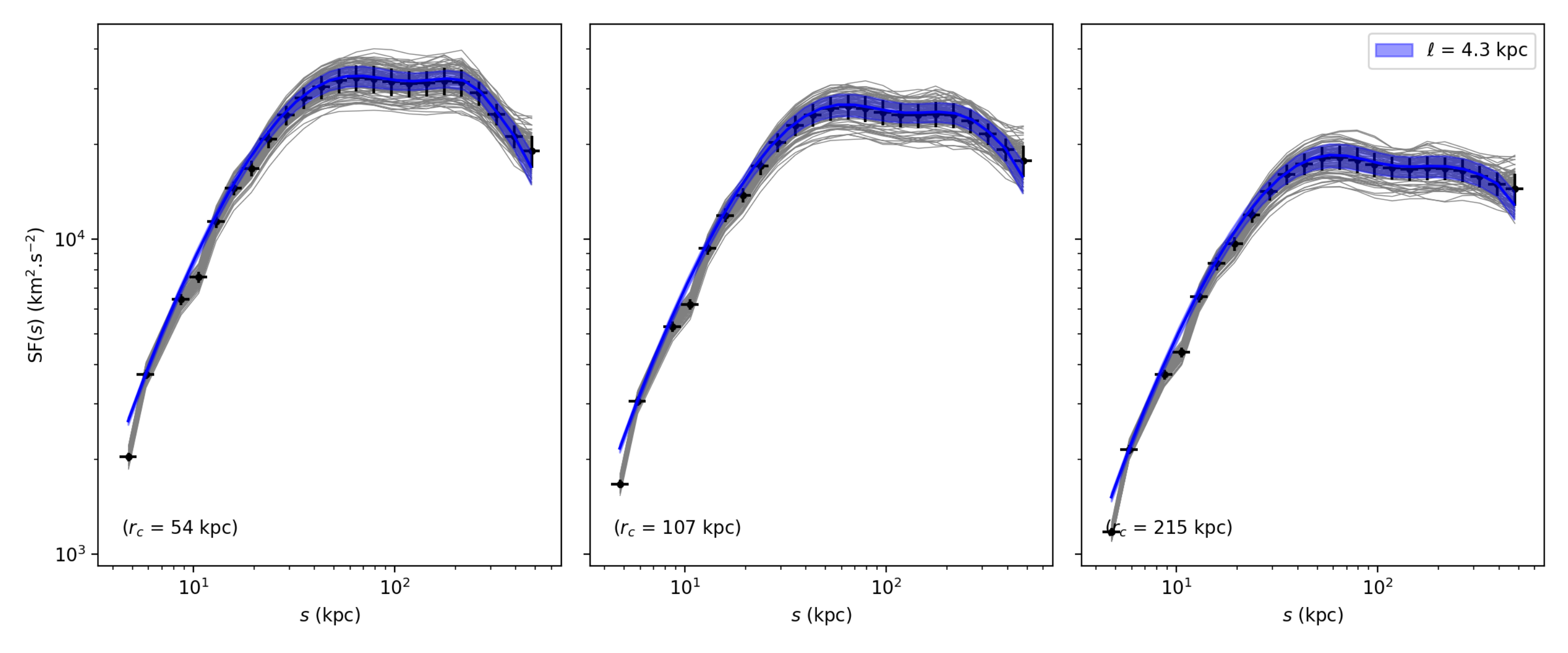}
   	\caption{Comparison of the simulated and calculated structure function in a similar way as Fig.~\ref{fig:sf_compa_Xbeta}, except the emissivity model is of type \emph{beta} (spherical $\beta$-model gas density). This induces non-stationarity of the projected velocity field, noticeable by the drop at large $\sepa$. The coloured curves represent the analytical calculation of the mean structure function following Eqs.~\ref{eq:sf_average} and~\ref{eq:sf_average_corrected}. For simplicity, the variance remains calculated according to Eq.~\ref{eq:varsf_xbeta_fovlarge}, i.e.~assuming \emph{Xbeta} emissivity with an effective core-radius $c=\sqrt{r_c^2+\theta_\mathrm{eff}^2}$ with $\theta_\mathrm{eff}=80$~kpc.}
    \label{fig:sf_compa_beta}
\end{figure*}

%% New:
We finally relax the assumption of a constant emissivity and we show in Fig.~\ref{fig:sf_compa_beta} a comparison of the structure functions obtained for a spherical $\beta$-model density (\emph{beta} emissivity model). The analytical formula Eq.~\ref{eq:sf_average} recovers the mean structure function, despite the spatial non-stationarity of the projected velocity field. We corrected for border effects using Eq.~\ref{eq:sf_average_corrected}, the region analysis being a circle of diameter 520~kpc. As highlighted in App.~\ref{app:filter_field}, we are not able to use the simple prescription for significantly large pixel binnings. Moreover we did not carry the full evaluation of the sample variance using Eq.~\ref{eq:covar_sf} for this figure. Instead, we computed the variance according to Eq.~\ref{eq:varsf_xbeta_fovlarge} assuming an effective core radius $c=\sqrt{r_c^2+\theta_\mathrm{eff}}$. Such approximation of the complex emissivity field by an effective emissivity extracted at a radius $\theta_\mathrm{eff}=80$~kpc from the cluster centre makes the calculation more tractable. It shows a good agreement with results obtained from the Monte-Carlo simulations.

%%%%%%%%%%%%%%%%%%%%%%%%%%%%%%%%%%%%%%%%%%%%%%%%%%%%%%%%%%%%%%%%%
	\section{Discussion}
	\label{sect:discussion}

This work provides an extension of earlier studies, among others \citet{inogamov2003}, \citet{churazov2012}, \citet{zhuravleva2012} and~\citet{zuhone2016}. We extended the formalism presented in these papers by: i)~addressing the case of an arbitrary emissivity field; ii)~computing second order statistics beyond expected values (i.e.~the sample variance) and iii)~identifying limiting cases in which these studies coincide.

		\subsection{Validity of hypotheses and range of applicability}
Our study remain formal and rely on strong hypotheses such as uniform, ergodic and isotropic turbulent velocity fields throughout the intra-cluster medium, whose physics is encapsulated in a universal Kolmogorov power-spectrum, meaning that turbulence follows an identical physical description from cluster to cluster. The latter assumption is often implicitly made and mirrors an intent to concentrate all the unknown physics of turbulence in a single mathematical description. It is clear that this hypothesis may fail if widely distinct mechanisms produce turbulent motions: for instance large-scale matter accretion and central AGN feedback.

More specifically, our assumption of isotropic turbulent motions may break down in the stratified intra-cluster medium where buoyancy-restoring forces tend to suppress motions along the radial direction. Numerical simulations indicate a change in the morphology of turbulent fields in regions showing strong density gradients \citep[e.g.][]{shi2019}, even in cluster cores \citep{valdarnini2019}. According to these findings, one can therefore expect our model to become less representative as larger and larger cluster radii enter the emission line analysis, or in presence of strong cool-core clusters. A study of anisotropic turbulence is out of the scope of this paper: for instance, one would undertake similar derivation steps as shown in appendices, dropping the assumption $P_{3D}(\kthree) = P_{3D}(k)$.

The existence of several (two) drivers of turbulence acting at different scales may change the shape of the velocity power-spectrum and more generally the statistical relations between Fourier coefficients of the velocity field. \citet{zuhone2016} proposed to rewrite the resulting $P_{3D}$ as a sum of two Kolmogorov-like power-spectra with different injection scales, based on the simulations and results of \citet{yoo2014}. Such a prescription enters the framework presented in this paper, because our results are independent on the exact shape of $P_{3D}$. However, it remains to be checked whether the decomposition of $\langle V_{\jthree} V_{\kthree} V_{\lthree} V_{\mthree} \rangle$ proposed in App.~\ref{app:sf_covariance} holds under such conditions, which most likely can be addressed through numerical simulations.

Our hypothesis also assumes full decoupling of the gas emissivity and the local behaviour of turbulent motions. This simplifying assumption may largely fail if gas motions are induced by merging of an external galaxy group which shows high emissivity in its vicinity. An interesting perspective of the present calculations would be the coupling between $P(k)$, the turbulent power spectrum and $\emithree$, the emissivity; however it is likely that calculations would become more complex and the gain over Monte-Carlo simulations would become less obvious. Moreover, density fluctuations, hence emissivity fluctuations, are thought to be directly linked to the turbulent power spectrum based on theoretical grounds \citep{churazov2012}. At first order though, the broad-scale emissivity of the galaxy cluster gas is the dominant component modulating the Doppler shift in the integrated line profiles and our approach remains a reasonable one in this regard.

Finally, our work deliberately neglects measurement uncertainties and instrumental noises. We address this assumption in a subsequent study (paper II, Cucchetti et al., in press) by propagating the impact of measurement uncertainties on the line diagnostics, in particular the structure function. An interesting conclusion of this study is that sample variance effects dominate on large scales the error budget for observations based on next-generation X-ray instruments such as \emph{Athena}/X-IFU, while statistics dominate at small scales.

		\subsection{An application: forecasting line shift and width profiles}

\begin{figure*}
    \centering
    	\includegraphics[width=\linewidth]{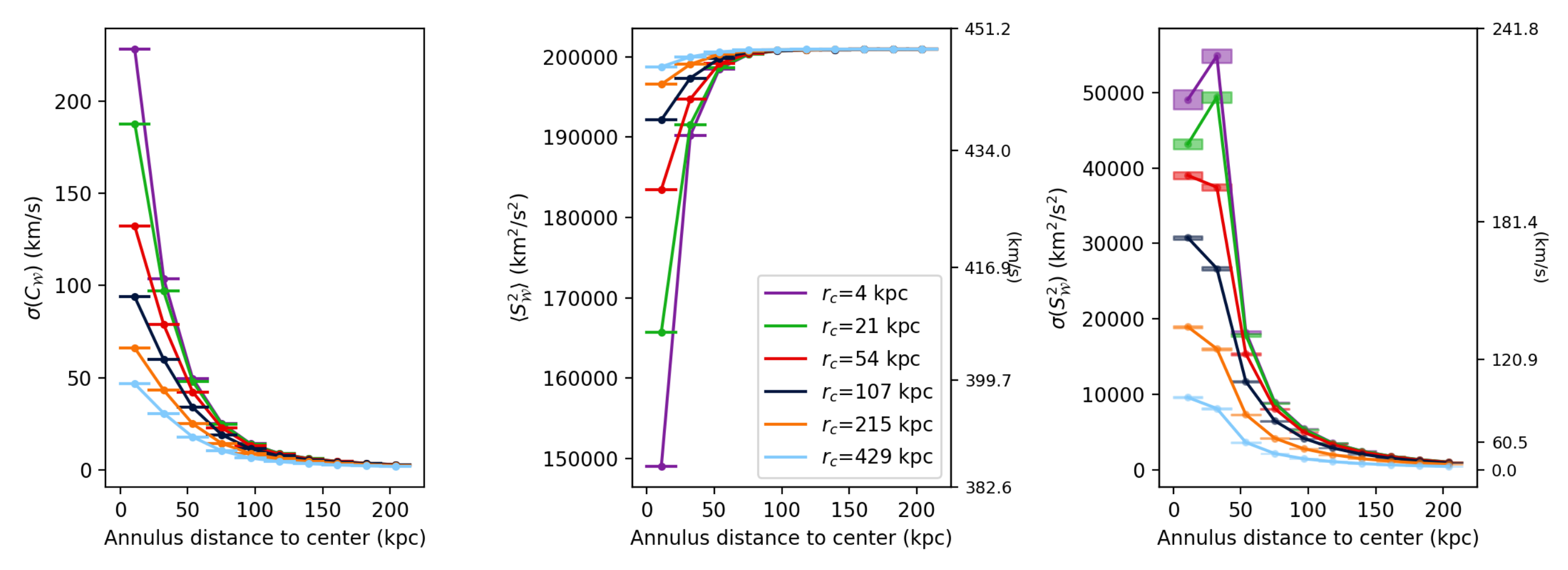}
   	\caption{Model predictions for radial profiles of line properties, i.e~measurements in spectra collected in circularly concentric annuli of equal width (21~kpc). Shown are the sample variance of the centroid shift (left panel), the sample average of the line broadening (middle panel) and the sample variance of the line broadening. The injection scale is $L_{inj}=100$~kpc, the emissivity model is a spherical $\beta=2/3$ model with core radii indicated in legend. Shaded rectangles indicated bin widths (horizontally) and numerical uncertainties (vertically).}
    \label{fig:annuli_variance}
\end{figure*}

The formulas derived in Sect.~\ref{sect:2d_statistics} provide the sample mean and variance of both the line centroid shift and width in arbitrary apertures. As such, they can be used to predict measurements in concentric annuli centred on a galaxy cluster, i.e.~a radial profile. Formally, an annular aperture mask is defined as the difference between two concentric circular apertures. Thanks to the linear behaviour of the Fourier transform, the coefficient $c_{\emithree.\mathcal{W}}$ is also the difference between the two corresponding coefficients, both easily computed following App.~\ref{app:fourier_eta}. Interestingly, the emissivity in each annulus can be considered as constant, i.e.~$\emithree(x, \vec \theta) = \epsilon(x)$, if the gas density shows spherical symmetry and the annuli are thin enough. As already noted, this property drastically reduces the computing time needed to integrate the equations. An example of the profiles of centroid shift variance, line width average and line width variance are displayed on Fig.~\ref{fig:annuli_variance} for a turbulent power spectrum with injection scale 100~kpc and $\sigma_{turb}=448$~km\,s$^{-1}$. No thermal broadening is included in this exercise. As expected, the centroid shift (whose average value is zero) shows larger variance in the central bins than the outskirts and the larger the core radius, the smaller the effect. The average broadening shows the reverse behaviour with smaller widths in the central parts and reaching a plateau (corresponding to $\sigma_{turb}$) in the outskirts. The line width variance shows diverse behaviours but here again, the general trend is a decrease towards the outskirts.

		\subsection{Forecasting the structure function from upcoming instrumentation}

\begin{figure*}
    \centering
    	\includegraphics[width=0.9\linewidth]{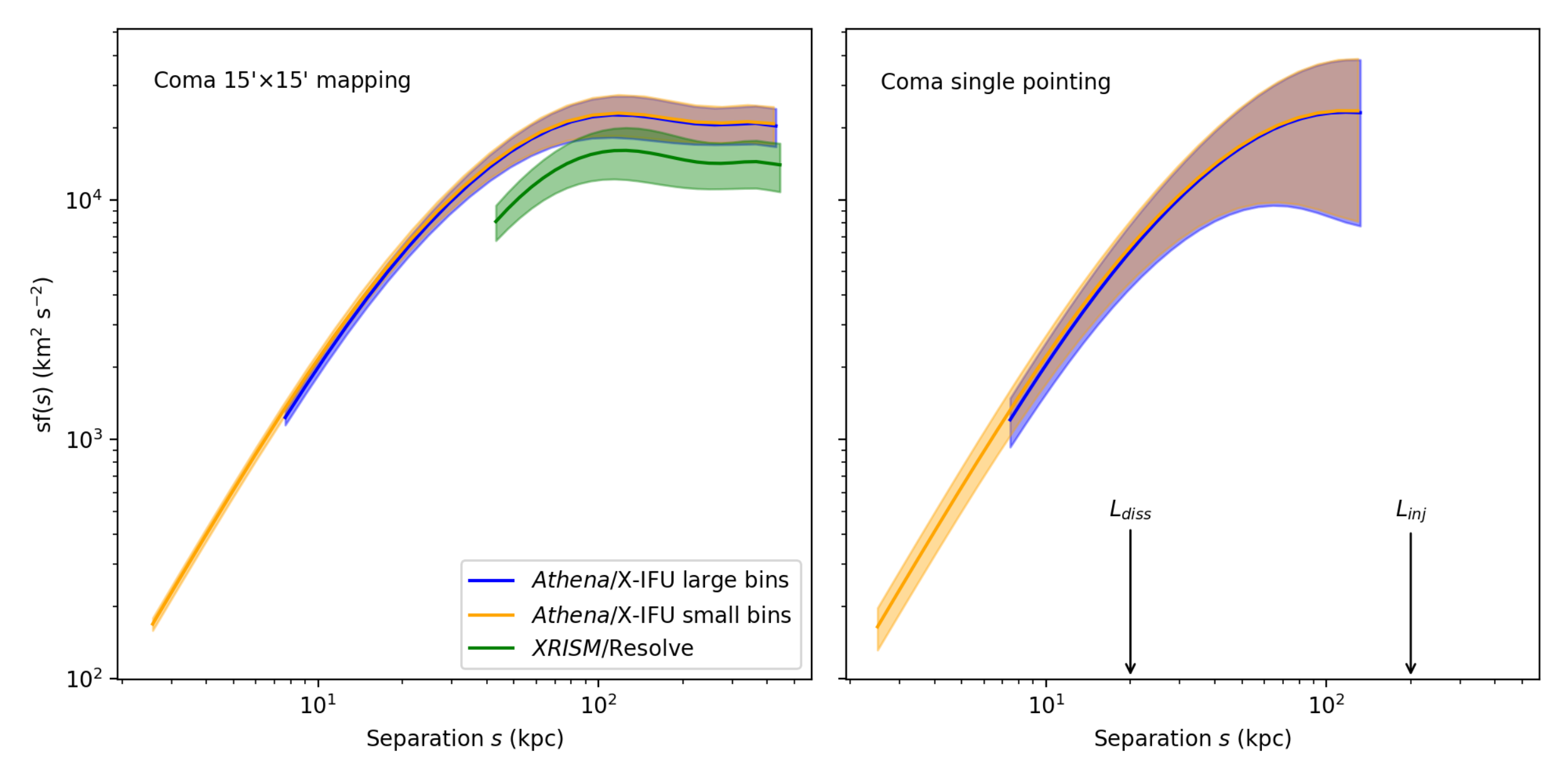}
   	\caption{Model predictions for structure functions and their associated sample variances under two instrumental setups: \emph{XRISM}/Resolve (assuming $1.5\arcmin$ resolution elements) and \emph{Athena}/X-IFU (assuming $5\arcsec$ and $15\arcsec$ resolution elements for high and low signal to noise ratios respectively). The left panel shows predictions for a $\sim 15\arcmin \times 15\arcmin$ contiguous mapping of the Coma cluster while the right panel shows the result for a single X-IFU pointing. A single Resolve pointing ($3\arcmin$ on a side) would be too small for a useful derivation of the structure function. See text for details on the input turbulent power spectrum and gas density model.}
    \label{fig:athenaxrism}
\end{figure*}

One particularly interesting perspective consists in inverting the formulas presented here to evaluate the power of future astronomical X-ray micro-calorimeters in constraining the nature of turbulent motions in galaxy clusters. By properly selecting the samples (typically, the number of objects and their core-radii and distances) and the observing strategy (mapping, exposure times, etc.) one is able to focus the constraints on, e.g.~the slope of the power-spectrum or the injection scale. This assumes that turbulence has identical characteristics throughout the sample considered, which hopefully is a reasonable guess.
We postpone the complete exercise to later investigation. Rather we compute the expected structure functions for a simplified Coma-like galaxy cluster, following a setup similar to \citet{zuhone2016} and the associated uncertainties due to sample variance only (statistical errors are disregarded). We consider two instruments: i)~\emph{XRISM}/Resolve with a resolution element of $1.5\arcmin$ and a field-of-view of $3.4\arcmin$ equivalent diameter and ii)~\emph{Athena}/X-IFU with a resolution element of $5\arcsec$ and a field-of-view $5\arcmin$ equivalent diameter \citep{barret2018}. We consider two observing strategies: either one single pointing towards the cluster centre, or the mapping of a $\sim 15\arcmin \times 15\arcmin$ area with multiple pointings. We also consider the case of a $15\arcsec$ pixelization rebinned images for \emph{Athena}/X-IFU, such that the signal-to-noise ratio of each spectrum is increased \citep[e.g.][also Cucchetti et al., in press]{roncarelli2018}. At the redshift of Coma, $1\arcmin$ on sky corresponds roughly to 27~kpc physical separation. We consider a turbulent power spectrum with $\sigma_{turb}=438$~km/s, $\alpha=-11/3$, injection scale at 200~kpc and dissipation scale at 20~kpc. Given the proximity of Coma and its apparent size, using the \emph{Xbeta} emissivity model is amply justified, as already noted by \citet{churazov2012, zuhone2016}. Figure~\ref{fig:athenaxrism} shows the result outcome of our model.
For identical sky coverages, X-IFU provides smaller relative variance in comparison to Resolve, thanks to its better angular resolution. Even in one single pointing X-IFU can provide a measurement of the structure function up to $\sim 100$~kpc separation scales. The associated variance is larger though, due to a smaller number of pairs entering the structure function.

This example provides the basis in view of optimising an observational strategy for a given instrumental setup. Our formalism involves Fourier transforms of window functions (denoted $\mathcal{A}$ and $\mathcal{W}$) and therefore accounts for arbitrary instrumental shapes and pointing strategies, by taking advantage of standard properties of the Fourier transform. For instance, a window function made of multiple non-overlapping pointings can be considered as a sum of identical, translated window functions; linearity then makes the computation of its Fourier transform straightforward.

%%%%%%%%%%%%%%%%%%%%%%%%%%%%%%%%%%%%%%%%%%%%%%%%%%%%%%%%%%%%%%%%%
	\section{Conclusions}
	\label{sect:conclusion}

In this paper we have derived analytical expressions for the sample mean and variance of three indicators of turbulence in X-ray emitting, optically-thin, plasmas under the hypothesis of homogeneous and isotropic Kolmogorov turbulence. These are the line centroid shift $C$, the line broadening $S$ and the structure function $SF$.

\begin{enumerate}
\item We obtained exact expressions for the mean and variance of $C$ and $S$ obtained from single line-of-sight measurements through arbitrary gas emissivity: Eqs.~\ref{eq:c1d_mean}, \ref{eq:c1d_var}, \ref{eq:s1d_mean} and~\ref{eq:s1d_var}. We numerically validated the results with Monte-Carlo simulations of turbulent velocities with Gaussian or constant amplitudes.
\item We generalised these expressions for measurements in apertures of arbitrary shapes and sizes and for arbitrary 3-dimensional emissivity fields: Eqs.~\ref{eq:avg_centroidshift3d}, \ref{eq:var_centroidshift3d}, \ref{eq:avg_broadening3d} and~\ref{eq:var_broadening3d}. We provided in App.~\ref{app:fourier_eta} useful formulas for the common $\beta$-model and for circular apertures. We numerically validated the formulas using Monte-Carlo simulations of 3-d velocity fields in a range of emissivity and power-spectrum configurations.
\item We derived an expression for the mean structure function under the assumption of negligible border effects (Eq.~\ref{eq:sf_average}). Notably, this formula does not assume constant ('flat') emissivity in the plane-of-sky direction. It involves a specific definition for the two-dimensional power-spectrum of the projected velocity field, introduced in App.~\ref{app:p2d_general}. We provided in App.~\ref{app:fourier_rho} useful formulas for the common $\beta$-model and for circular domains of analysis.
\item In App.~\ref{app:sf_general} we provided a correction formula for border effects (Eq.~\ref{eq:sf_average_corrected}) valid for non-binned maps of the projected velocity and for domains of arbitrary shapes. We explicitly computed the case of a circular field-of-view.
\item We provided in App.~\ref{app:filter_field} a simple prescription to account for binning (or pixelisation) on the mean structure function. It is valid as long as pixels are smaller than the typical scale of flux variations and much smaller than the domain of analysis.
\item We derived a fairly generic expression for the sample variance of $SF$ under assumption of negligible border effects and for arbitrary emissivity fields (Eq.~\ref{eq:covar_sf}). This equation takes a tractable form in case of flat emissivity and very large domain of analysis (Eq.~\ref{eq:varsf_xbeta_fovlarge}).
\item We numerically validated our results for the sample mean and variance of $SF$ in the case of 'flat' emissivity fields ($\beta$-models with a range of core radii) and various binnings.
\item We numerically validated our results for the sample mean of $SF$ in the case of non-flat emissivity fields (spherical $\beta$-models with a range of core radii) and negligible binning.
\item We discussed our results and presented forecasts for observations of the core of the Coma cluster with the integral field units X-ray calorimeters planned to embark onboard \emph{XRISM} (Resolve) and \emph{Athena} (X-IFU).
\end{enumerate}

%%%%%%%%%%%%%%%%%%%%%%%%%%%%%%%%%%%%%%%%%%%%%%%%%%%%%%%%%%%%%%%%%

\begin{acknowledgements}
The authors thank the referee for useful comments that helped to improve the quality and perspectives of this work. The authors thank D.~Barret for fruitful discussions that led to improve this manuscript. NC thanks A.~Marin-Lafl{\`e}che for help in the selection of an FFT algorithm suited to this research, A.~Proag for useful discussions on geometrical calculations and L.~Le~Briquer for commenting a preliminary version of this manuscript.
This research made use of Astropy,\footnote{http://www.astropy.org} a community-developed core Python package for Astronomy \citep{astropy:2013, astropy:2018}. This research made use of Matplotlib \citep{hunter2007}.
\end{acknowledgements}

\bibliographystyle{aa} % style aa.bst
\bibliography{myreferences} % your references Yourfile.bib

%%%%%%%%%%%%%%%%%%%%%%%%%%%%%%%%%%%%%%%%%%%%%%%%%%%%%%%%%%%%%%%%%
%%%%%%%%%%%%%%%%%%%%%%%%%%%%%%%%%%%%%%%%%%%%%%%%%%%%%%%%%%%%%%%%%
\onecolumn
    \appendix

\section{Calculations in the 1-dimensional case}
	\label{app:1d_calc}
	
This Appendix details the calculation leading to results presented in Sect.~\ref{sect:1d_statistics}, namely the expected values for the centroid shift $C$ and line broadening $S^2$ and their variances, in the case of measurements along a single line-of-sight (Eq.~\ref{eq:c1d_mean}, \ref{eq:c1d_var}, \ref{eq:s1d_mean} and~\ref{eq:s1d_var}). These calculations are generalised in Sect.~\ref{sect:2d_statistics} for 2-dimensional diagnostics of the velocity field.
	
	\subsection{Statistics of the centroid}

The finding $\langle C \rangle = 0$ is a direct consequence of random uncorrelated phases.

We calculate $\mathrm{Var}(C)= \langle C^2 \rangle$ by noting that:
\begin{equation}
    \langle C^2 \rangle = F^{-2} \iint \epsilon(x) \epsilon(x^{\prime}) \langle v(x) v(x^{\prime}) \rangle \dd x \dd x^{\prime}
\end{equation}

From eq.~\ref{eq:coeffaverage}, the term within brackets reads:
\[
\langle v(x) v(x^{\prime}) \rangle = \sum_{k} P(k) \exp{\left( i k \omega(x^{\prime}-x)\right)}
\]

It is then easily shown that:
\[
\langle C^2 \rangle = F^{-2} \sum_k P(k) \left| \int \epsilon(x) \exp\left( i k \omega x \right) \dd x \right|^2
\]

The term within modulus is $\widetilde{\epsilon}(k)$, namely the $k^{th}$ Fourier coefficient of $\epsilon$. This identification leads to the expression shown in Eq.~\ref{eq:c1d_var}.

    \subsection{Statistics of the dispersion}

The variations of $S^2$ are due to the second term of equation~\ref{eq:dispersion}, therefore we will focus now on studying the statistics of the double integral $\iint G$.

        \subsubsection{Average of $\iint G$}

First we write:
\[
A \equiv \langle \iint G \rangle = \iint \langle G \rangle = \iint \epsilon(x) \epsilon(x^{\prime}) \langle \left[v(x)-v(x^{\prime})\right]^2\rangle
\]

And:
\[
v(x)-v(x^{\prime}) = \sum_k V_k \left( e^{i k \omega x} - e^{i k \omega x^{\prime}} \right)
\]

Therefore, using Eq.~\ref{eq:coeffaverage}:
\[
\left\langle \left[ v(x)-v(x^{\prime}) \right]^2 \right\rangle =  \sum_k P(k) \left| e^{i k\omega x} - e^{i k\omega x^{\prime}} \right|^2 = 2 \sum_k P(k) \left[ 1- \cos(k \omega (x^{\prime}-x)) \right] 
\]

This leads to:
\[
A = 2 \sum_k P(k) \iint \dd x \dd x^{\prime} \epsilon(x) \epsilon(x^{\prime}) \times \left[1- \cos(k \omega x^{\prime}) \cos(k \omega x) - \sin(k \omega x^{\prime}) \sin(k \omega x) \right]
\]

This expression again can be rewritten using the power-spectrum of the emissivity:
\begin{equation}
\label{eqn:result_A}
A = 2 \sum_k P(k) \left[ F^2 -  P_{\epsilon}(k) \right]
\end{equation}

The average of the measured line width then reads:
\[
\langle S^2 \rangle = \frac{1}{F} \int \epsilon(x) \sigma_{th}^2(x) \dd x + \sum_k P(k) \left[ 1 - \frac{P_{\epsilon}(k)}{F^2} \right]
\]

Which we write under the simple form:
\[
\langle S^2 \rangle = \overline{\sigma_{th}^2} + \sigma_{turb}^2 - F^{-2} \sum_k P(k) P_{\epsilon}(k)
\]
where an horizontal bar denotes averaging of the thermal component along the line-of-sight.

        \subsubsection{Variance of $\iint G$}

We define $B$ such that $\mathrm{Var}(S^2) = F^{-4} (B^2 - A^2)/4$.
We note that:
\begin{equation}
      \left \langle \left( \iint G \right)^2 \right\rangle = B^2 = \int \epsilon(x) \epsilon(y) \epsilon(z) \epsilon(t) \left \langle \left[ v(x)-v(y) \right]^2 \left[ v(z)-v(t) \right]^2 \right \rangle \dd x \dd y \dd z \dd t
\end{equation}

The term within brackets reads:
\begin{equation}
\langle [.]^2 \times [.]^2 \rangle = \sum_{j,k,l,m} \langle V_j V_ k V_l V_m \rangle \times \left( e^{i k \omega x} - e^{i k \omega y} \right) \left( e^{i j \omega x} - e^{i j \omega y} \right) \times \left( e^{i l \omega z} - e^{i l \omega t} \right) \left( e^{i m \omega z} - e^{i m \omega t} \right)
\end{equation}

Since phases are two-by-two independent, we assume the simplest possible expression for the 4-term product of Fourier coefficients $V_k$, namely:
\[
    \langle V_j V_k V_l V_m \rangle = \left\{
    \begin{array}{cl}
    P(k) P(l) & \mathrm{if \ } (k=-j) ; (l=-m) ; (k \neq \pm l) \, \{A\}\\
    P(k) P(j) & \mathrm{if \ } (k=-l) ; (j=-m) ; (k \neq \pm j) \, \{B\}\\
    P(k) P(j) & \mathrm{if \ } (k=-m) ; (j=-l) ; (k \neq \pm j) \, \{C\}\\
    \langle |V_k|^4 \rangle & \mathrm{if \ } (k = -j = l = -m) \, \{D\}\\ 
    \langle |V_k|^4 \rangle & \mathrm{if \ } (k = -j = -l = m) \, \{E\}\\
    \langle |V_k|^4 \rangle & \mathrm{if \ } (k = j = -l = -m) \, \{F\}\\
    0 & \mathrm{else}\\
    \end{array} \right.
\]

All cases are mutually exclusive. Remarking that conditions $B$ and $C$ lead to identical expressions under transformation $l \leftrightarrow m$, and similarly for $D$ and $E$, we can rewrite the sum, hence the triple integral, with a sum of 4 terms:
\[
B^2 = b_A + 2 b_B + 2 b_D + b_F
\]

\paragraph{First term:}
If $(k=-j)$ and $(l=-m)$ the bracket writes:
\begin{equation*}
\langle . \rangle_A  =  \sum_{k \neq \pm l} P(k) P(l) \left| e^{i k \omega x} - e^{i k \omega y} \right|^2   \left| e^{i l \omega z} - e^{i l \omega t} \right|^2
\end{equation*}

The integration over $x,y,z$ and $t$ provides after some algebra:
\begin{equation}
b_A = 4 \sum_{j \neq \pm k} P(k) P(j) \left[ F^4 - F^2 P_{\epsilon}(k) - F^2 P_{\epsilon}(j) + P_{\epsilon}(k) P_{\epsilon}(j) \right] = 4 \sum_{j \neq \pm k} P(k) P(j) \left[ F^2 - P_{\epsilon}(k) \right] \left[ F^2 - P_{\epsilon}(j) \right]
\end{equation}
where we have used the same trigonometric decomposition as for deriving Eq.~\ref{eqn:result_A}.

\paragraph{Second term:}
The symetries in the expression lead to:
\[
b_B = \sum_{k \neq \pm j} P(k) P(j) \left| \iint \epsilon(x) \epsilon(y) f_k(x,y) f_j(x,y) \dd x \dd y \right|^2
\]
introducing the complex function:
\[
f_k(x,y) = f_{-k}^*(x,y) = e^{i k \omega x} - e^{i k \omega y}
\]

Developing the product $f_k f_j$ we can rewrite the term under the modulus as:
\[
\left|\iint ... \right|^2 = 4 \left|  F \widetilde{\epsilon}(j+k) - \widetilde{\epsilon}(j) \widetilde{\epsilon}(k) \right|^2
\]

\paragraph{Third term:}
The term within brackets is rewritten as:
\[
\langle . \rangle_D = \sum_k \langle |V_k|^4 \rangle \left| e^{ik \omega x}- e^{i k \omega y} \right|^2 \left| e^{ik \omega z}- e^{i k \omega t} \right|^2
\]
which, using similar calculations as for $b_A$, leads to:
\[
b_D = 4 \sum_k \langle |V_k|^4 \rangle \left[F^2 - P_{\epsilon}(k) \right]^2
\]

\paragraph{Fourth term:}
Similarly as for the computation of $b_F$ above, we find:
\[
b_F = 4 \sum_k \langle |V_k|^4 \rangle \left| F \widetilde{\epsilon}(2k) - \widetilde{\epsilon}(k)^2 \right|^2
\]

Therefore, combining previous expressions we obtain:
\begin{multline*}
\mathrm{Var}(S^2) = 2 \sum_{j \neq \pm k} P(k)P(j) \left| \frac{\widetilde{\epsilon}(j+k)}{F}-\frac{\widetilde{\epsilon}(j) \widetilde{\epsilon}(k)}{F^2} \right|^2 +  \sum_k \langle |V_k|^4 \rangle \left| \frac{\widetilde{\epsilon}(2k)}{F} - \frac{\widetilde{\epsilon}(k)^2}{F^2} \right|^2 \\
+ 2 \sum_k \left( \langle |V_k|^4 \rangle - P(k)^2 \right) \left[ 1- \frac{P_{\epsilon}(k)}{F^2} \right]^2
\end{multline*}
which is rearranged so to provide Eq.~\ref{eq:s1d_var}.

%%%%%%%%%%%%%%%%%%%%%%%%%%%%%%%%%%%%%%%%%%%%%%%%%%%%%%%%%%%%%%%%%
%%%%%%%%%%%%%%%%%%%%%%%%
    \section{Fourier transform of the emissivity field $\emithree$ ($\beta$-model)}
		\label{app:fourier_eta}

We provide here calculations of $c_{\emithree.\mathcal{W}}$, the 3-d power-spectrum of the emissivity $\emithree(x,y,z)$ in a case of a $\beta$-model, seen through a sky aperture $\mathcal{W}(y,z)$. This is particularly useful for deriving the line centroid and broadening statistics. We assume that $\emithree \propto n_e^2$ where $n_e$ is the gas density and effectively follows a $\beta$-model profile, as in a isothermal, isometallic intra-cluster medium.

		\subsection{Spherical model}

For a spherical $\beta$-model density with core-radius $r_c$ centred on $\vec \theta = 0$, the emissivity is expressed as:
\[
\emithree(x,\vec \theta)= \emithree(0) \left( 1+ \frac{x^2 + \theta^2}{r_c^2} \right)^{-3 \beta}
\]

The flux integrated along the line-of-sight writes:
\begin{align}
\label{eq:flux_beta2d}
F(\vec \theta) & = \emithree(0) r_c u_{\beta} \left( 1+ \frac{\theta^2}{r_c^2} \right)^{1/2-3\beta}\\
	\mathrm{with:\ } u_{\beta} & = 2 \int_{0}^{\pi/2} \cos^{6\beta-2}(t) \dd t = \sqrt{\pi} \frac{\Gamma(3\beta-1/2)}{\Gamma(3 \beta)} \nonumber .
\end{align}

The value of $c_{\emithree.\mathcal{W}}$ is defined as follows:
\[
c_{\emithree.\mathcal{W}}(k_x, \vec \xi) = \iint \mathcal{W}(\vec \theta) \emithree(x,\vec \theta) e^{-i \omega (k_x x + \vec \xi \cdot \vec \theta)} \dd x \dd \vec \theta = \emithree(0)  \int \dd \vec \theta \mathcal{W}(\vec \theta) e^{-i \omega \vec \xi \cdot \vec \theta} \int \dd x e^{-i \omega k_x x }\left(1+ \frac{x^2+\theta^2}{r_c^2} \right)^{-3 \beta}
\]

This Fourier transform is calculated first along the x-axis:
\begin{align}
\int_{- \infty}^{+\infty} \left( 1+ \frac{x^2+\theta^2}{r_c^2} \right)^{-3\beta} e^{-i \omega k_x x} \dd x & = 2 r_c^{6 \beta} (\omega |k_x|)^{6 \beta-1} \int_0^{+ \infty} \frac{\cos (t) \dd t}{\left[ (\omega k_x)^2 (\theta^2 + r_c^2) + t^2\right]^{3 \beta}} \nonumber \\
	& =  \frac{2^{3/2-3\beta} \sqrt{\pi}}{\Gamma(3 \beta)} r_c^{6 \beta} \left( \frac{\sqrt{\theta^2 + r_c^2}}{\omega |k_x|} \right)^{1/2-3 \beta} K_{3\beta -1/2}\left(\omega |k_x| \sqrt{\theta^2+r_c^2}\right) \label{eq:tf_beta1d}
\end{align}
where $K_n$ is the modified Bessel function of the second kind\footnote{The case $k_x=0$ is recovered using $\lim_{x \rightarrow 0} x^n K_n(x) = \pi 2^{n-1}/ \left[ \sin(n \pi) \Gamma(1-n) \right]$ \citep{spiegel2003}.}. In the special case of $\beta=2/3$ this formula is equivalent\footnote{Because $\mathcal{F}_{3/2}(x) = x^{3/2} K_{3/2}(x) = \sqrt{\pi/2}(1+x) \exp(-x)$} to equation~23 in \citet{zuhone2016}, namely: $\widetilde{\epsilon}/F=(1+\omega |k_x| c) \exp(-\omega |k_x| c)$, with $c^2 = \theta^2+r_c^2$.

Introducing $\mathcal{F}_n(x) = x^{n} K_n(x)$, the integration over the plane-of-sky coordinates $\vec \theta$ provides:
\begin{equation}\label{eq:cetaw_beta}
c_{\emithree.\mathcal{W}}(k_x, \vec \xi) = \emithree(0) r_c \frac{2^{3/2-3\beta} \sqrt{\pi}}{\Gamma(3 \beta)}   \int \left(1+\frac{\theta^2}{r_c^2} \right)^{1/2-3\beta} \mathcal{F}_{3\beta-1/2} \left( \omega |k_x| \sqrt{\theta^2+r_c^2} \right) \mathcal{W}(\vec \theta) e^{-i \omega \vec \xi \cdot \vec \theta} \dd \vec \theta
\end{equation}

The unknown normalization factor $\emithree(0)$ is unimportant in this paper, since the Fourier transform always appears divided by the aperture flux $F_{\mathcal{W}}$ defined by:
\[
F_{\mathcal{W}} = \int F(\vec \theta) \mathcal{W}(\vec \theta) \dd \vec \theta = \emithree(0) r_c u_{\beta} \int \left(1+ \frac{\theta^2}{r_c^2} \right)^{1/2-3\beta} \mathcal{W}(\vec \theta) \dd \vec \theta
\]

\begin{figure}
    \centering
	    \includegraphics[width=\linewidth]{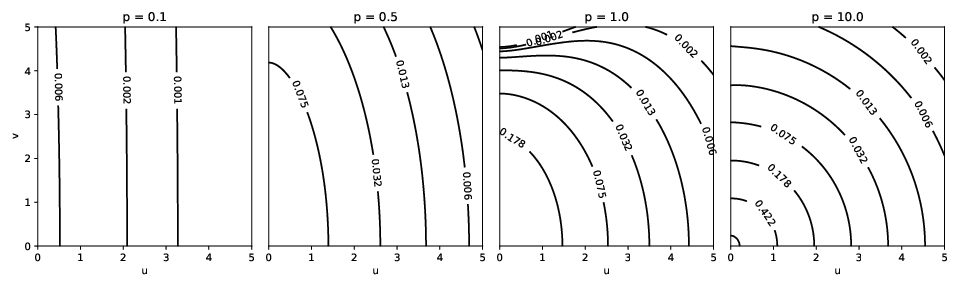}
    \caption{Numerical calculations of $\mathcal{I}_{p;n}(u,v)$ for various values of $p$ and $n=3/2$. Logarithmically spaced contours (identical in all panels) indicate the value of the function. This function is involved in the calculation of the Fourier transform of a spherical $\beta$-model ($n=3\beta-1/2$) observed through a concentric circular aperture of radius $p$ times the core radius.}
    \label{fig:ipn_abaque}
\end{figure}

An usual practical case is for a circular aperture $\mathcal{W}$ of radius $R_{ap}$ centred on $\vec \theta = 0$. For this particular case:

\begin{equation*}
\frac{c_{\emithree.\mathcal{W}}}{F_{\mathcal{W}}}(k_x, \vec \xi) = \frac{2^{5/2-3\beta} (3\beta-3/2)}{\Gamma(3\beta-1/2)} \left( 1- \left[ 1+ \frac{R_{ap}^2}{r_c^2} \right]^{3/2-3\beta} \right)^{-1} \times \mathcal{I}_{(R_{ap}/r_c); (3\beta-1/2)}\left(\omega |k_x|r_c, \omega \xi r_c\right)
\end{equation*}
which uses the special integral defined below and represented in Fig.~\ref{fig:ipn_abaque} for $n=3/2$ ($\beta=2/3$):
\[
\mathcal{I}_{p;n}(u,v) =  \int_{0}^{p} \frac{t J_0(v t)}{(1+t^2)^n} \mathcal{F}_{n}\left(u \sqrt{1+t^2} \right) \dd t
\]

		\subsection{Plane-constant model}

The integral~\ref{eq:cetaw_beta} can be simplified if the core-radius $r_c$ is much larger than the typical size of the window function $\mathcal{W}$. In such case, it is equivalent to consider an emissivity that is independent of the line-of-sight direction $\vec \theta$, i.e.~$\emithree(x,y,z) = \epsilon(x)$. An effective impact parameter $\theta_\mathrm{eff}$ is introduced so that:
\[
\epsilon(x) = \epsilon(0) \left(1+ \frac{x^2+\theta_\mathrm{eff}^2}{r_c^2}\right)^{-3 \beta}
\]

The calculations above then become:
\begin{align}
c_{\emithree . \mathcal{W}}(k_x, \vec \xi) & =  F\left( \theta_\mathrm{eff}\right)\frac{2^{3/2-3\beta}}{\Gamma(3 \beta-1/2)} \mathcal{F}_{3\beta-1/2}\left(\omega |k_x| c\right) \widehat{\mathcal{W}}(\vec \xi)\\
F_{\mathcal{W}} & = \mathcal{S}_{\mathcal{W}} F\left(\theta_\mathrm{eff} \right) \nonumber
\end{align}
where we introduced $c^2 = \theta_\mathrm{eff}^2 + r_c^2$ and $\mathcal{S}_{\mathcal{W}} = \int \mathcal{W}$ is the area of the aperture on sky and $\widehat{\mathcal{W}}$ its (2-d) Fourier transform.

An usual practical case is for a circular aperture $\mathcal{W}$ of radius $R_{ap}$ and an emissivity $\epsilon(x)$ in form of a $\beta=2/3$-model independent of the line of sight. For this particular case:
\[
\frac{c_{\emithree.\mathcal{W}}(k_x, \vec \xi)}{F_{\mathcal{W}}} = 2 e^{- \omega c |k_x|} \left( 1 + \omega c |k_x| \right) \frac{J_1(\omega \xi R_{ap})}{\omega \xi R_{ap}}
\]
with $J_1$ the Bessel function of the first kind and order 1.

%%%%%%%%%%%%%%%%%%%%%%%%%%%%%%%%%%%%%%%%%%%%%%%%%%%%%%%%%%%%%%%%%
\section{Two-dimensional power-spectrum for generic emissivity field}
    \label{app:p2d_general}

For a pencil-beam aperture $\mathcal{W}(\vec \theta^{\prime}) = \delta(\vec \theta - \vec \theta^{\prime})$, the expression for the centroid shift writes:
\[
C(\vec \theta) = \int \rho(x, \vec \theta) v(x, \vec \theta) \dd x
\]

Since $\chi^{\vec \theta}(x) = \rho(x,\vec \theta)$ we obtain:
\begin{equation}
	C(\vec \theta) = \sum_{\kthree} V_{\kthree} e^{i \omega \vec \xi \cdot \vec \theta} \widetilde{\chi^{\vec \theta}}(k_x)
\end{equation}
where tilde indicates one-dimensional Fourier transform along direction $x$. Indeed,
\begin{equation}
\label{eq:chitilde}
\widetilde{\chi^{\vec \theta}}(k_x) = \int \dd x e^{i \omega k_x x} \rho(x, \vec \theta) = \left(\frac{\omega}{2 \pi} \right)^2 \int  \dd \vec \xi^{\prime}  e^{-i \omega \vec \xi^{\prime}\cdot \vec \theta} \widetilde{\rho}(k_x,\vec \xi^{\prime})
\end{equation}
with $\widetilde{\rho}$ the (3D) Fourier transform of $\rho$. The last equality derives from the definition of the inverse 3-dimensional Fourier transform.

We remark that $\rho(x,y,z)=\emithree/F$ is defined in a domain of space $\mathcal{A}$ (area $\fovarea$) where the flux $F(y,z)$ of the source is non-zero (which in practice is a bounded region). If the source is infinitely extended (as in the formal case of a $\beta$-model) the boundary is imposed by the domain of analysis $\mathcal{A}$, e.g.~the instrument field of view. We therefore consider that $\rho$ is defined over the entire 3-dimensional space by filling regions outside of the bounded domain with zeros, which ensures the existence of $\widetilde{\rho}$. The 2D Fourier transform $\widehat{C}(\vec \xi)$ of $C(\vec \theta)$ is used to define:
\[
P_{2D}(\vec \xi) = \frac{1}{\fovarea} \left\langle \left| \widehat{C}(\vec \xi) \right|^2 \right\rangle
\]
The weighting by the total area ensures that the total 'energy' does not diverge as $\mathcal{A}$ becomes large. We provide later the expression for the limiting case of an infinitely extended analysis domain.

Equation~\ref{eq:chitilde} shows that at any given $k_x$, $\widetilde{\chi^{...}}(k_x)$ is the 2-dimensional inverse Fourier transform of $\widetilde{\rho}(k_x,...)$ and then:
\[
\int_{\vec \theta} \widetilde{\chi^{\vec \theta}}(k_x) e^{i \omega \vec \xi \cdot \vec \theta} \dd \vec \theta = \widetilde{\rho}(k_x,\vec \xi)
\]

Therefore:
\begin{equation*}
	\widehat{C}(\vec \xi) = \sum_{\kthree= (k_x, \vec \alpha)} V_{\kthree} \int_{\vec \theta} e^{i \omega (\vec \alpha + \vec \xi) \cdot \vec \theta} \widetilde{\chi^{\vec \theta}}(k_x) \dd \vec \theta
\end{equation*}
which leads to:
\begin{equation*}
\left| \widehat{C}(\vec \xi) \right|^2 = \sum_{\kthree_1, \kthree_2} V_{\kthree_1} V^*_{\kthree_2} \int_{\vec \theta_1, \vec \theta_2} e^{i \omega \vec \xi \cdot \left( \vec \theta_1 - \vec \theta_2 \right)} e^{i \omega \left( \vec \alpha_1 \cdot \vec \theta_1 - \vec \alpha_2 \cdot \vec \theta_2 \right)} \widetilde{\chi^{\vec \theta_1}}(k_{x_1}) \widetilde{\chi^{\vec \theta_2}}^*(k_{x_2}) \dd \vec \theta_1 \dd \vec \theta_2
\end{equation*}

Averaging over all possible realisations provides:
\begin{align*}
P_{2D}(\vec \xi) & = \frac{1}{\fovarea} \sum_{\kthree} P_{3D}(k) \int_{\vec \theta_1, \vec \theta_2} e^{i \omega \left( \vec \xi + \vec \alpha \right) \cdot \left( \vec \theta_1 - \vec \theta_2 \right)} \widetilde{\chi^{\vec \theta_1}}(k_{x}) \widetilde{\chi^{\vec \theta_2}}^*(k_{x}) \dd \vec \theta_1 \dd \vec \theta_2 \\
	& = \frac{1}{\fovarea} \sum_{\kthree} P_{3D}(k) \left| \int_{\vec \theta} \widetilde{\chi^{\vec \theta}}(k_{x}) e^{i \omega (\vec \xi + \vec \alpha) \cdot \vec \theta}  \dd \vec \theta \right|^2 \\
	& = \frac{1}{\fovarea} \sum_{\kthree= (k_x, \vec \alpha)} P_{3D}(k) P_{\rho}\left(k_x,\vec \alpha + \vec \xi \right)
\end{align*}
Which is equivalent to:

\begin{equation}
\label{eq:p2d_calcul}
P_{2D}(\vec \xi) = \frac{1}{\fovarea} \sum_{k_x, \vec \xi^{\prime}} P_{3D}\left(\sqrt{k_x^2 + |\vec \xi^{\prime}|^2}\right) P_{\rho} \left(k_x, \vec \xi - \vec \xi^{\prime} \right)
\end{equation}

We note that $P_{2D}$ is in general non isotropic, as the emissivity and the shape of the analysis domain are arbitrary.
However, in the particular case where the normalized line-of-sight emissivity is independent of the line-of-sight, i.e.~following our previous notations $\chi^{\vec \theta}(x) \equiv \epsilon(x)/F$, we find that:
\[
P_{\rho}(k_x, \vec \xi) = \frac{1}{F^2} P_{\epsilon}(k_x) P_{\mathcal{A}}(\vec \xi)
\]
with:
\[
\int P_{\mathcal{A}}(\vec \xi) \dd \vec \xi = \left( \frac{2\pi}{\omega}\right)^2 \fovarea
\]

This leads to the following expression:
\[
P_{2D}(\vec \xi) = \frac{1}{\fovarea} \sum_{\vec \xi^{\prime}} P_{\mathcal{A}}(\vec \xi-\vec \xi^{\prime}) \sum_{k_x} \frac{P_{\epsilon}(k_x)}{F^2}  P_{3D}\left(\sqrt{k_x^2 + |\vec \xi^{\prime}|^2}\right)  = \frac{1}{\fovarea} \left( \frac{\omega}{2\pi}\right)^2 \left( P_{\mathcal{A}} \otimes P_{2D}^{\infty}\right)(\vec \xi)
\]
with $\otimes$ representing the discrete convolution product. The power-spectrum $P_{2D}^{\infty}$ is defined such as it matches $P_{2D}$ for an extremely large domain of analysis. Indeed, $P_{\mathcal{A}}$ then becomes a very peaked function around $\vec \xi = 0$ and we obtain \citep[see also][]{zhuravleva2012}:
\begin{equation}
\label{eq:p2dinfini_calcul}
P_{2D}^{\infty}(\vec \xi) \equiv \lim_{\mathcal{A} \rightarrow \infty} P_{2D}(\vec \xi)  \simeq \left( \frac{2\pi}{\omega}\right)^2 \sum_{k_x} P_{3D}\left(\sqrt{k_x^2 + \xi^2}\right)\frac{P_{\epsilon}(k_x)}{F^2}
\end{equation}

%%---------%%
\section{Structure function for generic emissivity field}
    \label{app:sf_general}
 
	\subsection{Formal derivation neglecting border effects}
	
For convenience, we introduce the $\mathcal{W}$-normalized emissivity: $\chi^{\mathcal{W}}(\xthree)= F_{\mathcal{W}}^{-1} \emithree(\xthree)$. By extension, we define $\chi^{\vec \theta}(x) = \emithree(x, \vec \theta)/F(\vec \theta) = \rho(x, \vec \theta)$. We recall that $\rho=0$ outside of the domain of analysis by construction, this is equivalent to imposing the centroid shift to vanish outside of this region. 
Using the decomposition of the velocity field in Fourier series and the definition of the velocity power spectrum, one obtains:
\[
\left\langle \left| C_{\mathcal{W}}- C_{\mathcal{W}^{\prime}} \right|^2 \right\rangle = \sum_{\kthree} P_{3D}(| \kthree |) \left| \int \dd \xthree \,  e^{i \omega \kthree \cdot \xthree} \left( \mathcal{W}(\vec \theta) \chi^{\mathcal{W}}(\xthree) - \mathcal{W}^{\prime}(\vec \theta) \chi^{\mathcal{W}^{\prime}}(\xthree) \right) \right|^2
\]

which for the most common "pencil-beam" window function, $\mathcal{\vec \theta} = \delta(\vec \theta_0-\vec \theta)$, reduces to:
\begin{equation}
    \left \langle \left| C(\vec\theta_0+ \vec r) - C(\vec\theta_0) \right|^2 \right \rangle = \sum_{\kthree} P_{3D}(|\kthree|)  \left| C_{\vec \theta_0,\vec r}(\kthree)\right|^2
\end{equation}
with:
\[
C_{\vec \theta_0,\vec r}(\kthree) = \int \dd x e^{ik_x \omega x} \left[ \chi^{\vec \theta_0+ \vec r}(x) e^{i \omega \vec \xi \cdot \vec r} - \chi^{\vec \theta_0}(x) \right]
\]
  
The expected value for the structure function therefore writes:
\[
\mathrm{sf}(\sepa) = \frac{1}{N_p(\sepa)} \sum_{\kthree} P_{3D}(k) I_{\sepa}(\kthree)
\]
with:
\[
I_{\sepa}(\kthree) = \int_{\vec \theta, |\vec r|=\sepa} |C_{\vec \theta, \vec r}(\kthree)|^2
\]

Following notations in previous appendix, we can rewrite $C_{\vec \theta, \vec r}(\kthree)$ into:
\[
C_{\vec \theta, \vec r}(\kthree) = e^{i\omega \vec \xi \cdot \vec r} \widetilde{\chi^{\vec \theta+\vec r}}(k_x) - \widetilde{\chi^{\vec \theta}}(k_x)
\]
we then obtain:
\begin{equation*}
|C_{\vec \theta, \vec r}(\kthree)|^2 = |\widetilde{\chi^{\vec \theta}}(k_x)|^2 + |\widetilde{\chi^{\vec \theta + \vec r}}(k_x)|^2 
- 2 \times \mathrm{Re} \left[ e^{i\omega \vec \xi \cdot \vec r} \widetilde{\chi^{\vec \theta}}^{*}(k_x) \widetilde{\chi^{\vec \theta+ \vec r}}(k_x) \right]
\end{equation*}

In a first approximation, let us perform summation over all pairs, including those fully comprised within the domain of analysis $\mathcal{A}$ ("inner" pairs on Fig.~\ref{fig:sketch_geom}) and those with only one end in $\mathcal{A}$ ("Ext" pairs). By construction $\widetilde{\chi^{\vec \theta}}=0$ for $\vec \theta$ outside of $\mathcal{A}$. This approximation is equivalent to neglecting border effects and correction terms are discussed in the following subsection.

Using the relation between $\widetilde{\chi}$ and $\widetilde{\rho}$ identified previously, we obtain:
\[
\int_{p} |\widetilde{\chi^{\vec \theta}}(k_x)|^2 + |\widetilde{\chi^{\vec \theta + \vec r}}(k_x)|^2 = 2 \pi \int_{\vec \theta} |\widetilde{\chi^{\vec \theta}}(k_x)|^2 = 2 \pi \left(\frac{\omega}{2\pi}\right)^2 \int P_{\rho}(k_x,\vec \xi) \dd \vec \xi
\]

Using Eq.~\ref{eq:chitilde} and some algebra leads to:
\begin{equation*}
\widetilde{\chi^{\vec \theta}}^{*}(k_x) \widetilde{\chi^{\vec \theta+ \vec r}}(k_x) = \left(\frac{\omega}{2\pi}\right)^2 \int  \dd \vec\xi^{\prime} \dd \vec\xi^{\prime\prime} \widetilde{\rho}^*(k_x, \vec\xi^{\prime}) \widetilde{\rho}(k_x, \vec\xi^{\prime\prime}) e^{i\omega (\vec\xi^{\prime} - \vec\xi^{\prime\prime})\cdot \vec \theta} e^{-i \omega \vec\xi^{\prime\prime} \cdot \vec r}
\end{equation*}

Summing the term under the $\mathrm{Re}$ function over all pairs (without double-counting) we write:
\begin{align*}
\int_{p} \left[ e^{i\omega \vec \xi \cdot \vec r} \widetilde{\chi^{\vec \theta}}^{*}(k_x) \widetilde{\chi^{\vec \theta+ \vec r}}(k_x) \right] = & \frac{1}{2} \int_{\vec \theta} \int_{|\vec r|=\sepa} \left[ e^{i\omega \vec \xi \cdot \vec r} \widetilde{\chi^{\vec \theta}}^{*}(k_x) \widetilde{\chi^{\vec \theta+ \vec r}}(k_x) \right] \\
	= & \frac{1}{2}  \left(\frac{\omega}{2\pi}\right)^2 \int \dd \vec \xi^{\prime} P_{\rho}(k_x,\vec \xi^{\prime}) \int_{0}^{2\pi} e^{i \omega |\vec \xi + \vec \xi^{\prime}| \sepa \cos \phi} \dd \phi \\
= & \pi  \left(\frac{\omega}{2\pi}\right)^2 \int \dd \vec \xi^{\prime} P_{\rho}(k_x,\vec \xi^{\prime}) J_0(\omega |\vec \xi + \vec \xi^{\prime}| \sepa)
\end{align*}

We finally obtain:
\begin{equation}
\label{eq:i_delta}
I_{\sepa}(\kthree) = 2 \pi \left(\frac{\omega}{2\pi}\right)^2 \int \dd\vec \xi^{\prime} P_{\rho}(k_x,\vec \xi^{\prime}) \left[ 1 - J_0\left(\left| \vec \xi + \vec \xi^{\prime} \right| \omega \sepa \right) \right]
\end{equation}

Dividing by the total number of pairs $N_p^{tot}(\sepa) \simeq \frac{1}{2} N_p(\vec \theta) \times N_p(r = \sepa) = \pi \fovarea$ provides the general expression for the expected value of the structure function (see Eq.~\ref{eq:sf_average}). 
These calculations assume that integration over all pairs $(\vec \theta, \vec r)$ is continuous. Appendix~\ref{app:filter_field} describes the effect of pixelized and filtered data.

	\subsection{Finite-size effects (circular domain of analysis)}

\begin{figure}
    \centering
  \includegraphics[width=0.5\linewidth]{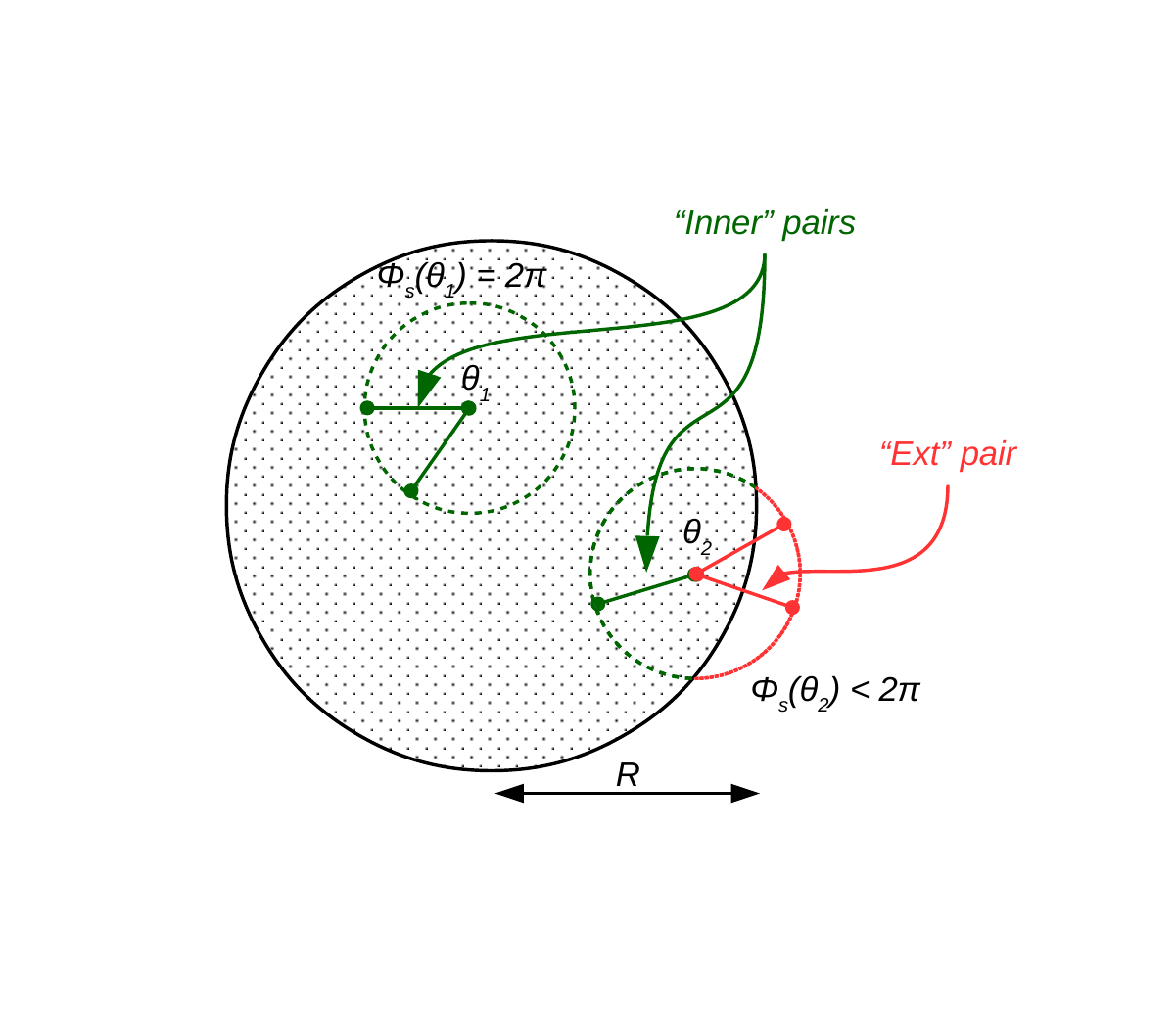}
      \caption{Sketch illustrating the counting of pairs within a circular domain of analysis of radius $R$ represented by the large black circle. Within this domain, the centroid shift $C(\vec \theta)$ takes values determined by the stochastic turbulent field, while we set $C=0$ outside. Counting inner pairs (materialized with green and red sticks) separated by a distance $\sepa$ is performed by computing the range of accessible angles $\phi_{\sepa}(\vec \theta)$ for a given position in the domain, then dividing by two. External pairs have only one end within the domain of analysis and the range of accessible angles is $2\pi -\phi_{\sepa}$ at a given position.}
    \label{fig:sketch_geom}
\end{figure}

Previous calculation neglects border effects in the integration over pairs of points. We have set $\rho = 0$ outside of the domain of analysis, which implies $C=0$. Consequently a number of extra pairs are erroneously included in this derivation, translating into extra terms $\langle |C(\vec \theta + \vec r) - C(\vec \theta)|^2\rangle = \langle|C(\vec \theta)|^2 \rangle$ in the numerator $I_{\sepa}$ and the number of pairs entering the denominator $N_p(\sepa)$ needs to be corrected (see Fig.~\ref{fig:sketch_geom}).

The integrals shown previously run over all pairs separated by $\sepa$ with at least one extremity within the field of view. There are $N_p^{tot}(\sepa)$ such pairs and $N_p^{ext}(\sepa)$ pairs with only one end within the field of view. Naturally we denote $N_p^{in}=N_p^{tot}-N_p^{ext}$ the pairs fully comprised within the domain analysis. A given point $\vec \theta$ in the analysis domain belongs to $\phi_{\sepa}(\vec \theta) \in [0, 2\pi]$ pairs in the field of view.
Let us first compute the exact number of pairs, assuming an infinitely fine tessellation:
\[
N_p^{tot}(\sepa) = \pi \fovarea + \frac{1}{2} N_p^{ext}(\sepa)
\]
We write:
\[
\left( N_p^{tot}(\sepa) - \frac{1}{2} N_p^{ext}(\sepa) \right) \mathrm{sf}(\sepa)  = \int_{p} \langle . \rangle = \int_\mathrm{in} \langle . \rangle^{in} + \int_\mathrm{ext} \langle . \rangle^\mathrm{ext} = N_p^{in} \mathrm{sf}^{\rm corr}(\sepa) + \int_\mathrm{ext} \langle . \rangle^{ext}
\]
denoting by $\mathrm{sf}^{\rm corr}(\sepa) = 1/N_p^{in} \int_\mathrm{in} \langle . \rangle$ the value of the structure function corrected from finite-size effects.
We obtain:
\[
\int_\mathrm{ext} \langle . \rangle^{ext} = \int_{\vec \theta \in \mathcal{A}} (2 \pi - \phi_{\sepa}(\vec \theta)) \langle | C(\vec \theta)|^2 \rangle \dd \vec \theta
\]
As demonstrated in Sect.~\ref{sect:1d_statistics}:
\[
\langle | C(\vec \theta)|^2 \rangle = \sum_{\kthree} P_{3D}(\kthree) \left|\widetilde{\chi^{\vec \theta}}(k_x) \right|^2
\]

Reassembling terms, we obtain the corrected mean structure function $(\mathrm{sf}^\mathrm{corr})$:
\begin{equation}\label{eq:sf_average_corrected}
\mathrm{sf}^\mathrm{corr}(\sepa) = \left( \frac{N_p^{ext}(\sepa)}{2N_p^{in}(\sepa)} +1 \right) \mathrm{sf}(\sepa) - \frac{1}{N_p^{in}(\sepa)} \sum_{\kthree} P_{3D}(\kthree) \int_{\vec \theta \in \mathcal{A}} (2 \pi - \phi_{\sepa}(\vec \theta)) \left|\widetilde{\chi^{\vec \theta}}(k_x) \right|^2 \dd \vec \theta
\end{equation}

The correction term depends both on the number extra pairs and on their separation $\sepa$ relative to the size of velocity fluctuations.

We have:
\[
N_p^{in}(\sepa) = \frac{1}{2} \int_{\vec \theta \in \mathcal{A}} \phi_{\sepa}(\vec \theta) \dd \vec \theta
\]
\[
N_p^{ext}(\sepa) = \int_{\vec \theta \in \mathcal{A}} (2 \pi - \phi_{\sepa}(\vec \theta)) \dd \vec \theta
\]

\begin{figure}
    \centering
    \begin{tabular}{cc}
  \includegraphics[width=0.5\linewidth]{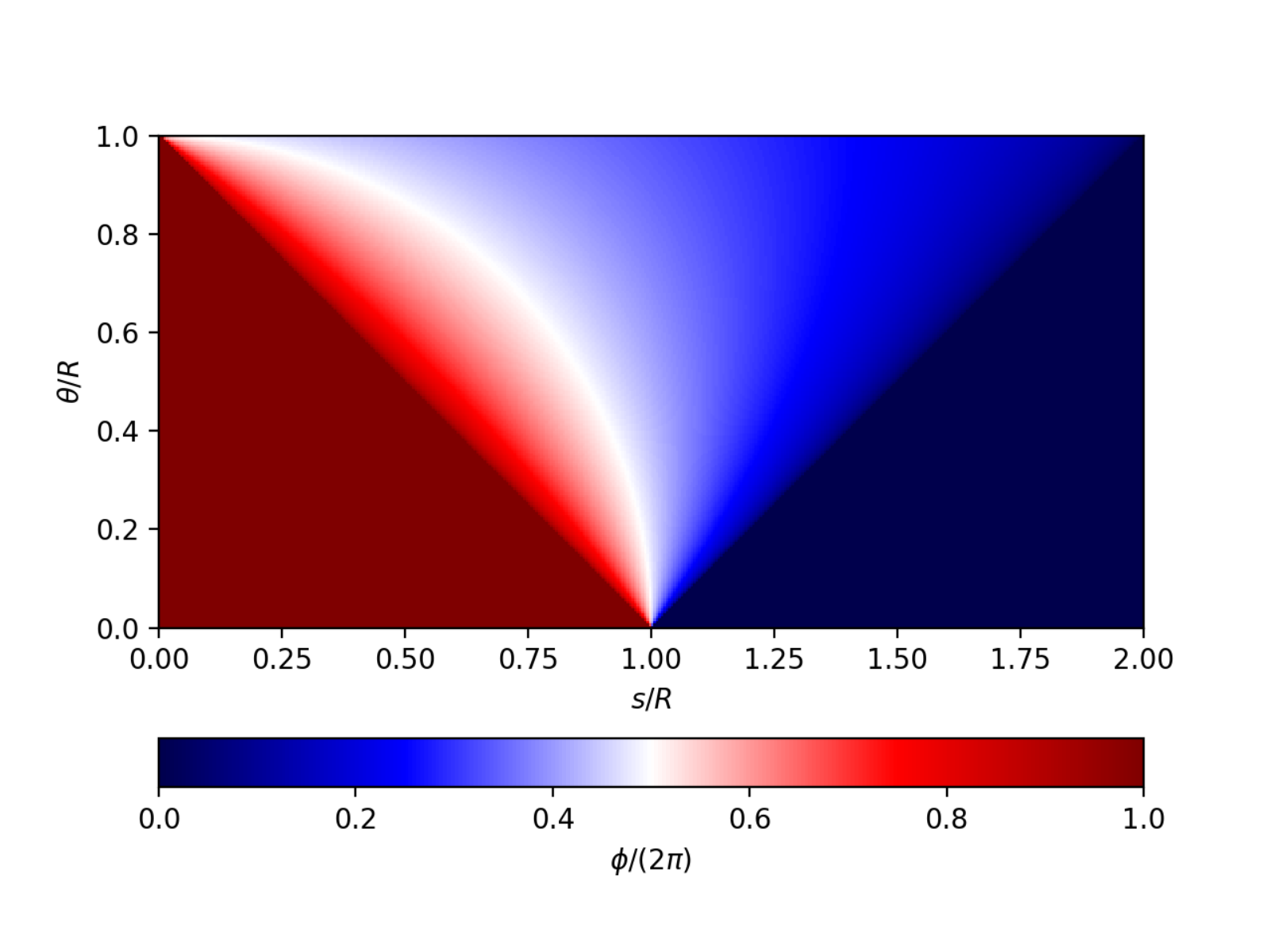} & \includegraphics[width=0.5\linewidth]{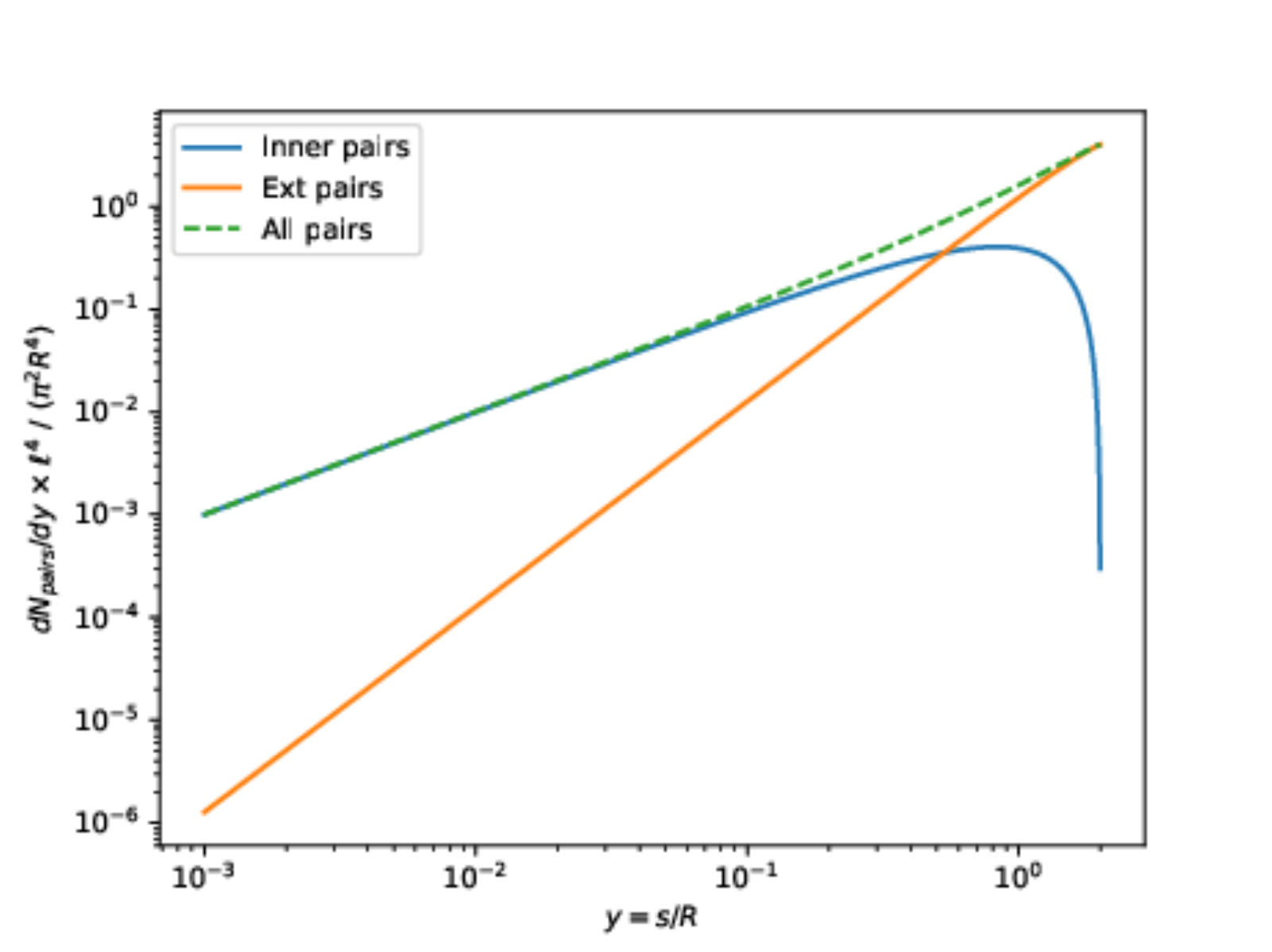}
    \end{tabular}
    \caption{\emph{Left: } two-dimensional representation of the function $\phi_{\sepa}(\vec \theta)$ for various values of the separation $\sepa$ and the position $\theta$ in a circular field of view of radius $R$. \emph{Right:} scaling of $N_p(\sepa)$, the number of pairs of points separated by a distance $\sepa$ within a circular field-of-view of radius $R$. "Inner" concerns those pairs integrally contained within the circular domain, while "Ext" concerns those pairs with only one end within the domain. "All" is the sum of the two numbers. The curves actually show $\delta(\sepa N_p(s))/\delta \sepa$, that is the differential number of pairs per interval of $\sepa$ expressed in units of the radius $R$. Here $\ell \ll R$ is the side length of an elementary pixel, so that the total number of pixels in the circle is $\pi R^2/\ell^2$.}
    \label{fig:npairs_scaling}
\end{figure}

Estimating $\phi$ is easy under the assumption of a circular analysis region of radius $R$. We find the following expressions, graphically represented in Fig.~\ref{fig:npairs_scaling}, left:
\[
\phi_{\sepa}(\vec \theta) = \left\{
	\begin{array}{cl}
	2 \pi 			& 		\mathrm{if \ } \theta < R-\sepa \mathrm{\ and \ } \sepa < R	\\
	0 					&	\mathrm{if \ } \theta < \sepa -R \mathrm{\ and \ } \sepa > R	\\
	2 \mathrm{Arccos}\left( \frac{\theta^2 + \sepa^2 -R^2}{2 \theta \sepa} \right)	&	\mathrm{if \ } (R-\sepa < \theta < R \mathrm{\ and \ } \sepa < R) \mathrm{\ or \ } (\sepa - R < \theta < R  \mathrm{\ and \ } \sepa > R)\\
	\end{array}\right.
\]

We therefore rewrite $N_p^{in}(\sepa) = \sepa^2 F(R/\sepa)$ and $N_p^{out}(\sepa) = \sepa^2 G(R/\sepa)$ with:
\[
F(x) = \left\{
    \begin{array}{cl}
    \pi^2 (x-1)^2 + 2 \pi \int_{x-1}^{x} \gamma(u;x) u \dd u & \mathrm{if \ } x>1 \\
2 \pi \int_{1-x}^x  \gamma(u;x) u \dd u & \mathrm{if \ } 0.5 < x < 1 \\
    \end{array}\right.
\]
and
\[
G(x) = \left\{
    \begin{array}{cl}
    4 \pi \int_{x-1}^{x} \left[ \pi - \gamma(u;x) \right] u \dd u & \mathrm{if \ } x>1 \\
	2 \pi^2 (1-x)^2 + 4 \pi \int_{1-x}^x \left[ \pi - \gamma(u;x) \right] u \dd u & \mathrm{if \ } 0.5 < x < 1 \\
    \end{array}\right.
\]
having introduced $\gamma(u;x) = \mathrm{Arccos}\left( \frac{u^2-x^2+1}{2 u} \right)$.

The expressions for the number of pairs as a function of the separation distance are represented on Fig.~\ref{fig:npairs_scaling}. As expected, the number of extra pairs is negligible for small pair separations. It equals the number of regular ("inner") pairs for $\sepa \simeq 0.5 R$ and becomes dominant past this value.

\begin{figure}
    \centering
  \includegraphics[width=0.9\linewidth]{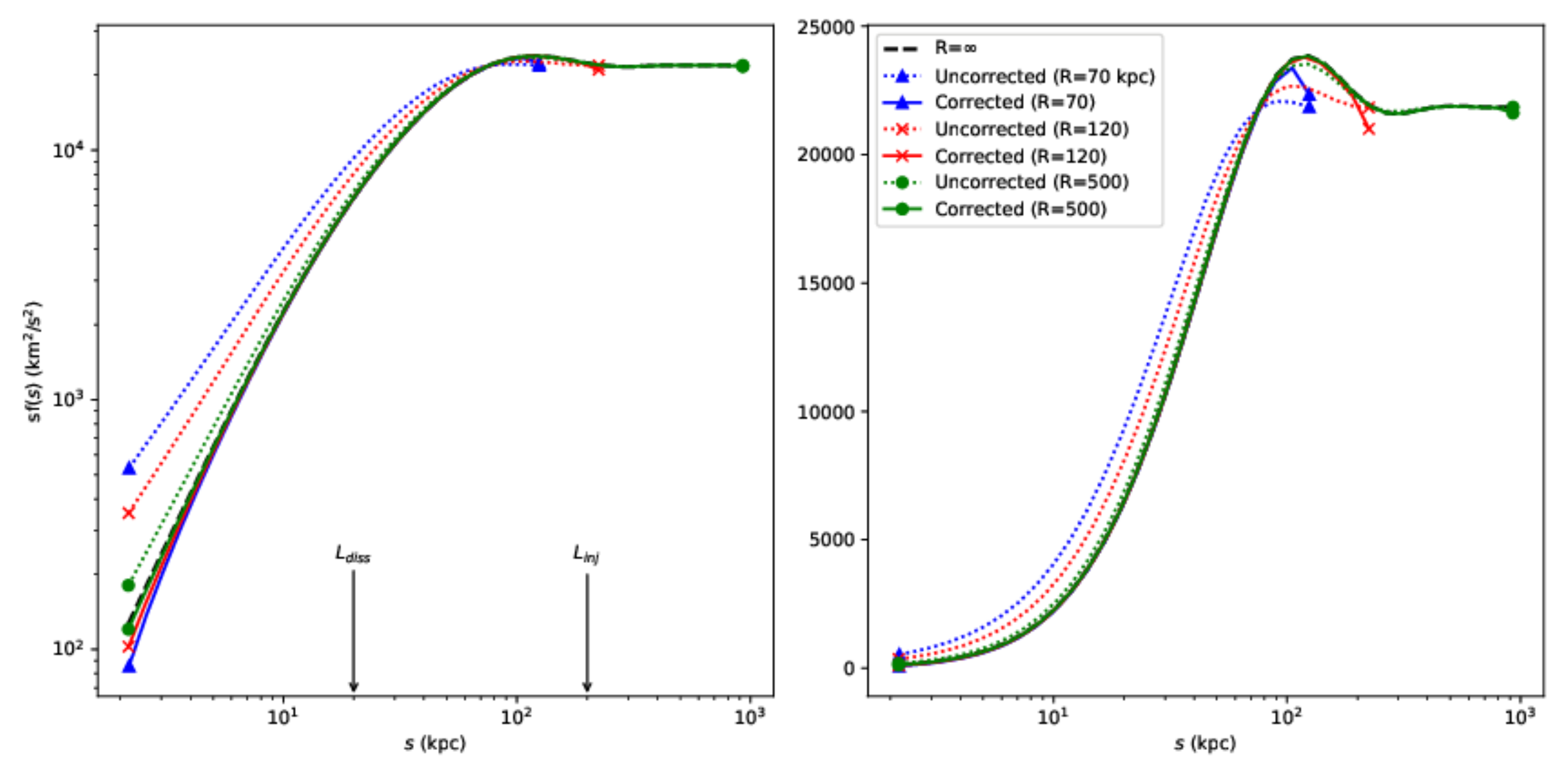}     
  \caption{Impact of finite-size corrections on the mean structure functions computed according to our model. The domain of analysis is a circle of radius $R$. Both panels show the same data, in logarithmic or linear scales. The emissivity model is of type \emph{Xbeta} with a core radius $r_c=400$~kpc. The turbulent power spectrum has injection and dissipation scales $L_{inj}$ and $L_{diss}$ respectively. The thick dashed line is barely visible and shows the exact result obtained assuming an infinitely extended analysis domain. Points at large separations $\sepa$ are subject to slight numerical instabilities.}
    \label{fig:fov_correction_sf}
\end{figure}

Finally, we note that for the flat emissivity field ($\rho(x, \vec \theta) = \epsilon(x)/F$) we have $|\widetilde{\chi^{\vec \theta}}|^2 = P_{\epsilon}/F^2$ and then:
\begin{equation}\label{eq:sf_average_plane_corrected}
\mathrm{sf}^\mathrm{corr}(\sepa) =  \left( \frac{N_p^{ext}(\sepa)}{2 N_p^{in}(\sepa)} + 1 \right) \mathrm{sf}(\sepa) - \frac{N_p^{ext}(\sepa)}{N_p^{in}(\sepa)} \times \left( \frac{\omega}{2\pi} \right)^2\sum_{\vec \xi} P_{2D}^{\infty} (\vec \xi)
\end{equation}
Since in this case the projected velocity field is stationary and isotropic, it may be easier and more exact to compute the mean structure function using $P_{2D}^{\infty}$ in Eq.~\ref{eq:sf_average}, instead of involving $P_{2D}$ and applying this correction formula. This property is used to check the validity of the correction formula in Eq.~\ref{eq:sf_average_plane_corrected}. Figure~\ref{fig:fov_correction_sf} shows the result of our calculation for a flat emissivity field with core radius $r_c=400$~kpc and a turbulent power spectrum with injection scale $L_{inj} = 10 L_{diss} = 200$~kpc. The domain of analysis is circular of radius $R=70, 120$ or 500~kpc. The result for an unbounded domain is also shown.
For small analysis domains ($R=70$~kpc) the uncorrected formula induces discrepancies at small separations, due to the high-pass behaviour of the mask $\mathcal{A}$. As the field-of-view increases ($R=500$~kpc), border effect become negligible and all structure functions match the exact one.
Finally, as $N_p^{in}$ approaches zero for separation length of size $\sepa=2R$, the correction formula becomes numerically unstable at large separation lengths.

%%%%%%%%%%%%%%%%%%%%%%%%
	\section{Structure function from pixelized and/or filtered data}
		\label{app:filter_field}

\begin{figure*}
    \centering
    	\includegraphics[width=\linewidth]{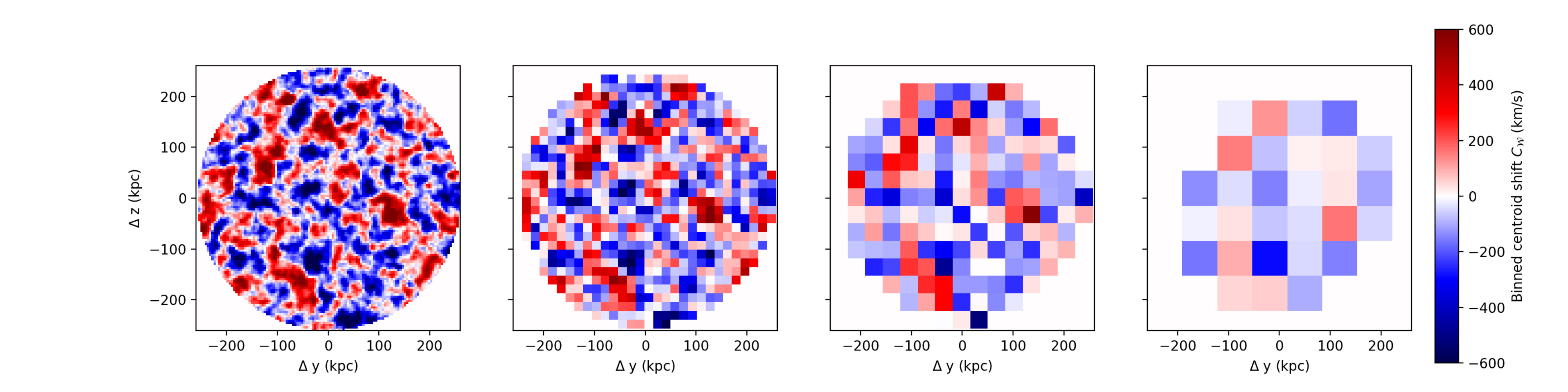}
   	\caption{Effect of a larger pixelization when computing the structure function. Pixel size ranges from $\ell=4,17,34,69$~kpc (from left to right). All four panels represent the same projected velocity field (injection scale at 100~kpc) in a galaxy cluster represented by a \emph{Xbeta} model of core-radius 21~kpc. In general, larger pixels reduce the power in the 2-dimensional velocity fluctuations and roughly act like a smoothing convolution filter on the high-resolution centroid map.}
    \label{fig:cmap_binning}
\end{figure*}

Previous derivations assume that the centroid shift can be measured along every line of sight. In general, real datasets are convolved by an instrumental point-spread function and a pixel design is effectively grouping line-of-sights within a single spectral line measurement. Both processes are formally close to each other, since pixelization along a regular grid can be reformulated as a top-hat filtering followed by the selection of points at the centre of each pixel (Fig.~\ref{fig:cmap_binning}).

We define a new map $D(\vec \theta)$ as:
\begin{equation}
D = \frac{\mathcal{F}_{\ell} * \left(FC\right)}{\mathcal{F}_{\ell} * F}
\end{equation}
where $*$ represents the usual convolution product, $F(\vec \theta)$ and $C(\vec \theta)$ are respectively the flux and centroid maps as defined in Sect.~\ref{sect:3dfield}. The filter $\mathcal{F}_{\ell}$ may represent the instrumental point-spread function, or the pixel window function (previously noted $\mathcal{W}$) or a combination of the two, we assume its characteristic scale is $\ell$ (e.g.~instrument FWHM, pixel size, etc.) and it is normalized to 1 by integrating over all values of $\vec \theta$.
It is clear that $D(\vec \theta)$ is the value of the centroid shift measured after the filtering process, resulting from a flux-weighted average of individual centroid shifts.

This formula reduces to $D(\vec \theta) \simeq C(\vec \theta)$ for components of $C$ varying on scales much larger than $\ell$ (equivalently, for very sharp filters). For components of $C$ oscillating on tiny scales (much smaller than the filter size) we find $D \simeq 0$: as expected the pixelization or filtering process suppresses information on small scales.

\begin{figure}
    \centering
    \includegraphics[width=0.5\linewidth]{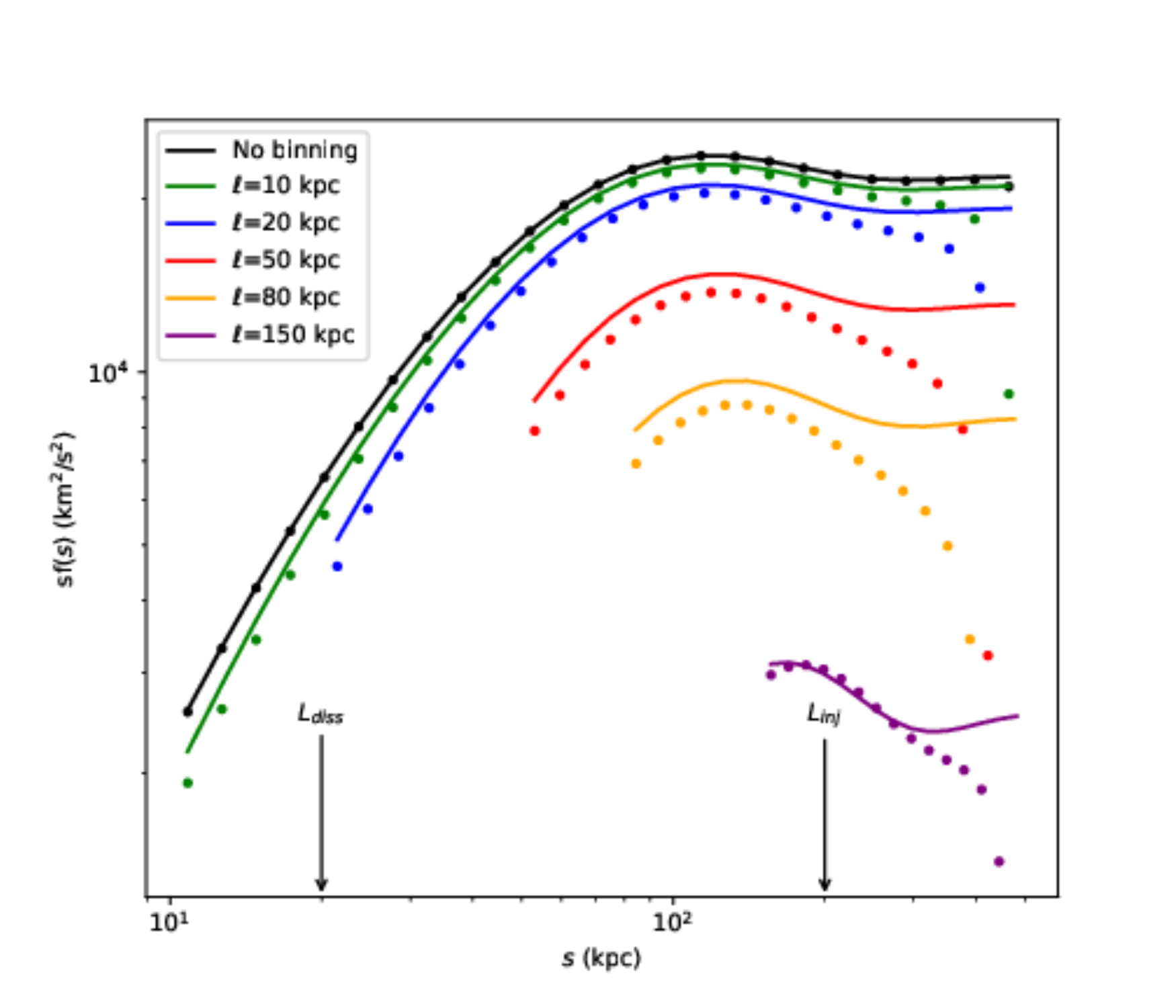}
    \caption{Analytical implementation of binning with pixels of size $\ell$ in the computation of the average structure function. The turbulent velocity field and the cluster emissivity of type \emph{Xbeta} are both identical to Fig.~\ref{fig:fov_correction_sf}. Plain lines show exact results using the stationary, unbounded ($R=\infty$) velocity field, effectively replacing $P_{2D}$ by $P_{\ell} P_{2D}^{\infty}$ in Eq.~\ref{eq:sf_average}. Dots are obtained by combining the correction formula Eq.~\ref{eq:sf_average_plane_corrected} for a circular domain of radius $R=250$~kpc, with the prescription $P_{2D} \rightarrow P_{\ell} P_{2D}$. Since the latter is only valid for slowly-varying flux maps, it fails at reproducing the true structure function if pixels have sizeable length with respect to $R$.}
    \label{fig:sf_corr_bin_limitation}
\end{figure}

A useful derivation can be carried out in case of a smooth flux map, varying on scales much larger than the filter size $\ell$. Then $F$ can be considered constant in the convolution products and one obtains: $D \simeq \mathcal{F}_{\ell} * C$.
All previous calculations now must incorporate this convolution product. For instance the following replacement takes place:
\[
C_{\vec \theta_0, \vec r} \rightarrow \int \dd \vec \mu \mathcal{F}_{\ell}(\vec \mu) e^{i \omega \vec \mu \cdot \vec \xi} C_{\vec \theta_0 - \vec \mu, \vec r}
\]
The calculation steps are similar to previous case, thanks to permutations of the integrals over $\vec \mu$ and other integrals. It leads to the expected value of the structure function at each $\sepa$:

\begin{align*}
\mathrm{sf}(\sepa) & =  2 \sum_{\kthree} P_{3D}(k)  \int \dd \vec \xi^{\prime} P_{\rho}(k_x,\vec \xi^{\prime}) P_{\ell}(\vec \xi + \vec \xi^{\prime}) \left[ 1 - J_0\left(\left| \vec \xi + \vec \xi^{\prime} \right| \omega \sepa \right) \right] \\
	& = 2 \int \left[ 1 - J_0\left( \omega \left| \vec \xi \right| \sepa \right)\right] P_{D}(\vec \xi) \dd \vec \xi
\end{align*}

where $P_D = P_{\ell} P_{2D}$ is the power-spectrum of the map $D(\vec \theta)$ and $P_{\ell}$ is the power-spectrum of the filter. Therefore, in the case of small filter sizes (relative to the flux variation scale) it is legitimate to replace in Eq.~\ref{eq:sf_average} the power-spectrum of the centroid map, $P_{2D}$, by the power-spectrum of the filtered centroid map, $P_{D}$. In case of larger pixels, this is generally no longer valid. This is critical in presence of a finite domain of analysis of size comparable to the pixel size, since then border effects must be treated more carefully. Figure~\ref{fig:sf_corr_bin_limitation} shows the result of applying the simple prescription $P_{2D} \rightarrow P_{\ell} P_{2D}$ to Eq.~\ref{eq:sf_average_plane_corrected}, with a similar parametric setup as in Fig.~\ref{fig:fov_correction_sf}. Since in this case the velocity field is stationary, we also have an exact computation of the structure function obtained by neglecting the finite-size domain, i.e.~by using $P_{\ell} P_{2D}^{\infty}$ in Eq.~\ref{eq:sf_average}. It is then obvious that, strictly speaking, the correction formula in Eq.~\ref{eq:sf_average_plane_corrected} is valid only for unbinned data.

Finally, in addition to this 'smoothing' effect, pixelization induces a discretization effect, or 'aliasing'. We do not develop a calculation for this effect. For a given 2-dimensional frequency $\omega \vec \xi$, aliasing arises for very specific combinations of separations $\sepa$ and pixel sizes, matching integer multiples of the associated spatial scales. It therefore strongly depends on the exact definition of the pixel grid. Since the power-spectrum is continuous in the inertial range, aliasing effects are smoothly distributed across the range of separations $\sepa$ between the injection and dissipation scales. It is expected to mostly affect the structure function $SF(\sepa)$ at separations close to the dissipation and the injection scales, where significant discontinuities show up in the power-spectrum.

%%%%%%%%%%%%%%%%%%%%%%%%
    \section{Derivation of the variance of the structure function}
		\label{app:sf_covariance}

		\subsection{General expression}

Following similar notations, for a given pair indexed by $i$, we write $n_i = N_p(\sepa_i)$ and we introduce:
\[
U_i = \left| C\left(\vec \theta_i + \vec r_i\right) - C\left(\vec \theta_i \right) \right|^2 = \sum_{\kthree, \kpthree} V_{\kthree} V_{\kpthree} e^{i \omega (\vec \xi+\vec \xi^{\prime}) \cdot \vec \theta_i} C_{\vec \theta_i, \vec r_i} (\kthree) C_{\vec \theta_i, \vec r_i}(\kpthree)
\]

We note in particular that:
\begin{equation}
\langle U_i \rangle = \sum_{\kthree} P_{3D}(k) \left| C_{\vec \theta_i, \vec r_i}(\kthree) \right| ^2
\end{equation}

Even in case of an infinitely dense grid of pixels, there is a source of uncertainty arising from the stochastic nature of the velocity field itself. To see this more clearly, we compute:
\[
\left\langle SF(\sepa_i) SF(\sepa_j) \right\rangle = \left \langle \frac{1}{n_i n_j} \int_{\vec \theta_i, \vec r_i, \vec \theta_j, \vec r_j} U_i U_j \right\rangle = \frac{1}{n_i n_j} \int_{\vec \theta_i, \vec r_i, \vec \theta_j, \vec r_j} \langle U_i U_j \rangle
\]

Neglecting border effects, $n_i = n_j \propto 2\pi\fovarea$.

\[
\langle U_i U_j \rangle = \sum_{\kthree, \kpthree, \lthree, \lpthree} \langle V_{\kthree} V_{\kpthree} V_{\lthree} V_{\lpthree}\rangle e^{i \omega (\vec \xi+\vec \xi^{\prime}) \cdot \vec \theta_i} e^{i \omega (\vec \chi+\vec \chi^{\prime}) \cdot \vec \theta_j} C_{\vec \theta_i, \vec r_i}(k_x,\vec \xi) C_{\vec \theta_i, \vec r_i}(k_x^{\prime},\vec \xi^{\prime}) C_{\vec \theta_j, \vec r_j}(l_x,\vec \chi) C_{\vec \theta_j, \vec r_j}(l_x^{\prime},\vec \chi^{\prime})
\]

We stress that $|\vec r_i| = \sepa_i$ and that $\sepa_i$ and $\sepa_j$ are not necessarily equal. Integration runs over all pairs indexed by $i$ and $j$ within the field. Similarly to the 1-dimensional case we assume that:
\[
    \langle V_{\jthree} V_{\kthree} V_{\lthree} V_{\mthree} \rangle = \left\{
    \begin{array}{cl}
    P_{3D}(k) P_{3D}(l) & \mathrm{if \ } (\kthree=- \jthree) ; (\lthree=- \mthree); (\kthree \neq \pm \lthree) \, \{A\} \\
    P_{3D}(k) P_{3D}(j) & \mathrm{if \ } (\kthree=- \lthree) ; (\jthree=- \mthree); (\kthree \neq \pm \jthree) \, \{B\} \\
    P_{3D}(k) P_{3D}(j) & \mathrm{if \ } (\kthree=- \mthree) ; (\jthree=- \lthree); (\kthree \neq \pm \jthree) \, \{C\} \\
    \langle | V_{\kthree} |^4 \rangle  & \mathrm{if \ } (\kthree=- \jthree = \lthree = - \mthree) \{D\} \, \\
    \langle | V_{\kthree} |^4 \rangle  & \mathrm{if \ } (\kthree=- \jthree = -\lthree = \mthree) \{E\} \, \\
    \langle | V_{\kthree} |^4 \rangle  & \mathrm{if \ } (\kthree= \jthree = - \lthree = - \mthree) \{F\} \, \\
    0 & \mathrm{else}\\
    \end{array} \right.
\]

The decomposition of the product in brackets therefore involves six terms $b_A, b_B, b_C, b_D, b_E, b_F$ with $b_B=b_C$ and $b_D=b_E$.

\paragraph{Computation of $b_A$:} It corresponds to a case $\kthree = -\kpthree$, $\lthree =-\lpthree$ and $\kthree \neq \pm \lthree$.
It writes:
\begin{align*}
b_A  & = (n_i n_j)^{-1} \sum_{\kthree \neq \pm \lthree} P_{3D}(k) P_{3D}(l) \int \left| C_{\vec \theta_i, \vec r_i}(\kthree) \right|^2 \left| C_{\vec \theta_j, \vec r_j}(\lthree) \right|^2 \\
	 & = (n_i n_j)^{-1} \left[ \sum_{\kthree} P_{3D}(k) I_{\sepa_i}(\kthree) \right] \left[ \sum_{\lthree} P_{3D}(l)  I_{\sepa_j}(\lthree) \right] - 2 (n_i n_j)^{-1} \sum_{\kthree}P_{3D}(k)^2 I_{\sepa_i}(\kthree) I_{\sepa_j}(\kthree)
\end{align*}
with $I_{\sepa}$ already defined in App.~\ref{app:sf_general}. Therefore, the first term is simply $\mathrm{sf}(\sepa_i) \mathrm{sf}(\sepa_j)$.

\paragraph{Computation of $b_B$:} It corresponds to a case $\kthree = -\lthree$, $\kpthree =-\lpthree$ and $\kthree \neq \pm \kpthree$. It writes:
\begin{align*}
b_B & = (n_i n_j)^{-1} \sum_{\kthree \neq \pm \kpthree} P_{3D}(k) P_{3D}(k^{\prime}) \int e^{i \omega(\vec \xi + \vec \xi^{\prime}) \cdot (\vec \theta_i - \vec \theta_j)} C_{\vec \theta_i, \vec r_i}(k_x,\vec \xi) C_{\vec \theta_i, \vec r_i}(k_x^{\prime},\vec \xi^{\prime}) C_{\vec \theta_j, \vec r_j}^*(k_x,\vec \xi) C_{\vec \theta_j, \vec r_j}^*(k_x^{\prime},\vec \xi^{\prime}) \\
	& = (n_i n_j)^{-1} \left( \sum_{\kthree, \lthree} P_{3D}(k) P_{3D}(l) J_{\sepa_i}(\kthree, \lthree) J_{\sepa_j}^*(\kthree, \lthree) - \sum_{\kthree} P_{3D}(k)^2 J_{\sepa_i}(\kthree, \kthree) J_{\sepa_j}^*(\kthree, \kthree) - \sum_{\kthree} P_{3D}(k)^2 I_{\sepa_i}(\kthree) I_{\sepa_j} (\kthree) \right)
\end{align*}

This involves the function:
\[
J_{\sepa}(\kthree, \lthree) = \int_{\vec \theta, |\vec r|=\sepa}e^{i \omega (\vec \xi+\vec \chi) \cdot \vec \theta} C_{\vec \theta,\vec r}(\kthree) C_{\vec \theta,\vec r}(\lthree)  \dd \vec \theta \dd \vec r 
\]
By developing the expression of $C_{\vec \theta, \vec r}$ and using Parseval's theorem, we find:
\begin{equation}
\label{eq:j_delta}
J_{\sepa}(\kthree,\lthree) = 4\pi \left(\frac{\omega}{2 \pi}\right)^2 \int  \widetilde{\rho}(k_x, \vec \xi+\vec \kappa) \widetilde{\rho}(l_x, \vec \chi - \vec \kappa) \left( 1- J_0(\omega \kappa \sepa) \right) \dd \vec \kappa
\end{equation}

\paragraph{Computation of $b_D$:} It corresponds to $\kthree = -\kpthree = \lthree = -\lpthree$. It writes:
\[
b_D = (n_i n_j)^{-1} \sum_{\kthree} \left\langle \left| V_{\kthree} \right|^4 \right\rangle I_{\sepa_i}(\kthree) I_{\sepa_j}(\kthree)
\]

\paragraph{Computation of $b_F$:} It corresponds to $\kthree = \kpthree = -\lthree = -\lpthree$. It writes:
\[
b_F = (n_i n_j)^{-1} \sum_{\kthree} \left\langle \left| V_{\kthree} \right|^4 \right\rangle J_{\sepa_i}(\kthree, \kthree) J_{\sepa_j}^*(\kthree, \kthree)
\]

Reassembling all terms together we obtain the covariance term defined by:
\[
 \Sigma_{ij}  = \left\langle SF(\sepa_i) SF(\sepa_j) \right\rangle - \left\langle SF(\sepa_i) \right\rangle \left\langle SF(\sepa_j) \right\rangle
\]
and which writes:
\begin{equation}
\label{eq:covar_sf}
\Sigma_{ij} = \frac{1}{(2 \pi \fovarea)^2} \left[ 2 \sum_{\kthree, \lthree} P_{3D}(k) P_{3D}(l)   J_{\sepa_i}(\kthree, \lthree) J_{\sepa_j}^*(\kthree, \lthree) - \sum_{\kthree} R_{\kthree} \times \left\{ 2 I_{\sepa_i}(\kthree) I_{\sepa_j} (\kthree) + J_{\sepa_i}(\kthree, \kthree) J_{\sepa_j}^*(\kthree, \kthree) \right\} \right]
\end{equation}

The expressions for $I$ and $J$ are given by equations~\ref{eq:i_delta} and~\ref{eq:j_delta} respectively. Notably, the second term within brackets vanishes for Rayleigh-distributed coefficients.
In principle, finite-size corrections must also apply, similarly as for the expected value of the structure function. We do not provide such corrections here and keep in mind that our variance estimate neglects border effects.

		\subsection{Case of an emissivity independent of the line-of-sight}

The expression above can be further simplified if the emissivity is independent of the line-of-sight direction. Introducing $\widehat{\mathcal{A}}$ the Fourier transform of the analysis region and $P_{\mathcal{A}}=|\widehat{\mathcal{A}}|^2$ its power-spectrum, we obtain $\widetilde{\rho}(k_x,\vec \xi) = \widetilde{\epsilon}(k_x) \widehat{\mathcal{A}}(\vec \xi)/F$. We define the following functions, whose principal interest resides in the fact that they only depend on the definition of the analysis region and can be precomputed numerically for any given instrumental field-of-view:
\begin{align*}
\mathcal{U}_{\mathcal{A}}(\vec \xi, \vec \chi ; \sepa) & = K \int \widehat{\mathcal{A}}(\vec \xi + \vec \kappa) \widehat{\mathcal{A}}(\vec \chi - \vec \kappa) \left(1-J_0(\omega \kappa \sepa)\right) \dd \vec \kappa \\
\mathcal{T}_{\mathcal{A}}(\vec \xi ; \sepa ) & = \mathcal{U}_{\mathcal{A}}(\vec \xi, -\vec \xi ; \sepa) = K \int P_{\mathcal{A}}(\vec \kappa) \left( 1- J_0(\omega |\vec \xi + \vec \kappa| \sepa ) \right) \dd \vec\kappa
\end{align*}
The normalization constant is $K=(\omega/2\pi)^2/\fovarea$ throughout this paper.

Using these functions the variance then writes:
\begin{multline}
\label{eq:varsf_xbeta}
\Sigma_{ij} = 8 \left(\frac{\omega}{2\pi}\right)^4 \sum_{\vec \xi, \vec \chi} P_{2D}^{\infty}(\xi) P_{2D}^{\infty}(\chi) \mathcal{U}_{\mathcal{A}}(\vec \xi, \vec \chi ; \sepa_i) \mathcal{U}_{\mathcal{A}}^{*}(\vec \xi, \vec \chi ; \sepa_j) \\
- 4 \left(\frac{\omega}{2\pi}\right)^4 \sum_{\vec \xi} Q_{2D}^{\infty}(\xi)^2 \left\{ 2 \mathcal{T}_{\mathcal{A}}(\vec \xi ; \sepa_i ) \mathcal{T}_{\mathcal{A}}(\vec \xi ; \sepa_j ) + \mathcal{U}_{\mathcal{A}}(\vec \xi, \vec \xi ; \sepa_i) \mathcal{U}_{\mathcal{A}}^{*}(\vec \xi, \vec \xi ; \sepa_j) \right\}
\end{multline}
where we introduced $P_{2D}^{\infty}$ the 2D power-spectrum of the centroid map over an infinitely extended domain (App.~\ref{app:p2d_general}) and
\[
Q^{\infty}_{2D}(\xi)^2 = \left(\frac{2\pi}{\omega} \right)^4 \sum_{k_x}  \frac{P_{\epsilon}(k_x)^2}{F^4} R_{k_x,\vec \xi}
\]
If the moduli are Rayleigh-distributed it is evident that $Q_{2D}=0$ and the expression for the variance (Eq.~\ref{eq:varsf_xbeta}) depends only on the 2D power-spectrum of the velocity. As already noted, the terms encapsulating the field-of-view geometry ($\mathcal{U}_{\mathcal{A}}, \mathcal{T}_{\mathcal{A}}$) are factored out from the emissivity and turbulence part ($P_{2D}, Q_{2D}$).

Further simplification can be made when the analysis region is extremely wide compared to the separations $\sepa$ and to the largest fluctuation scale of the velocity map (still assuming a constant emissivity on sky). The limit $\mathcal{A} \rightarrow \infty$ then applies and $\widehat{\mathcal{A}}$ becomes a strongly peaked function around 0. The above functions rewrite:
\[
\mathcal{U}_{\mathcal{A}}(\vec \xi, \vec \chi ; \sepa) \simeq K \left(1-J_0(\omega \xi \sepa)\right)  \int \widehat{\mathcal{A}}(\vec \xi + \vec \kappa) \widehat{\mathcal{A}}(\vec \chi - \vec \kappa)  \dd \vec \kappa = K \left(1-J_0(\omega \xi \sepa)\right) \left\{ (\widehat{\mathcal{A}} * \widehat{\mathcal{A}} )  (\vec \xi + \vec \chi)\right\}
\]

The sign $*$ indicates the convolution product. Since $\mathcal{A}^2 = \mathcal{A}$ we obtain:
\[
\mathcal{U}_{\mathcal{A}}(\vec \xi, \vec \chi ; \sepa) \simeq K \left(1-J_0(\omega \xi \sepa)\right)  \left( \frac{2\pi}{\omega}\right)^2 \widehat{\mathcal{A}}(\vec \xi+\vec \chi)
\]
This automatically shows that $\mathcal{T}_{\mathcal{A}}(\vec \xi ; \sepa) = 1-J_0(\omega \xi \sepa)$ and $\mathcal{U}_{\mathcal{A}}(\vec \xi,\vec \xi ; \sepa) = 0$.
Moreover:
\begin{align*}
\sum_{\vec \chi}  P_{2D}^{\infty}(\chi) \mathcal{U}_{\mathcal{A}}(\vec \xi, \vec \chi ; \sepa_i) \mathcal{U}^*_{\mathcal{A}}(\vec \xi, \vec \chi ; \sepa_j) & 	\simeq \left( \frac{2 \pi}{\omega} \right)^4  K^2  \left(1-J_0(\omega \xi \sepa_i)\right) \left(1-J_0(\omega \xi \sepa_j)\right) \sum_{\vec \chi} P_{\mathcal{A}}(\vec \chi + \vec \xi) P_{2D}^{\infty}(\chi) \\
& = \left( \frac{2 \pi}{\omega} \right)^2 \fovarea^{-1} P_{2D}^{\infty}(\xi) \left(1-J_0(\omega \xi \sepa_i)\right) \left(1-J_0(\omega \xi \sepa_j)\right) 
\end{align*}

Grouping terms together in Eq.~\ref{eq:varsf_xbeta} leads to Equation~\ref{eq:varsf_xbeta_fovlarge}.

%%%%%%%%%%%%%%%%%%%%%%%%
    \section{Fourier transform of the normalized emissivity field $\rho$ (spherical $\beta$-model)}
    	\label{app:fourier_rho}

We provide useful calculations for the 3-d power-spectrum of the normalised emissivity $\rho$ in a case of a $\beta$-model. We also discuss the limiting case of very extended sources (equivalently, very small field-of-views).
    
    	\subsection{General case}
    
Calculation of the 2-dimensional power-spectrum involves the calculation of $P_{\rho} = | \widetilde{\rho}|^2$ with $\rho(x, \vec \theta) = \emithree(x,\vec \theta)/F(\vec \theta)$. As noted above (App.~\ref{app:p2d_general}), it is compulsory to fill $\rho$ with zeros outside of its domain of definition or outside of the analysis domain. We consider here an arbitrarily large circular analysis domain $\mathcal{A}$ (of radius $R$, centred on the source) to perform the following calculations.

Using the expression for $F(\vec \theta)$ derived in previous section (Eq.~\ref{eq:flux_beta2d}) we obtain:
\[
\rho(x, \vec \theta) = \frac{1}{r_c u_{\beta}} \left( 1+ \frac{\theta^2}{r_c^2} \right)^{3\beta-1/2}{\left( 1+ \frac{x^2+\theta^2}{r_c^2} \right)^{-3\beta}}
\]

\begin{figure}
    \centering
	    \includegraphics[width=\linewidth]{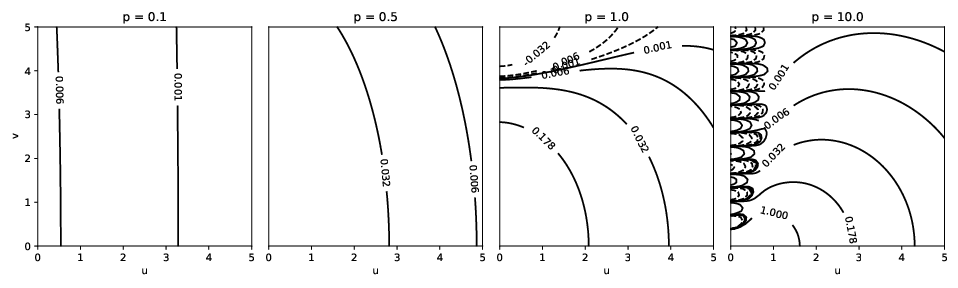}
    \caption{Numerical calculations of $\mathcal{H}_{p;n}(u,v)$ for various values of $p$ and $n=3/2$. Logarithmically spaced contours (identical in all panels, dashed lines for negative values) indicate the value of the function. This function is involved in the calculation of the Fourier transform of the normalized emissivity of a spherical $\beta$-model ($n=3\beta-1/2$) analysed over a concentric circular aperture of radius $p$ times the core radius. The strong oscillatory behaviour is particularly noticeable in the last panel.}
    \label{fig:hpn_abaque}
\end{figure}

The Fourier transform $\widetilde{\rho}(k_x, \vec \xi)$ is calculated in two steps. The integration over the $x$ axis is performed first and its result is already displayed in Eq.~\ref{eq:tf_beta1d}. Integration over the second axis runs for all $\vec \theta \in \mathcal{A}$ and we find:

\[
\widetilde{\rho}(k_x, \vec \xi) =  \frac{2^{5/2-3\beta}  \pi r_c^2 }{\Gamma(3 \beta-1/2)} \times \mathcal{H}_{(R/r_c);(3\beta-1/2)}\left(\omega |k_x| r_c, \omega\xi r_c\right)
\]
which uses the special integral defined below and represented in Fig.~\ref{fig:hpn_abaque} for $n=3/2$ ($\beta=2/3$):
\[
\mathcal{H}_{p;n}(u,v) = \int_{0}^p t J_0(vt) \mathcal{F}_{n}\left(u \sqrt{1+t^2}\right) \dd t
\]
recalling that $\mathcal{F}_n(x) = x^n K_{n}(x)$,

	\subsection{\label{sect:prho:limit}Limit for small analysis domains ($R \ll r_c$)}

If $R/r_c$ is small, those terms in $\sqrt{1+t^2} \simeq 1$ are approximatively constant under the integral and:

\[
P_{\rho} (k_x, \vec \xi) \simeq \frac{2^{3-6\beta}}{ \Gamma(3 \beta-1/2)^2} \left[ \mathcal{F}_{3\beta-1/2}\left( \omega |k_x| r_c \right) \right]^2 \times \left[ 2 \pi R^2 \frac{J_1(\omega \xi R)}{\omega \xi R} \right]^2
\]

The rightmost factor involving the order 1 Bessel function $J_1$ is the power spectrum of a circular pupil of radius $R$. Its integral over $\vec \xi$ equals $\pi R^2 (2\pi/\omega)^2$ and rapidly falls to zero for $\omega |\vec \xi| \gtrsim 3.83 R^{-1}$. 
This result can easily be generalised to an analysis domain of arbitrary shape, introducing its power-spectrum $P_{\mathcal{A}}$:
\begin{equation}
\label{eq:prho_smallA}
P_{\rho}(k_x, \vec \xi) \simeq \frac{P_{\epsilon}(k_x; \theta=0)}{F(0)^2} \times P_{\mathcal{A}}(\vec \xi)
\end{equation}

and we naturally recover the limiting case discussed several times in this paper where the emissivity $\emithree(x, \vec \theta) = \epsilon(x)$ does not depend on the line-of-sight $\vec \theta$ over the analysis domain. Finally, $R (\ll r_c)$ can still be very large and therefore $P_{\mathcal{A}}(\vec \xi) \rightarrow \left( \frac{2 \pi}{\omega}\right)^2 \fovarea \delta(\vec \xi)$: the component of $P_{\rho}$ in the $\vec \xi$ plane can then be seen as a sharp low-pass filter ; this corresponds to the case discussed in previous works \citep[e.g.][]{zuhone2016}.

%%%%%%%%%%%%%%
	\section{Numerical validation results, continued}
	\label{app:valid_200-300}
	
\begin{figure}[h]
    \centering
    \includegraphics[width=\linewidth]{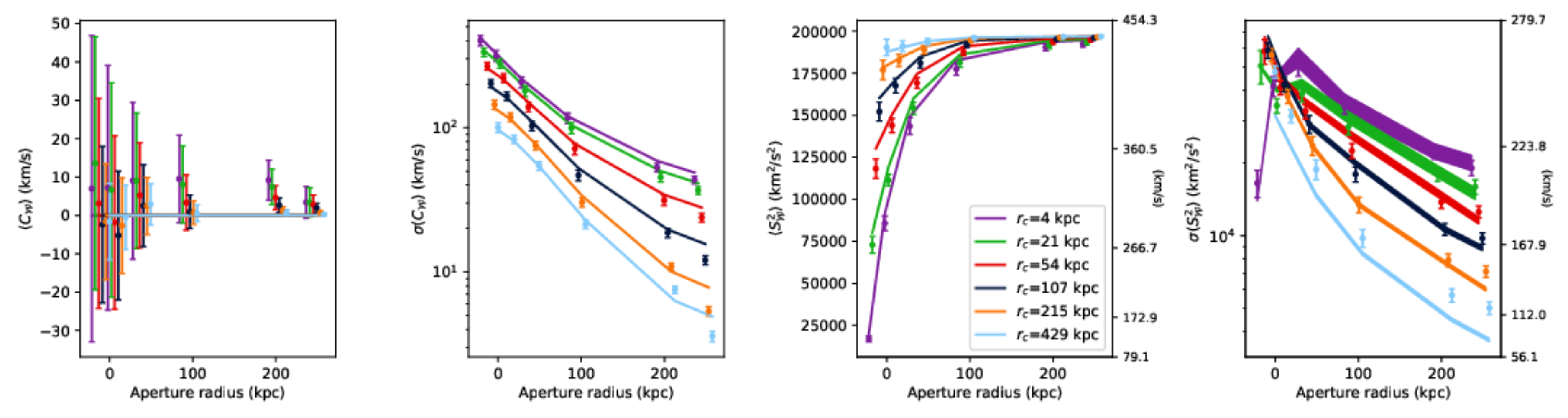}
    \caption{Figure similar to Fig.~\ref{fig:linestat_linj100_Xbeta} for the simulation with injection scale 200~kpc.}
    \label{fig:linestat_linj200_Xbeta}
\end{figure}

\begin{figure}[h]
    \centering
    \includegraphics[width=\linewidth]{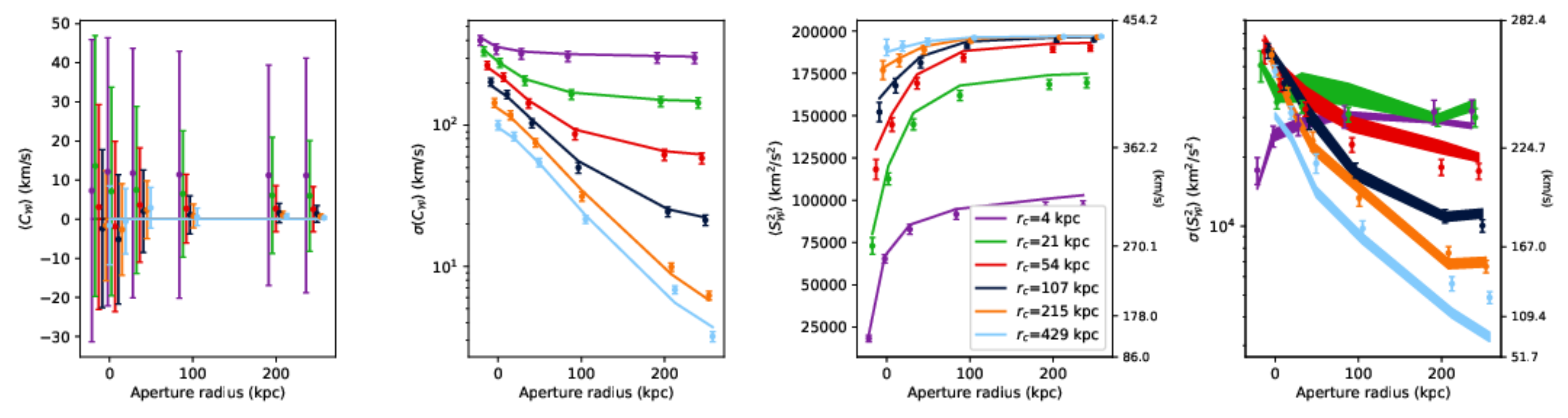}
    \caption{Figure similar to Fig.~\ref{fig:linestat_linj100_beta} for the simulation with injection scale 200~kpc.}
    \label{fig:linestat_linj200_beta}
\end{figure}

\begin{figure}[h]
    \centering
    	\includegraphics[width=\linewidth]{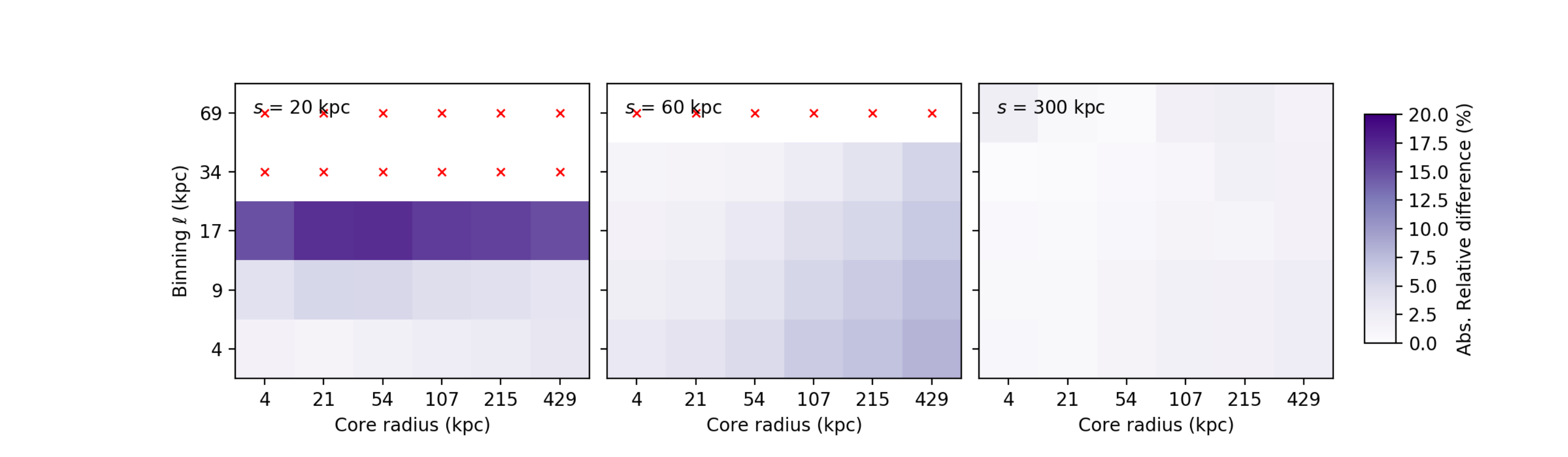}
   	\caption{Figure similar to Fig.~\ref{fig:reldiff_sf_mean} for the simulation with injection scale 200~kpc.}
    \label{fig:reldiff_sf_mean_linj200}
\end{figure}

\begin{figure}[h]
    \centering
    	\includegraphics[width=\linewidth]{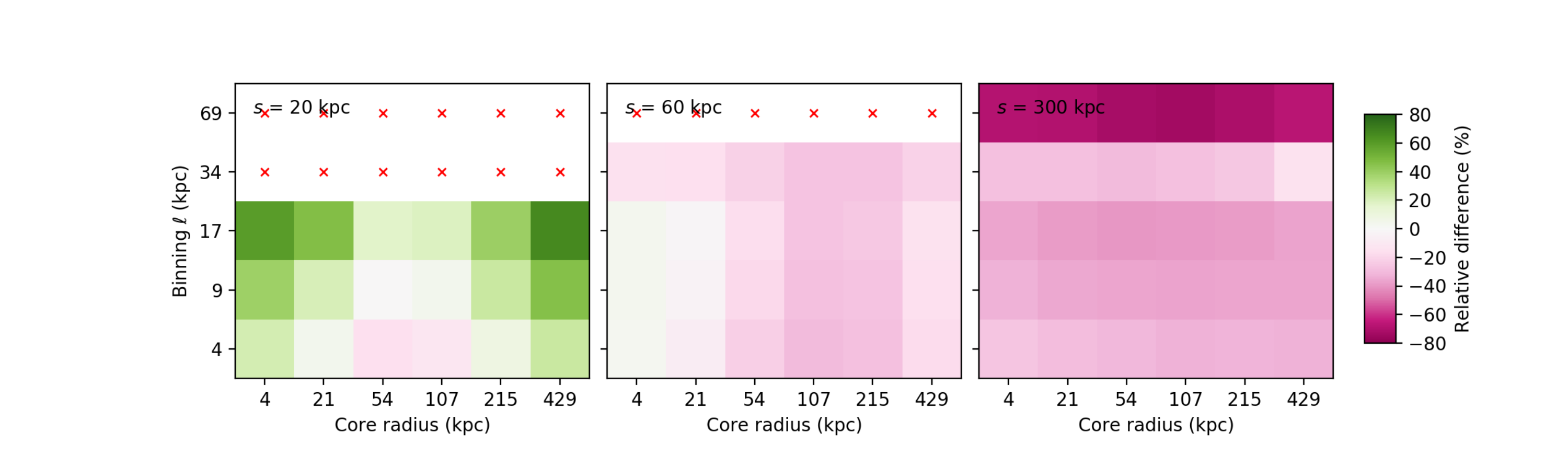}
   	\caption{Figure similar to Fig.~\ref{fig:reldiff_sf_var} for the simulation with injection scale 200~kpc.}
    \label{fig:reldiff_sf_var_linj200}
\end{figure}

\begin{figure}[ht]
    \centering
    \includegraphics[width=\linewidth]{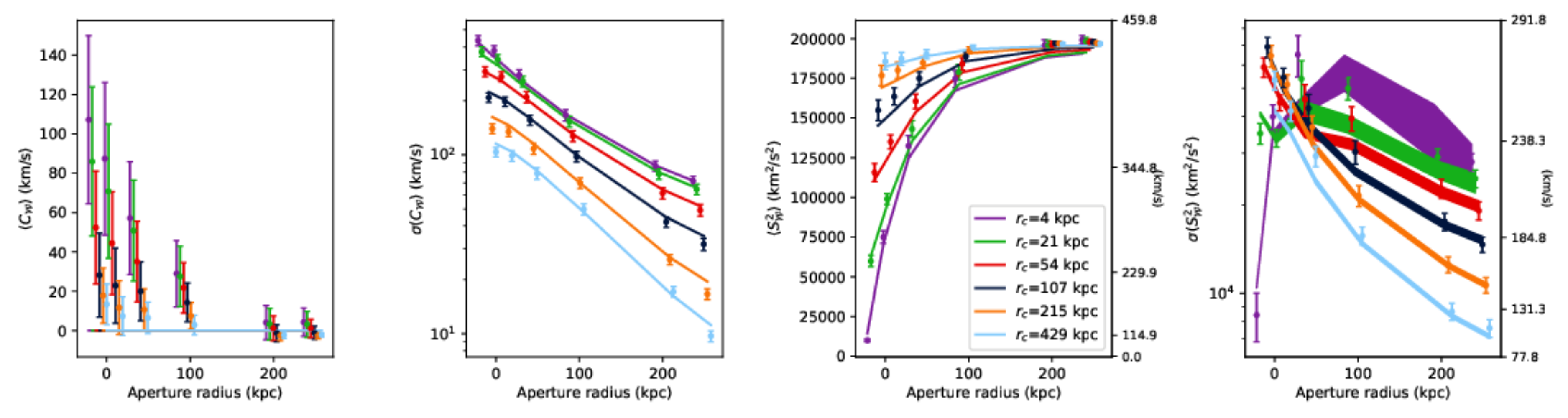}
    \caption{Figure similar to Fig.~\ref{fig:linestat_linj100_Xbeta} for the simulation with injection scale 300~kpc.}
    \label{fig:linestat_linj300_Xbeta}
\end{figure}

\begin{figure}[h]
    \centering
    \includegraphics[width=\linewidth]{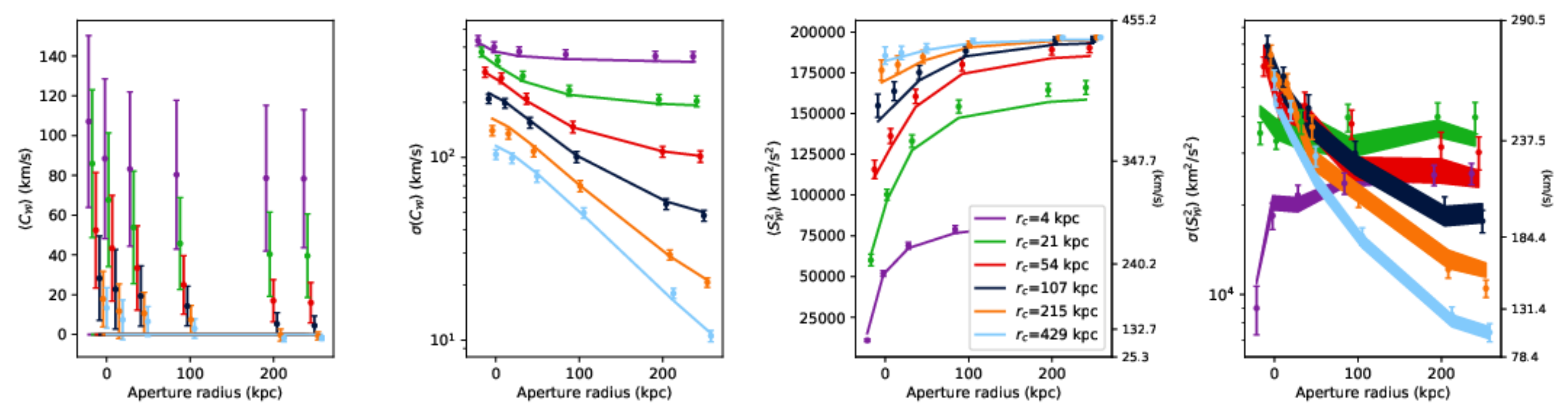}
    \caption{Figure similar to Fig.~\ref{fig:linestat_linj100_beta} for the simulation with injection scale 300~kpc.}
    \label{fig:linestat_linj300_beta}
\end{figure}

\begin{figure}[h]
    \centering
    	\includegraphics[width=\linewidth]{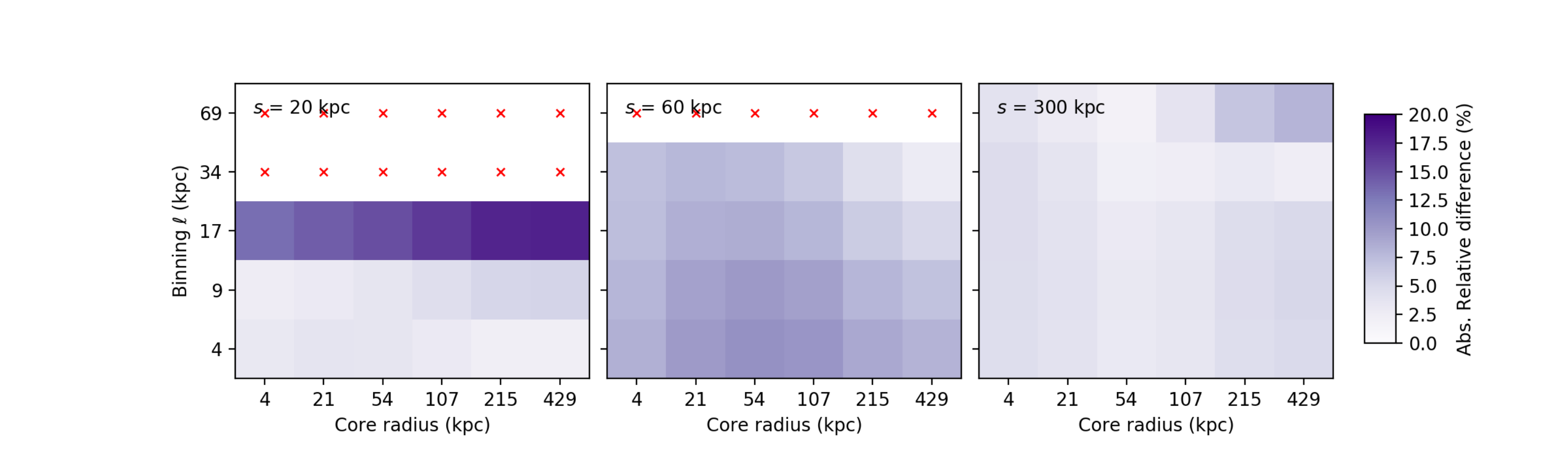}
   	\caption{Figure similar to Fig.~\ref{fig:reldiff_sf_mean} for the simulation with injection scale 300~kpc.}
    \label{fig:reldiff_sf_mean_linj300}
\end{figure}

\begin{figure}[h]
    \centering
    	\includegraphics[width=\linewidth]{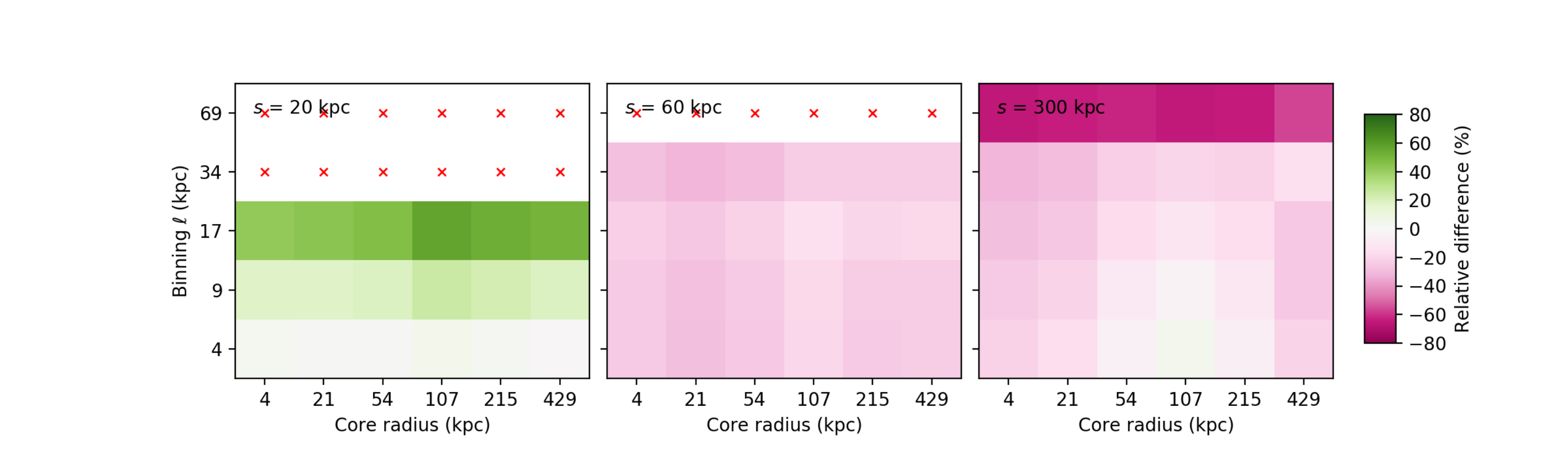}
   	\caption{Figure similar to Fig.~\ref{fig:reldiff_sf_var} for the simulation with injection scale 300~kpc.}
    \label{fig:reldiff_sf_var_linj300}
\end{figure}

We show here the equivalent of Figures~\ref{fig:linestat_linj100_Xbeta}, \ref{fig:linestat_linj100_beta}, \ref{fig:reldiff_sf_mean} and~\ref{fig:reldiff_sf_var} for the two other sets of 100 simulations (Table~\ref{table:simu_char}). All parameters remain the same apart from the injection scale, respetively $L_{inj}=200$~kpc (Figs.~\ref{fig:linestat_linj200_Xbeta}-\ref{fig:reldiff_sf_var_linj200}) and $L_{inj}=300$~kpc  (Figs.~\ref{fig:linestat_linj300_Xbeta}-\ref{fig:reldiff_sf_var_linj300}).
These comparisons still demonstrate a good match between the simulations and analytic results.

%- 200 -

%- 300-

\end{document}